\documentclass[
 aip,
 amsmath,amssymb,
 reprint,longbibliography,
]{revtex4-1}
\usepackage[utf8]{inputenc}
\usepackage[T1]{fontenc}
\usepackage{lmodern}  % polish letter
\usepackage{subcaption}  % letter enumeration
\usepackage{ragged2e} % for the \justifying macro
\DeclareCaptionJustification{justified}{\justifying}
\usepackage{graphicx}
\usepackage{bbold}
\usepackage{cases}
\usepackage{upgreek}
\usepackage{bm}
\usepackage[usenames, dvipsnames]{color}
\usepackage[colorlinks=true, citecolor=Red, linkcolor=RoyalBlue, urlcolor=OliveGreen ]{hyperref}
\usepackage{braket}
\usepackage[normalem]{ulem}
\usepackage{xfrac}

\usepackage{textgreek}
\usepackage{gensymb}
\usepackage{dcolumn}

\usepackage[ruled, linesnumbered, noline]{algorithm2e}

\newcommand \mc[1] { \mathcal{#1} }
\newcommand \bmc[1] { \bm{\mathcal{#1}} }
\newcommand \dd[1]  { \!\textrm d{#1} \,}   % infintesimal
\newcommand \rmm[1]  { \textrm{#1} }

\newcommand \ev[1] { \langle #1 \rangle }
\newcommand \tU {\tau_\mc{U}}

\makeatletter
\def\@email#1#2{%
 \endgroup
 \patchcmd{\titleblock@produce}
  {\frontmatter@RRAPformat}
  {\frontmatter@RRAPformat{\produce@RRAP{*#1\href{mailto:#2}{#2}}}\frontmatter@RRAPformat}
  {}{}
}%
\makeatother
\begin{document}

\preprint{AIP/123-QED}

\title{Short-lived memory in multidimensional spectra encodes full signal evolution}
\author{Thomas Sayer}
\thanks{These authors contributed equally to this work.}
\affiliation{Department of Chemistry, University of Colorado Boulder, Boulder, CO 80309, USA\looseness=-1} 
\affiliation{Department of Chemistry, Durham University, South Road, Durham, DH1 3LE, United Kingdom\looseness=-1} 

\author{Ethan H. Fink}
\thanks{These authors contributed equally to this work.}
\affiliation{Department of Chemistry, University of Colorado Boulder, Boulder, CO 80309, USA\looseness=-1} 

\author{Zachary R. Wiethorn}
\affiliation{Department of Chemistry, University of Colorado Boulder, Boulder, CO 80309, USA\looseness=-1} 
\author{Devin R. Williams}
\affiliation{Department of Chemistry, Colorado State University, Fort Collins, Colorado 80523-1872, USA\looseness=-1} 
\author{Anthony J.~Dominic III}
\affiliation{Department of Chemistry, University of Colorado Boulder, Boulder, CO 80309, USA\looseness=-1} 
\author{Luke Guerrieri}
\affiliation{Department of Chemistry, Colorado State University, Fort Collins, Colorado 80523-1872, USA\looseness=-1} 
\author{Yi Ji}
\affiliation{Department of Chemistry, Massachusetts Institute of Technology, Cambridge, Massachusetts 02139, USA\looseness=-1} 
\author{Veronica Policht}
\affiliation{Department of Physics, Colorado School of Mines, Golden, Colorado 80401, USA\looseness=-1} 
\author{Jennifer Ogilvie}
\affiliation{Department of Physics, University of Ottawa, 150 Louis-Pasteur Pvt., Ottawa, ON K1N 6N5, Canada\looseness=-1} 
\author{Gabriela Schlau-Cohen}
\affiliation{Department of Chemistry, Massachusetts Institute of Technology, Cambridge, Massachusetts 02139, USA\looseness=-1} 
\author{Amber Krummel}
\affiliation{Department of Chemistry, Colorado State University, Fort Collins, Colorado 80523-1872, USA\looseness=-1} 

\author{Andr\'{e}s Montoya-Castillo}
\homepage{Andres.MontoyaCastillo@colorado.edu}
\affiliation{Department of Chemistry, University of Colorado Boulder, Boulder, CO 80309, USA\looseness=-1}

%--------------------------------------------------------------------------------------------

\date{\today}

\begin{abstract}
Ultrafast multidimensional spectroscopies are powerful tools that can access charge and energy flow in complex materials, shifting chemical kinetics, and even many-body interactions in correlated matter. However, current implementations typically involve complex apparatuses and long averaging times. As a result, these methods have been limited to detailed mechanistic investigations of a few samples, precluding the broad characterization of molecular systems and/or the optimization of material ones. For example, converging the statistical noise in 2D spectra becomes exponentially expensive with increasing waiting times, and extended laser exposure heightens the probability of sample degradation. We address this fundamental challenge by developing a new technique, the spectral generalized master equation (GME), that enables one to employ short-waiting time 2D spectra to determine the full evolution of 2D spectra over arbitrary waiting times with high temporal resolution. In addition to reducing the cost of experiments by multiple orders of magnitude, our approach accurately removes statistical noise, reducing the need for time averaging, while circumventing the increasing convergence costs with longer waiting times. We provide a rigorous theoretical footing for the spectral GME and demonstrate its applicability on theoretically generated and experimentally measured 2D electronic and 2D infrared spectra. We anticipate that this advance has the potential to enable the investigation of systems that are too delicate for study with current multidimensional spectroscopies and accelerate the progress of 2D spectroscopy-based microscopies that can offer highly resolved excitation dynamics with spatial resolution over heterogeneous environments.
\end{abstract}

\maketitle

\section{Introduction}\label{sec:Intro}

Ultrafast two-dimensional (2D) spectroscopies like 2D terahertz, infrared, electronic, and their mixed-frequency combinations, probe the third-order polarization response in a time- and frequency-resolved manner, allowing the interrogation of otherwise congested and overlapping spectral features, excitation dynamics and dissipation, and chemical transformations across molecular, nanomaterial, and solid-state systems. These experiments have catalyzed important breakthroughs, enabling researchers to, for instance, interrogate the shifting nature of hydrogen bonds in solution,\cite{Dereka2021} quantify charge-phonon couplings responsible for transport in perovskites,\cite{Qu2025} disentangle the atomistic motions that accompany electron transfer in solution,\cite{Biasin2021} uncover how vibronic coherences can drive ultrafast photochemistry,\cite{Gaynor2019} unravel the impacts of evolutionary pressures in molecular paleontology,\cite{Heckmeier2024} and reveal accelerated energy exchange in vibrational polaritons.\cite{Chen2022}
Clearly, the applicable domain of 2D experiments is broad, ranging from chemical biology to solid-state physics. Yet, accessing such data requires the design, construction, and operation of complex instruments. The result is that 2D spectroscopy costs significant time and money---costs that can limit applications and deter scientific progress. % Hence, it is critical to reduce the cost and difficulty of 2D spectroscopy to enable the next generation of breakthroughs that it can offer.

\begin{figure*}
    \centering
    \includegraphics[width=\textwidth]{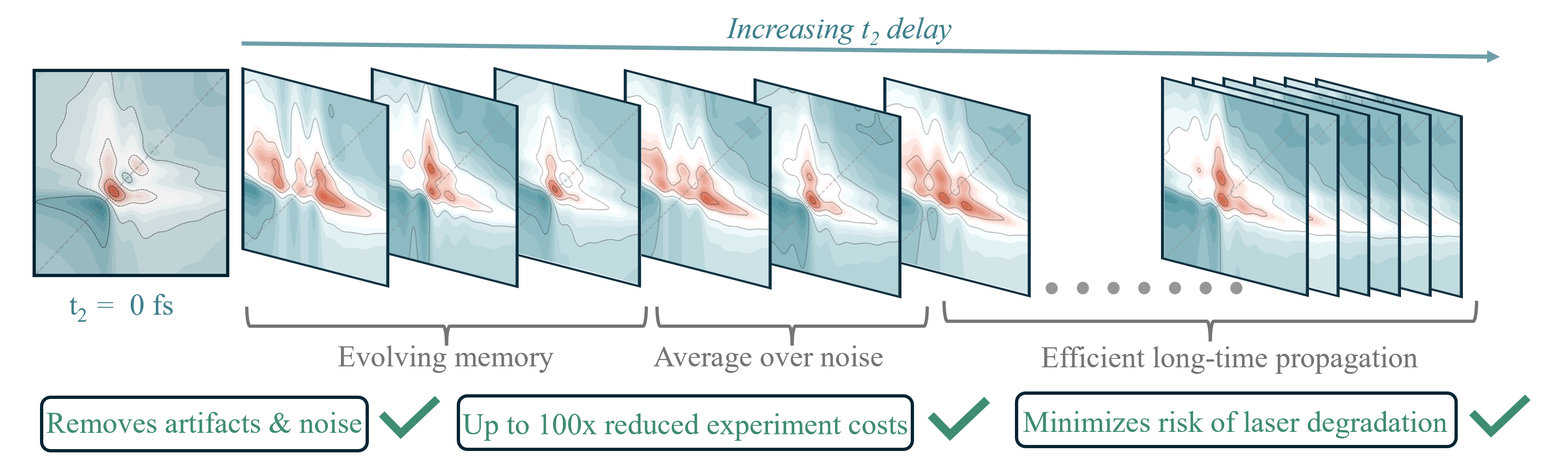}
    \caption{\label{fig:intro-schematic}Schematic representation of 2D spectroscopy with the spectrum divided across $t_2$ into regions according to the spectral GME. Our primary discovery is that 2D spectroscopy exhibits short-lived memory, that one can extract this memory from experiment \textit{directly} to construct the dynamical propagator, and that this propagator can be used to \textit{predict} the full evolution of the 2D spectrum using only its early-time data.}
\end{figure*}

To appreciate the cost of a 2D experiment, consider its output, the third-order response, 
\begin{equation}\label{eq:third-order-optical-response}
    R^{(3)} = {\rm Tr} \Big\{ \hat{\rho}_0 \tilde{\bm{\mu}}(t_3+t_2+t_1) \tilde{\bm{\mu}}(t_1+t_2) \tilde{\bm{\mu}}(t_1) \hat{\bm{\mu}}(0)\Big\},
\end{equation}
a 3D grid encompassing the signal arising from four light-matter interactions, expressed in terms of dipole moment operators.\cite{BookMukamel} This 3-time quantity is typically presented as a mixed time-and-frequency object where $t_1$ and $t_3$ are Fourier-transformed into $\omega_1$ and $\omega_3$ and slices along the untransformed $t_2$ axis are presented as 2D frequency maps. Experimentally, one measures the correlation between excitation ($\omega_1$) and emission ($\omega_3$) frequencies separated by some stage-controlled ($t_2$) `waiting time', repeating the experiment with a time resolution that covers the dynamics of interest. With the development of 100 kHz IR spectrometers,\cite{Luther2016} which currently achieve the fastest collection rates across the electromagnetic spectrum, a typical 2D~IR spectrum of 3000-4000 independent averages requires only around a minute of collection time for each waiting time. Therefore, a standard workflow with a few dozen $t_2$ points can be collected in under an hour. Despite these advances, signal-to-noise ratios become progressively worse with increasing $t_2$, meaning that signals become increasingly difficult to converge at long waiting times. Now suppose one aims to acquire additional information, such as the spatial resolution microscopy offers in heterogeneous materials.\cite{Dabrowski2017, Gross2023, Ostrander2016, Baiz2014, Tiwari2018, Guo2017, Guo2015} Here, a 2D grid of only 30 spatial points requires $\sim6$ hours to collect. For biological samples with an additional spatial dimension, collection may require days---often longer than the process under study. Moreover, such integration times under laser exposure lead to sample degradation, precluding successful data acquisition. Hence, reducing the cost, acquisition time, and signal-to-background sensitivity of 2D spectroscopic microscopy poses a fundamental challenge to its broad adoption. Surmounting it would, for instance, enable spatially resolved mapping of protein aggregation, biocondensate formation, and disease progression in live tissues; reveal membrane organization and ion channel transport dynamics; interrogate defect-mediated energy flow in nanomaterials; and expose the atomic-level restructuring that underlies battery failure.

Here, we introduce a simple yet rigorous method to reduce the data required to determine the \textit{full} evolution of 2D spectra, tame the impact of statistical noise in the measurements, and enhance the productivity of 2D spectroscopy setups. To achieve this, we develop a novel generalized master equation (GME) framework that leverages experimental 2D spectra acquired over short $t_2$ waiting times to deterministically predict the entire evolution of 2D spectra in $t_2$ indefinitely into the future (see the schematic in Fig.~\ref{fig:intro-schematic}). We verify the efficacy of our method by comparing measurement-predicted data to benchmark experiments, and demonstrate its robustness across a range of methods subject to distinct experimental challenges. Our work reveals an unexpected physical insight with important consequences for the feasibility of 2D spectroscopy and microscopy: early-time dynamics in 2D spectra encode the entire evolution of excitations over equilibration times that can be orders of magnitude longer, including even signatures of dark states that emerge well after the early-time limit. 

\section{Theoretical development}
\label{sec:theory}
%%%%%%%% GME context & setup

GMEs offer a powerful and versatile kinetic framework to account for memory effects that arise when evolution timescales overlap---a common occurrence---but their application to experimental spectra has proven difficult. Originally designed to facilitate the theoretical descriptions of dissipative systems\cite{Nakajima1958a, Zwanzig1960, Mori1965b} with applications in transport,\cite{Bhattacharyya2024b, Bhattacharyya2025b} chemical kinetics,\cite{Macnamara2008, Zhang2019, Shi2023} and computational spectroscopy,\cite{Ivanov2015, Sayer2024, Sayer2025} a GME offers a compact prescription to predict the evolution of one-time correlation functions, $\mathcal{C}(t)$, of observables in an open (quantum or classical) system. GMEs provide two key advantages: they facilitate interpretation by leveraging a states-and-rates picture, and lower the cost of encoding and predicting the evolution of these states. However, combining the GME framework and experimental spectroscopic data has proven difficult because the states necessary to build spectral responses that would underlie a spectroscopic GME (e.g., the individual transition dipoles) cannot be accessed from spectroscopic measurements. Instead, the observed response sums over all such transitions without the ability to resolve their individual contributions (see Supplementary Information (SI) Sec.~1). For a 2D spectrum, the problem is further complicated by the multitime nature of the GME.\cite{Ivanov2015, Sayer2024} Yet, we show that the multitime nature of the response function gives an alternative path to build a GME from the 2D spectrum directly. 

We argue that the 2D spectrum satisfies an effective one-time GME along the $t_2$-axis whose generator can be constructed solely from 2D spectra at early $t_2$. Specifically, we propose that the evolution of the 2D spectrum
\begin{equation}\label{eq:2d-spectrum}
    S(\omega_1, t_2, \omega_3) = \sum_{\pm} \iint_0^{\infty} \dd{t_1} \dd{t_3}\ e^{i\omega_3 t_3 \pm i\omega_1 t_1} R^{(3)},
\end{equation}
(see SI Sec.~2 for a detailed derivation) can be predicted with an \textit{integrated} time-local GME we call the spectral GME:
\begin{equation}
\label{eq:spectralUGME}
    \mathbf{S}(t_2+ \delta t_2) 
    = \bm{\mc{U}}(t_2)\mathbf{S}(t_2),
\end{equation}
where we reformat the 2D spectrum as a dynamical matrix, $S(\omega_1, t_2, \omega_3) \rightarrow \mathbf{S}(t_2)$, with $\omega_1$ and $\omega_3$ enumerating the rows and columns and $\delta t$ denoting the timestep. While Eq.~\eqref{eq:spectralUGME} may resemble a Markovian (memoryless) kinetic description of the 2D spectrum, the time-dependent $\bmc{U}(t_2)$ encapsulates all non-Markovian effects. These may be expected to disappear\cite{Sayer2023, Dominic2022, Dominic2023} when $\boldsymbol{\mc{U}}(t_2)$ reaches its equilibrium limit at $\tau_{\mc{U}}$. For $t_2 \geq \tau_\mc{U}$, we set $\bmc{U}_\infty(\delta t_2) \equiv \bmc{U}(t_2\geq \tau_\mc{U})$. However, the features in $\mathbf{S}(t_2)$ generally take much longer than $\tau_{\mc{U}}$ to exhibit a simple exponential decay. This mismatch in decay timescales is what enables the GME to reduce the cost of measuring 2D spectra. Instead of simply asserting the applicability of these equations using the resulting savings as post-hoc proof, we provide a rigorous theoretical footing for their validity and tests for their utility below.

Beyond noting that $\mathbf{S}(t_2)$ possesses a shape appropriate for a GME, we consider the requirements for a one-time GME. To derive a GME, one can employ the projection operator technique,\cite{Grabert2006} where the central object is the projection operator, $\mathcal{P} = |\mathbf{A})(\mathbf{A}|\mathbf{A})^{-1}(\mathbf{A}|$. The projection operator must also be linear and satisfy idempotency, $\mathcal{P}^2 = \mathcal{P}$, which is ensured by the factor $(\mathbf{A}|\mathbf{A})^{-1}$.\cite{BookChapterBerne} 
This implies that $(\mathbf{A}|\mathbf{A})$ must be an invertible matrix. One must also define the inner product, $(\mathbf{A}|\mathcal{O}|\mathbf{A})$, which can be tailored to a physical problem of interest. Having satisfied these requirements, one can build a correlation function, $\mathbf{\mathcal{C}}(t_2) \equiv (\mathbf{A}|e^{-i\mc{L}t_2}|\mathbf{A})(\mathbf{A}|\mathbf{A})^{-1}$, whose evolution can be described by a GME. This motivates the search for a rewriting of $\mathbf{S}(t_2)$ as the inner product of the members of a projector with the evolution operator, $e^{-i\mc{L}t_2}$.

We obtain the desired formulation by rewriting the 2D spectrum via a modified doorway-window representation,\cite{Yan1990,BookMukamel}
\begin{equation}\label{eq:S2-inner-product}
\begin{split}
    S(\omega_3, t_2, \omega_1) &= \mathrm{Tr}\{\hat{\bm{\omega}}_3^{\dagger} e^{-i\mc{L}t_2} \hat{\bm{\omega}}_1\},
\end{split}
\end{equation}
where
\begin{subequations}
\begin{align}
    \hat{\bm{\omega}}_3 &\equiv \tilde{\bm{\mu}}\tilde{\mathbf{R}}^{+}(-\omega_3)\hat{\bm{\mu}},\\
    \hat{\bm{\omega}}_1 &\equiv \sum_{\pm \omega_1} \tilde{\bm{\mu}} \tilde{\mathbf{R}}^{-}(\pm \omega_1)\tilde{\bm{\mu}}\hat{\rho}_0, 
\end{align}
\end{subequations}
and the resolvent is the Fourier-Laplace transform of the propagator,
\begin{equation}\label{eq:resolvent}
    \tilde{\mathbf{R}}^{\mp}(\pm \omega) \equiv
    \lim_{\epsilon \to 0} \int_0^\infty \dd{t} e^{i(\pm \omega + i\epsilon) t } e^{\mp i\mc{L} t} = \frac{i}{\pm \omega \mp \mc{L} }.
\end{equation}
We note that $\tilde{\mathbf{R}}^{\mp}(\pm \omega) = [\tilde{\mathbf{R}}^{\mp}(\mp \omega)]^{\dagger}$. Equation~\ref{eq:S2-inner-product} indicates that the 2D spectrum, $\mathbf{S}(t_2)$, is an infinite-dimensional dynamical matrix where the rows (columns) are indexed by continuous values of $\omega_3$ ($\omega_1$). In practice, however, $\mathbf{S}_{\omega_3, \omega_1}(t_2)$ is measured or theoretically generated with finite resolution in $\omega_1$ and $\omega_3$, making the 2D spectrum a finite-dimensional dynamical matrix. 

Our rewriting suggests using $\omega$ as the enumerator of the members of the projector, $\mathcal{P}_{\omega} = |\hat{\bm{\omega}}_1)(\hat{\bm{\omega}}_3|\hat{\bm{\omega}}_1)^{-1}(\hat{\bm{\omega}}_3|$. Here, the inner product of these high-dimensional vectors can incorporate a modified treatment of the vector that encodes the initial condition, 
\begin{equation}\label{eq:projector-inner-product}
\begin{split}
    (\hat{\bm{\omega}}_3| \hat{O}&|\hat{\bm{\omega}}_1) \\\equiv& \sum_{\pm \omega_1} \mathrm{Tr} \Big\{ [\tilde{\bm{\mu}}\tilde{\mathbf{R}}^+(- \omega_3)\hat{\bm{\mu}}]^{\dagger}
    \hat{O}
    [\tilde{\bm{\mu}} \tilde{\mathbf{R}}^-(\pm \omega_1)\tilde{\bm{\mu}}\hat{\rho}_0]
    \Big\},
\end{split}
\end{equation}
enabling us to recover Eq.~\eqref{eq:S2-inner-product} using the projector, $\mathcal{P}_{\omega}$. See SI Sec.~3 for further details on this formulation.

%%%%%%%%%%%%%%%%%%%%%%%%%%%%%%%%%%%%%%%%%%%%%%%%%%%%
%\section{Simulation Proof of Concept}
\section{Applications}
\label{sec:Applications}

Below, we illustrate the applicability of our spectral GME to both 2DES and 2DIR and its ability to reduce, in both cases, the amount of work (simulation or experiment) one must do to fully determine the full evolution of the 2D spectrum. In particular, we illustrate how to apply the spectral GME to the theoretically generated 2D spectra of a commonly invoked energy transfer dimer in the condensed phase. This proof-of-principle shows that our method works as intended and can lower the cost of generating long-$t_2$ 2D spectra. We then turn to experimental 2DES data for biologically relevant cyanine (Cy3-Cy5) dimers. Because experimental data is characterized by a threshold of statistical noise, we introduce an averaging procedure to tame its effects on our predictions. We show that our spectral GME can capture the 3500~ps-long evolution of the 2DES using only the first 150~ps of data, demonstrating a factor of 20 reduction in cost. Finally, we demonstrate that our method can be applied even to complex battery electrolytes consisting of glassy ionic liquids, where our spectral GME not only reduces the cost of collecting 2DIR by over an order of magnitude but also serves to remove statistical artifacts that obscure signals at long waiting times. We further show that our spectral GME accurately captures the evolution of the center line slope---a widely used diagnostic tool for spectral diffusion---and even predicts the emergence of a dark state at longer waiting times than those used to construct its propagator. 
\begin{figure*}[!t]
    \hspace{-20pt}
    \begin{subfigure}[b]{0.3\textwidth}
        \resizebox{1.\textwidth}{!}{\includegraphics{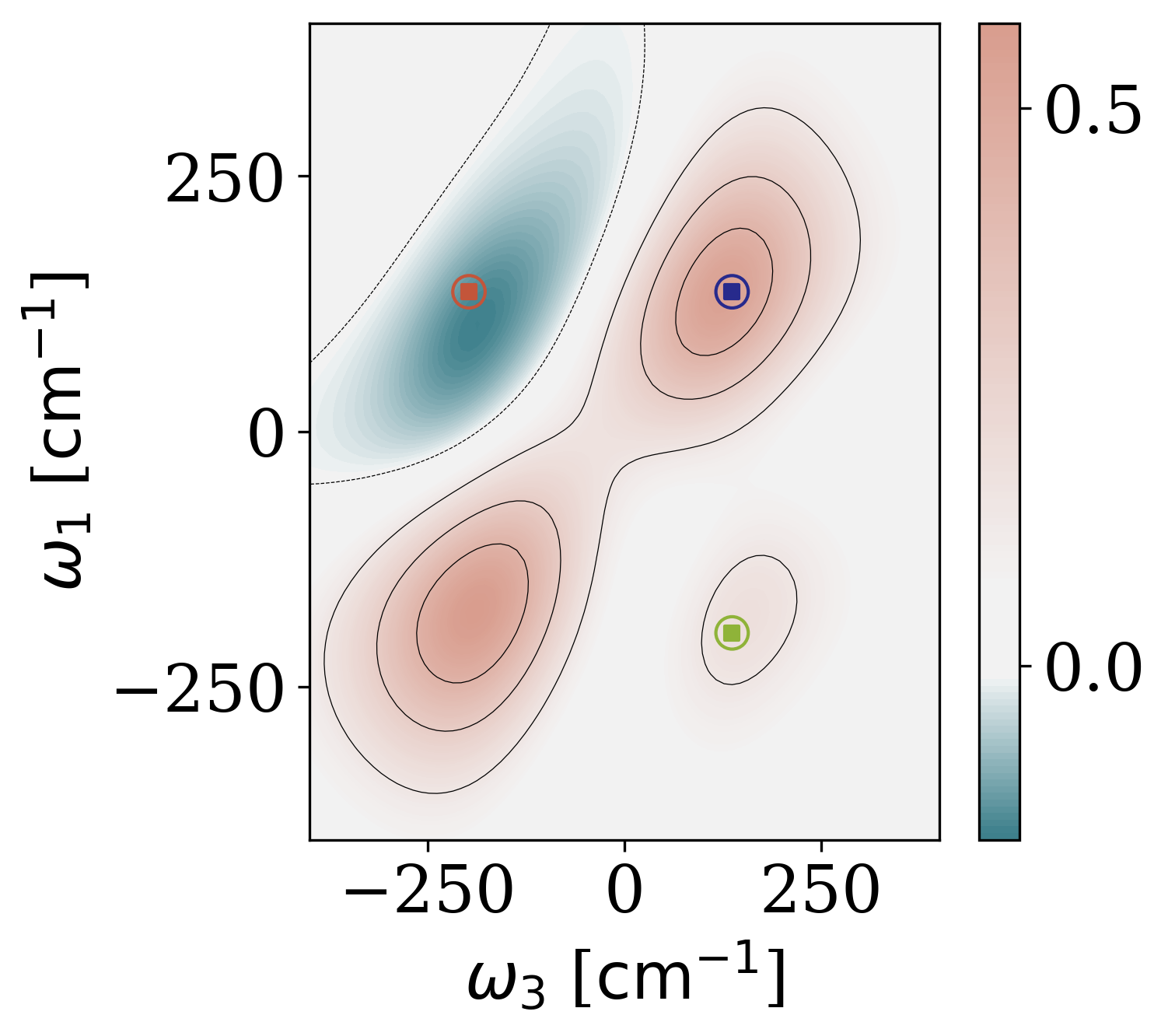}} 
        \vspace{0pt}
        \caption{}
        \label{fig:theory_t2-48}
    \end{subfigure}
    \begin{subfigure}[b]{0.45\textwidth}
        \resizebox{1.\textwidth}{!}{\includegraphics{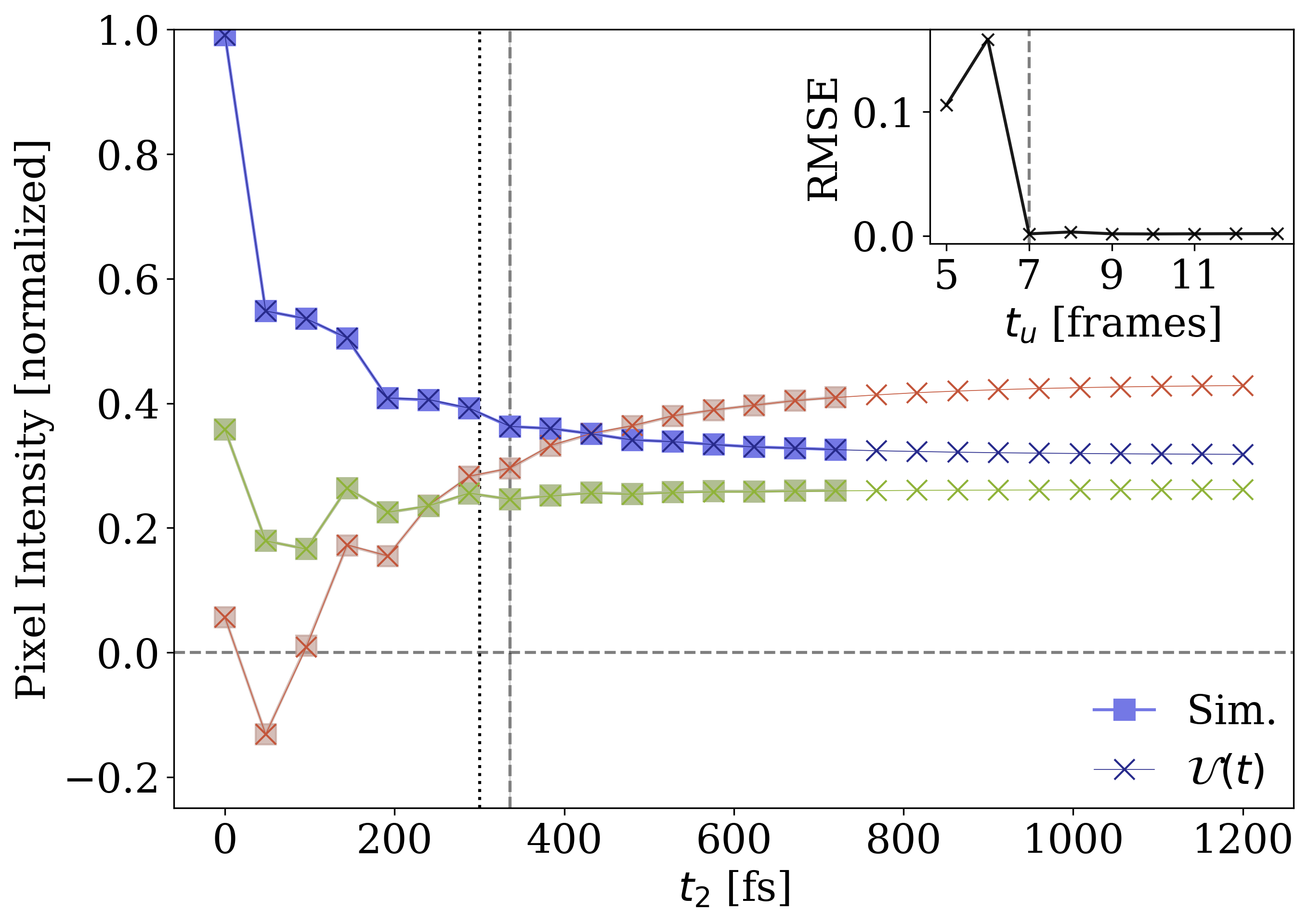}}
        \vspace{-15pt}
        \caption{}
        \label{fig:theory_pixels}
    \end{subfigure}
    \begin{subfigure}[b]{0.15\textwidth}
        \resizebox{.95\textwidth}{!}{\includegraphics{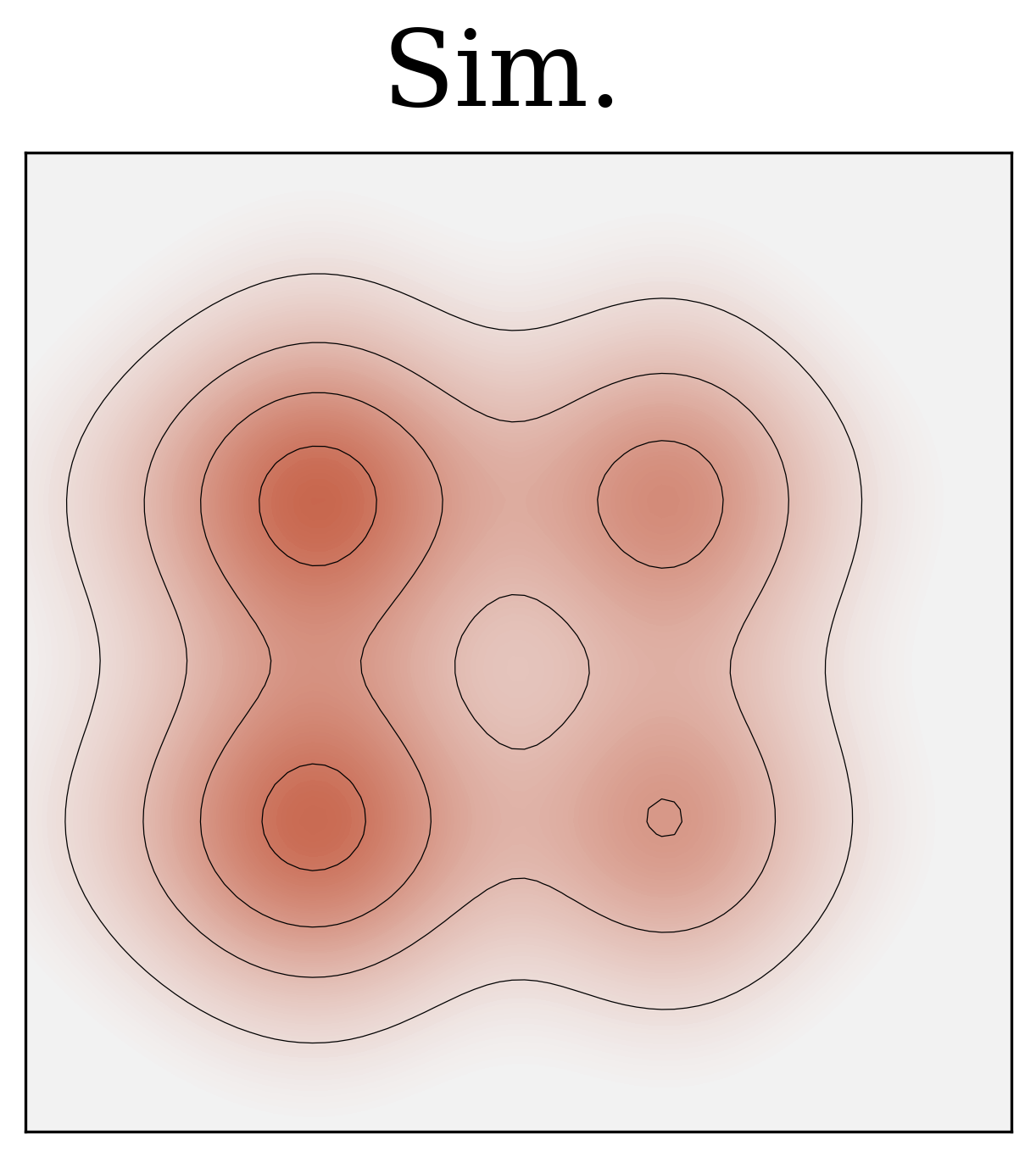}}
        \\
        \resizebox{.95\textwidth}{!}
        {\includegraphics{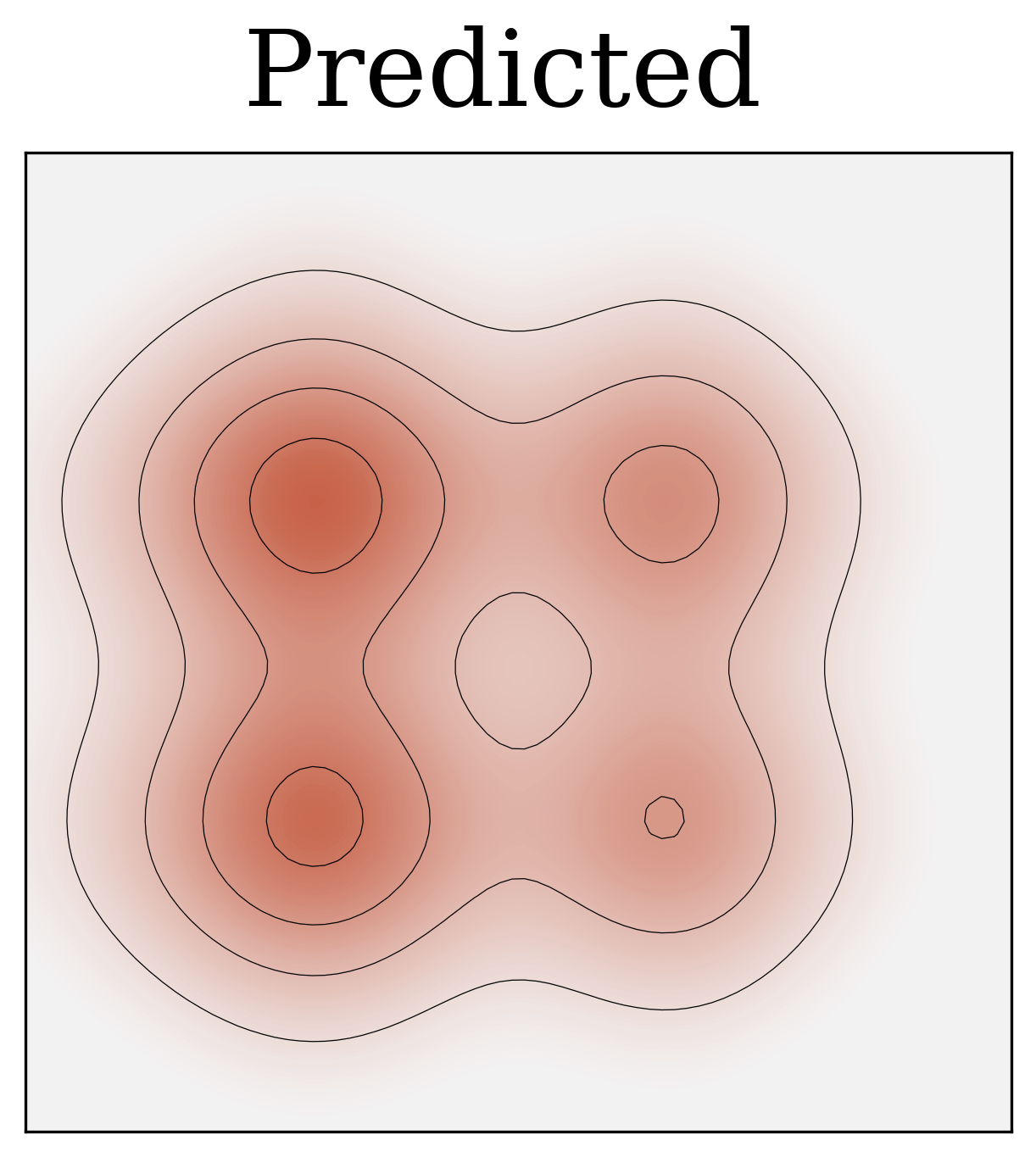}}
        \caption{}
        \label{fig:theory_prediction}
    \end{subfigure}
\caption{\label{fig:theorydata} Applicability of the spectral GME to simulated 2DES data of a model nonadiabatic energy transfer dimer. 2D spectra span a $80 \times 80$ spectral grid with $\delta \omega_1 = \delta \omega_2 = 10$~cm$^{-1}$ and were sampled at a timestep of $\delta t_2 = 48$~fs. \textbf{(a)} Snapshot of the 2D spectrum at $t_2=48$~fs highlighting the 3 pixels to visualize spectral evolution in \textbf{(b)}. We highlight pixels located at indices $[(20,53), (53, 20), (53, 53)]$ to monitor both diagonal and off-diagonal spectral features. \textbf{(b)} Spectral evolution of the selected pixels along $t_2$. Squares correspond to benchmark simulation data up to $t_2=720$~fs. The dashed grey line is the chosen GME cutoff, $\tau_\mc{U}$. The dotted grey line is the known time-nonlocal memory decay timescale for the multitime response function from previous work.\cite{Sayer2024} Inset shows the error plot which, together with the visual agreement contextualizing the error value, justifies $\tau_\mc{U} \simeq 336$~fs. We employ an SVD truncation threshold of $\xi = 10^{-10}$ and $\tU=t_u$ for this system. As such, the spectral GME spectrum contains 22 singular values at $t_2 = 0$~fs, and uses only the first 7~frames~up to the $\tau_\mc{U}=336$~fs cutoff. \textbf{(c)} Comparison of 2D spectra at $t_2=720$~fs obtained from the simulation benchmark (top) and spectral GME prediction (bottom).}
\end{figure*}

\subsection{Simulated data}
\label{ssec:simulated2DES}

While our formulation of the 2D spectrum in terms of the inner product in Eq.~\eqref{eq:projector-inner-product} suggests the viability of a one-time GME for the 2D spectrum, its validity requires the invertibility of $(\hat{\bm{\omega}}_3| \hat{\bm{\omega}}_1) = \mathbf{S}(t_2=0)$. This is no small requirement. Physically, this invertibility requires sufficient alignment of the states probed by pump ($\hat{\bm{\omega}}_1$) and probe ($\hat{\bm{\omega}}_3$) pulses. For example, if $\hat{\bm{\omega}}_1$ and $\hat{\bm{\omega}}_3$ probe energy scales with little to no overlap---as is the case for 2D spectroscopies that probe nearly or fully nonoverlapping regions of the electromagnetic spectrum,\cite{Song2019} such as 2D EV/VE,\cite{Oliver2014, Courtney2015} V-THz, etc.---then $(\hat{\bm{\omega}}_3| \hat{\bm{\omega}}_1)$ is not invertible, preventing one from proceeding with a GME. In contrast, when the 2D spectroscopy probes commensurate energy scales---as is the case in 2DES,\cite{Hybl1998} 2DIR,\cite{Asplund2000} 2DUV,\cite{West2012} 2D NMR,\cite{Aue1976, Derome2013} and 2DTHz\cite{Savolainen2013, Finneran2016}---$(\hat{\bm{\omega}}_3| \hat{\bm{\omega}}_1)^{-1}$ is more likely to exist. Nevertheless, this existence is not guaranteed. After all, the pseudo-continuous frequency basis likely leads to redundancy (i.e., linear dependence), resulting in noninvertibility. We overcome this difficulty with a standard Moore-Penrose pseudoinverse procedure, with the number of truncated singular values quantifying the extent of redundancy. We refer the reader to Sec.~\ref{ssec:BuildingU} for details on the construction of its generator, $\bm{\mathcal{U}}(t)$.

To test our spectral GMEs with simulation data free from noise and other artifacts, we begin with the 2DES spectrum of a model molecular dimer that exhibits dephasing, decoherence, energy transfer, and dissipation. We employ our previously published data\cite{Sayer2024} for the well-characterized electronic energy transfer dimer,\cite{Chen2010h} a paradigmatic two-site Frenkel exciton model that supports nonadiabatic energy flow whose dynamics we solved with the numerically exact Hierarchical Equations of Motion (HEOM).\cite{Tanimura1989, Ishizaki2005, Xu2007, Yan2021} Adopting this system allows us to directly compare our proposed one-time GME for $\mathbf{S}(t_2)$ with our \textit{multitime} decomposition,\cite{Sayer2024} which gives an estimate of the smallest amount of reference data required to capture the evolution of the 2D spectrum. From our multitime GME, we found that the memory decays within $300$~fs, which we posit should provide a guide for the generator lifetime in our spectral GME. 

\begin{figure*}[!t]
    \hspace{-65pt}
    \begin{subfigure}[b]{0.44\textwidth}
        \resizebox{.44\textwidth}{!}{\includegraphics{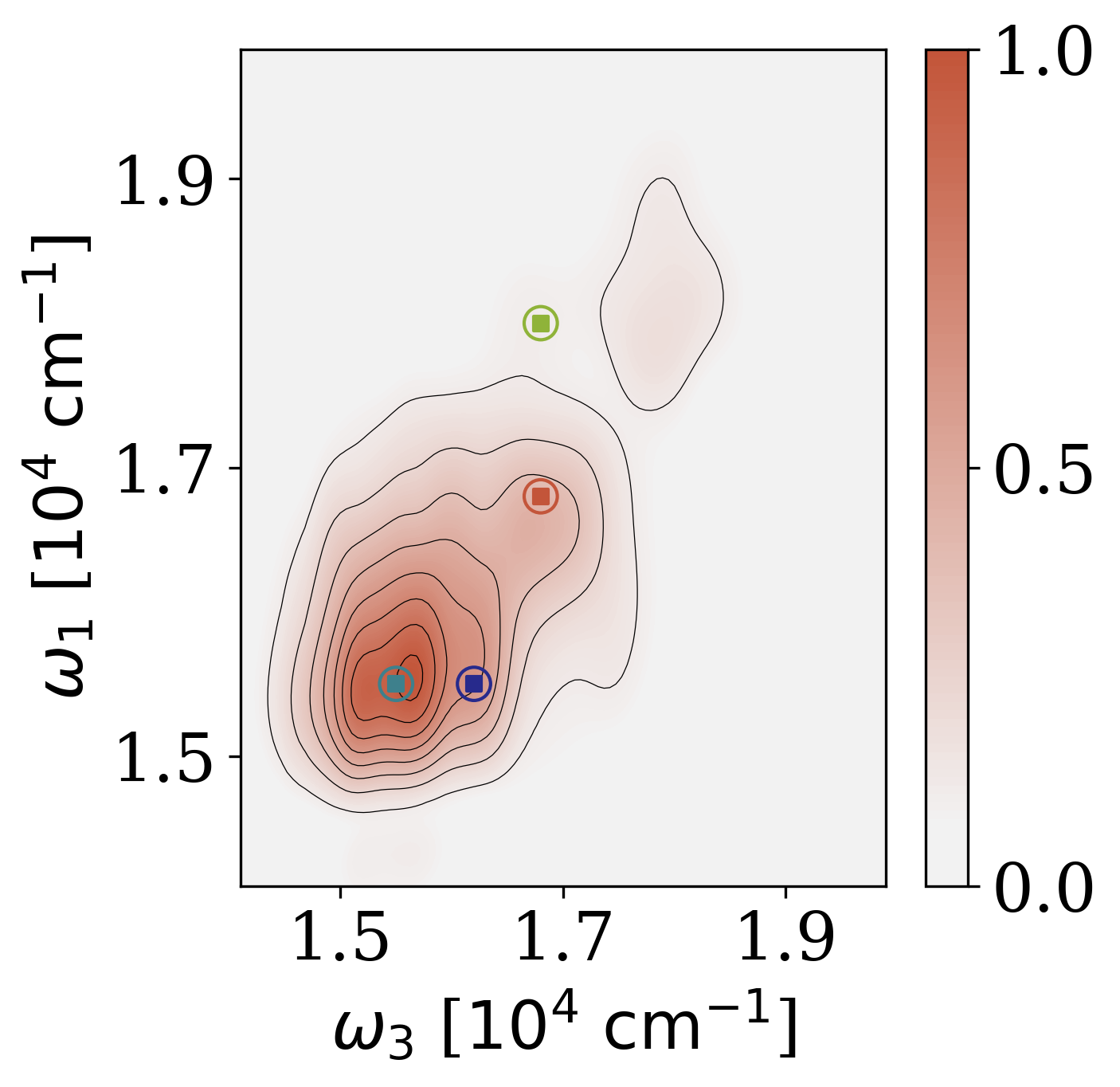}}
        \vspace{-5pt}\hspace{-15pt}
        \caption{}
        \label{fig:cy3_t2-0}
        \begin{subfigure}[b]{\textwidth}
        \vspace{-10pt}\hspace{15pt}
            \resizebox{.27\textwidth}{!}{\includegraphics{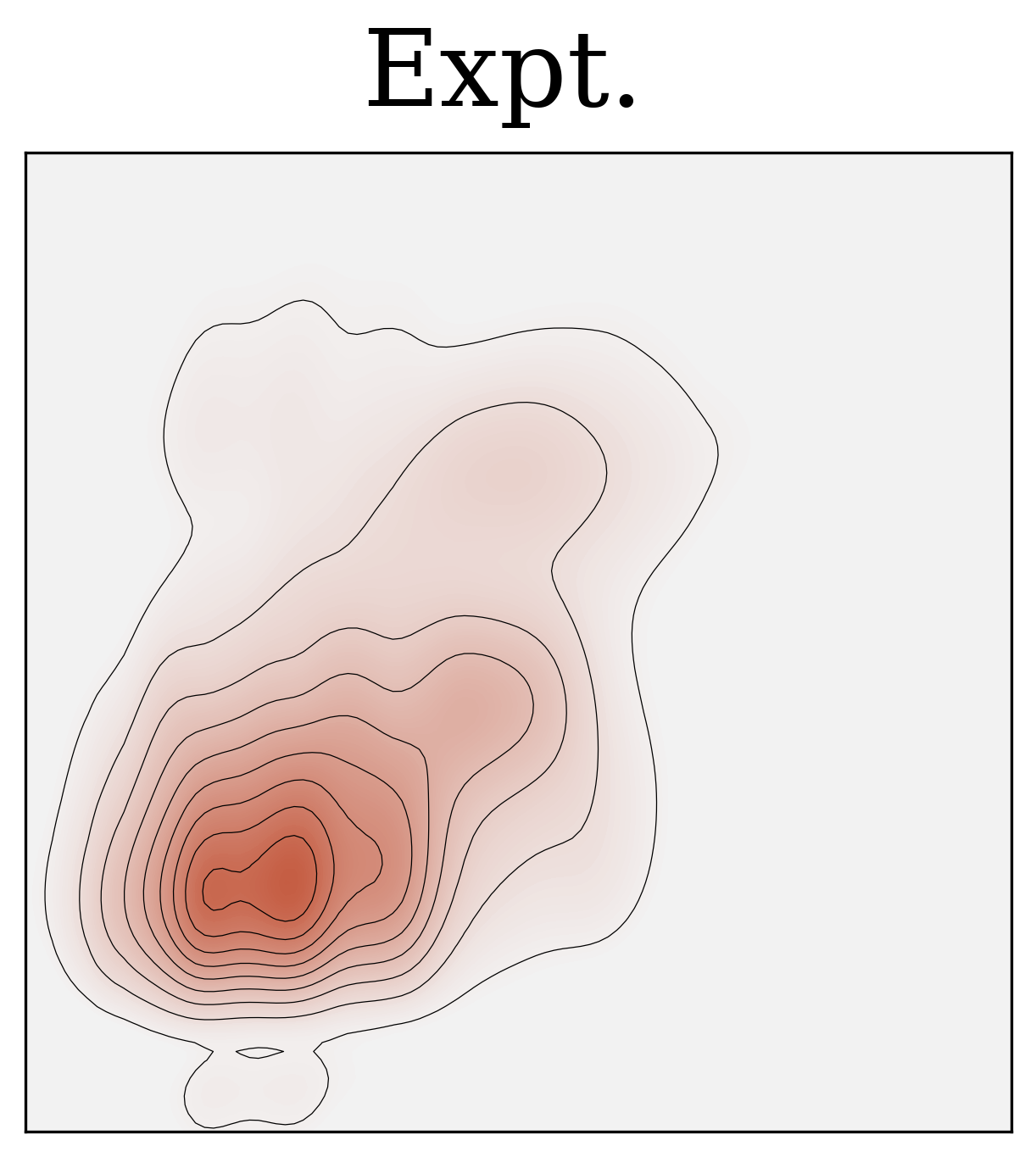}}
            \resizebox{.27\textwidth}{!}{\includegraphics{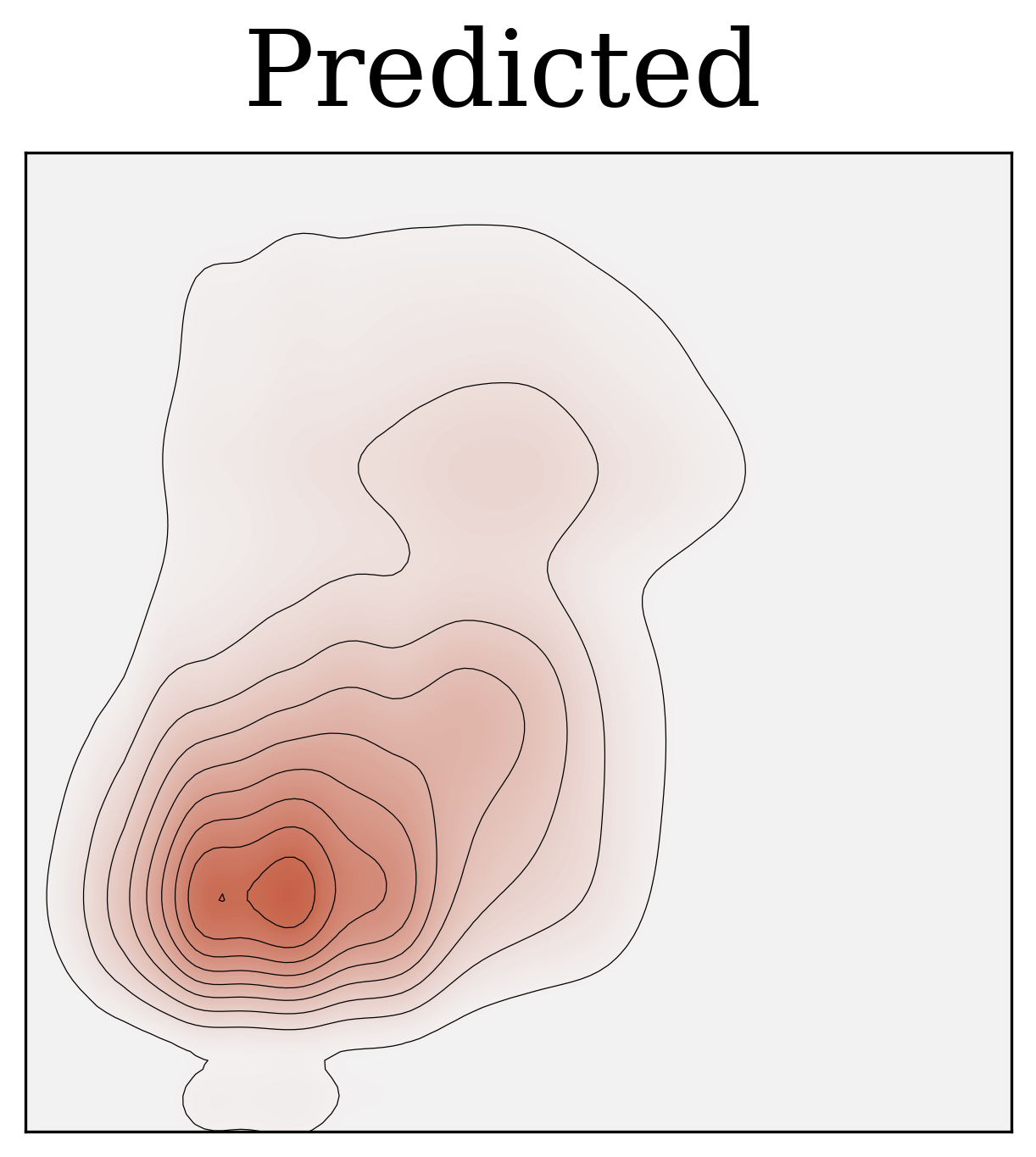}}
            \caption{}
            \label{fig:cy3_prediction}
        \end{subfigure}
    \end{subfigure}
    \hspace{-40pt}
    \begin{subfigure}[b]{0.46\textwidth}
        \resizebox{1.\textwidth}{!}{\includegraphics{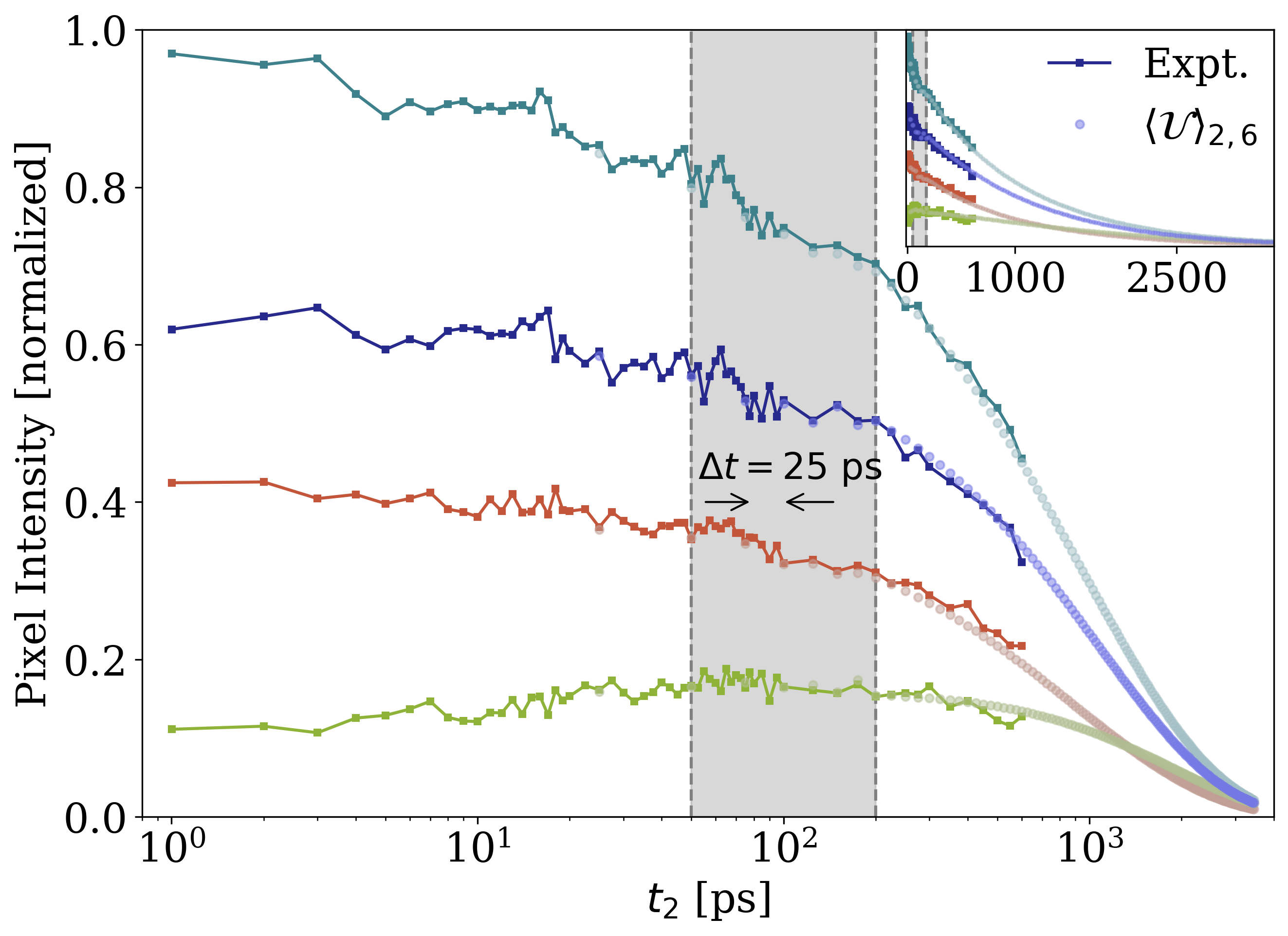}}
        \caption{}
        \label{fig:cy3_pixels}
    \end{subfigure}
    \hspace{-0pt}
    \begin{subfigure}[b]{0.26\textwidth}
        \vspace{-11pt}
        \resizebox{1.\textwidth}{!}{\includegraphics{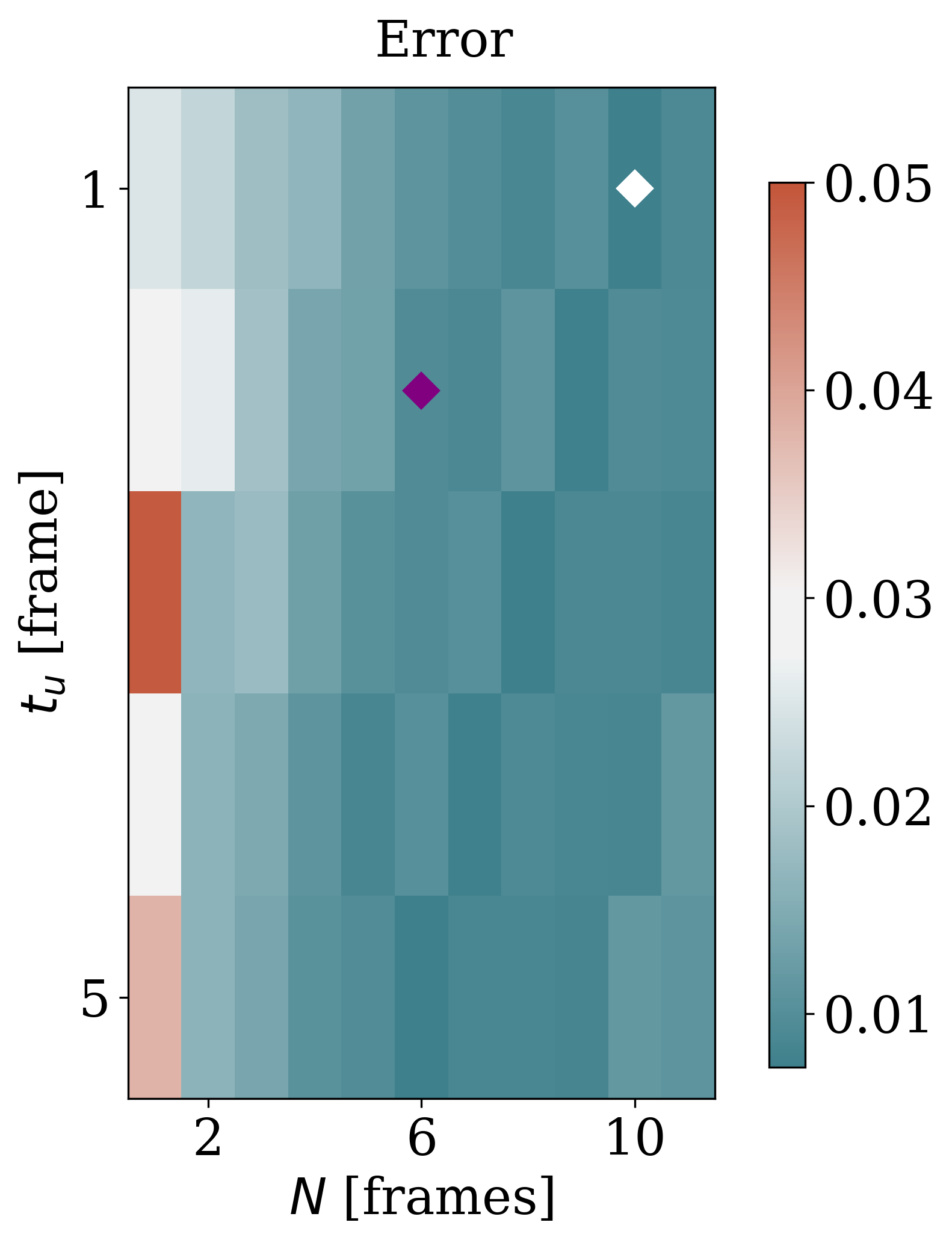}}
        \caption{}
        \label{fig:cy3_error}
    \end{subfigure}
\vspace{+5pt}
\caption{\label{fig:cy3} Applicability of the spectral GME to experimental 2DES of a Cy3-Cy5 dimer separated by six nucleotides on DNA origami and originally characterized experimentally in Ref.~\onlinecite{Hart2021}. 2D spectra span a $111 \times 111$ spectral grid with $\delta \omega_1 = \delta \omega_2 = 52.25$~cm$^{-1}$ and were sampled at variable timesteps, including $1$~ps, $2.5$~ps, $5$~ps, and $25$~ps. \textbf{(a)} 2D spectrum of Cy3-Cy5 in the visible range at $t_2=1$~ps. We highlight points at indices $[(27, 27), (51, 51), (51, 74), (40, 27)] $ as these span the full range of pixel intensities. \textbf{(b)} Comparison of 2D spectra at $t_2=600$~ps obtained from the experimental benchmark (left) and spectral GME prediction (right). \textbf{(c)} Spectral evolution of the selected pixels along $t_2$ on logarithmic time axis. Grey region represents the data used to construct $\langle\bm{\mc{U}}_\infty(\delta t_2)\rangle$, and circles are the spectral GME predictions, which we evolve out to $3500$~ps to capture the equilibration timescale. Inset: Spectral evolution on a linear time axis. \textbf{(d)} Root mean squared error heat map. The purple diamond represents the choice of parameters shown in~\textbf{(c)} while the white diamond denotes the global minimum error. Based on this plot, we choose $t_u=50$~ps, $\tU=200$~ps and $\xi=1$, and construct our spectral GME propagator using the data sampled with a timestep of $\delta t_2 = 25$~ps. We choose the parameterization $\langle\mc{U}\rangle_{2, 6}$ because it requires significantly less data while maintaining comparable accuracy to the global minimum.}
\end{figure*}

Our spectral GME recovers the correct dynamics of the 2D spectrum and recapitulates a generator lifetime similar to that of our multitime GME. Figure~\ref{fig:theory_pixels} illustrates the reference and the spectral GME evolution of three pixels highlighted in the HEOM-simulated 2D spectrum (Fig.~\ref{fig:theory_t2-48}) at $t_2=48$~fs corresponding to three distinct features in the exciton coupling signature. As Fig.~\ref{fig:theory_pixels} shows, the spectral GME quantitatively recovers the evolution of the signal intensities before and after the cutoff $\tau_{\mc{U}} = 336$~fs. The reference HEOM simulation of the 2D spectrum and the spectral GME reconstruction continue to display impressive agreement, even in the steady state of the 2D spectrum shown in Fig.~\ref{fig:theory_prediction}. Also interesting is the close agreement between the multitime GME decay timescale, $300$~fs, and that of the spectral GME, $\tau_{\mc{U}} = 336$~fs. Hence, as in our multitime GME for the prediction of 2D spectra,\cite{Sayer2024} the spectral GME can be expected to reduce the cost of measuring 2D spectra by multiple orders of magnitude, depending on the problem of interest. 

In our pseudoinverse approach to the isolation of $\bmc{U}(t_2)$ in Eq.~\eqref{eq:spectralUGME}, we converge a threshold (denoted as $\xi$) that ensures $\bmc{U}(t_2)$ reproduces the spectrum accurately and leads to a stable inverse. We discuss this procedure in Sec.~\ref{ssec:BuildingU}. This introduces an additional parameter into the workflow and, while choosing it in this way represents a principled choice free from user input, we show that one must allow some error when working with noisy spectra. Our error metric is the mean absolute deviation between the experimental spectrum and the spectrum predicted by the spectral GME as defined in Eq.~\eqref{eq:error} (see Sec.~\ref{ssec:errormetric} for additional details). Before the cutoff, where the error is negligible, SVD truncation is the only source of error. SI Secs.~4, 5 and~6 provide details on how to determine the SVD truncation threshold.

%%%%%%%%%%%%%%%%%%%%%%%%%%%%%%%%%%%%%%%%%%%%%%%%%%%%
\subsection{Experimental 2DES Data}
\label{sec:Expt2DES}

Having shown the applicability of our spectral GME to theoretically generated 2D electronic spectra free from statistical noise, we turn to experimental data and its accompanying uncertainties. We start with 2DES data taken on a Cy3-Cy5 dimer whose electronic-nuclear coupling and positions (and, hence, energy transfer-facilitating electronic couplings) have been engineered by attaching them with nanoscale precision to DNA chains.\cite{Hart2021} Such DNA-templated chromophores offer a powerful route to controlling exciton transport for applications in energy harvesting, transport, and quantum information processing.\cite{Hart2021} Cy3, Cy5, and other indole-based chromophores are often used as fluorescent markers for biological imaging and detection assays.\cite{Levitus2011} Figure~\ref{fig:cy3_t2-0} shows the 2D electronic spectrum at $t_2=1$~ps, spanning a 1.7--2.5~eV energy window, which is from just over 700~nm to 500~nm. Beyond their technological utility, excitations in cyanine dyes evolve over various connected timescales. For example, it supports long-lived vibrational modes that persist into the $>1$~ps timescale,\cite{Hart2020} undergoes a conformational change into a \textit{trans} intermediate state via a conical intersection mediated by coupled vibrational modes, and can exhibit H-, J-, or mixed-aggregate behavior depending on its orientation to other monomers.\cite{Hart2021} Due to these multiple timescales, the experimental data were collected using several different timesteps. We discuss the technical implications of this for our spectral GME in Sec.~\ref{ssec:preproc}. 

\begin{figure*}[!t]
    \hspace{-65pt}
    \begin{subfigure}[b]{0.46\textwidth}
        \resizebox{0.46\textwidth}{!}{\includegraphics{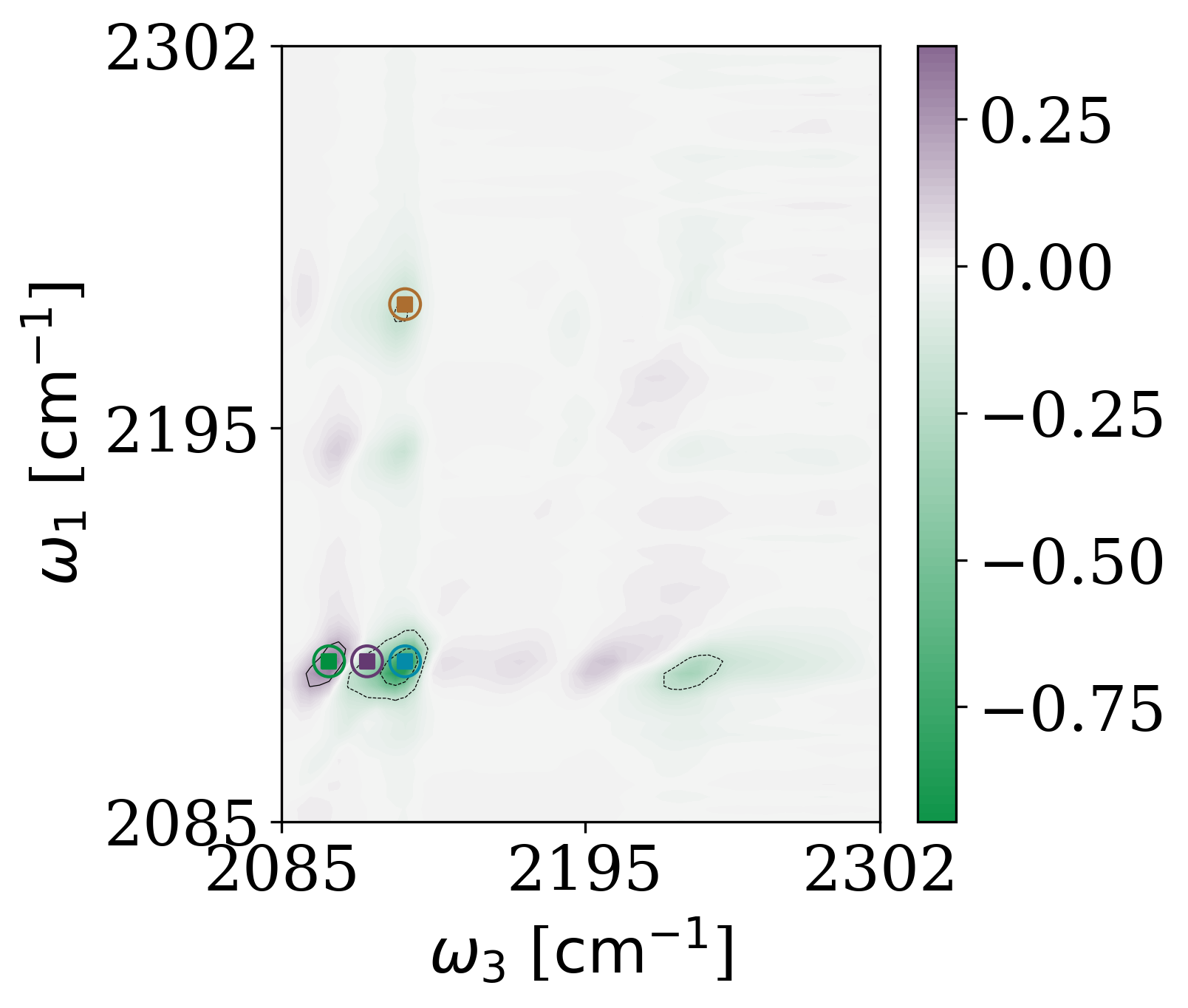}} 
        \vspace{-5pt}\hspace{-5pt}
        \caption{}
        \label{fig:DCA_t2-0}
        \begin{subfigure}[b]{\textwidth}
        \hspace{0pt}
           \resizebox{.23\textwidth}{!}{\includegraphics{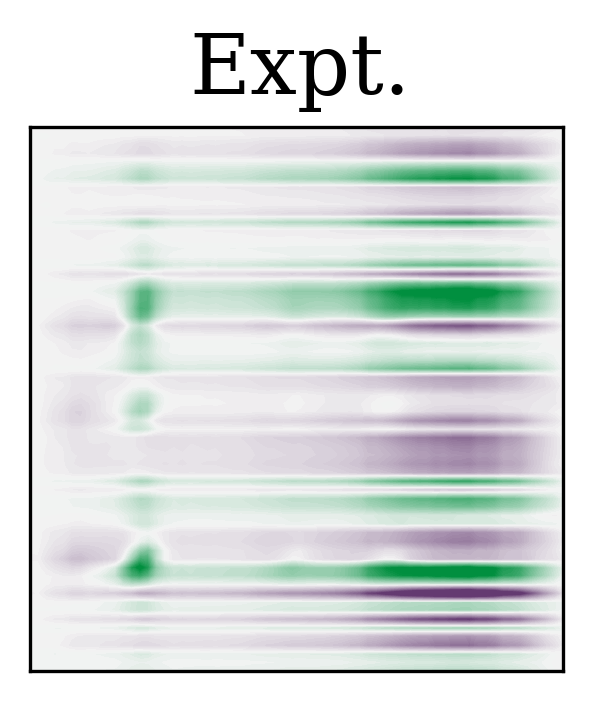}}
           \resizebox{.23\textwidth}{!}
           {\includegraphics{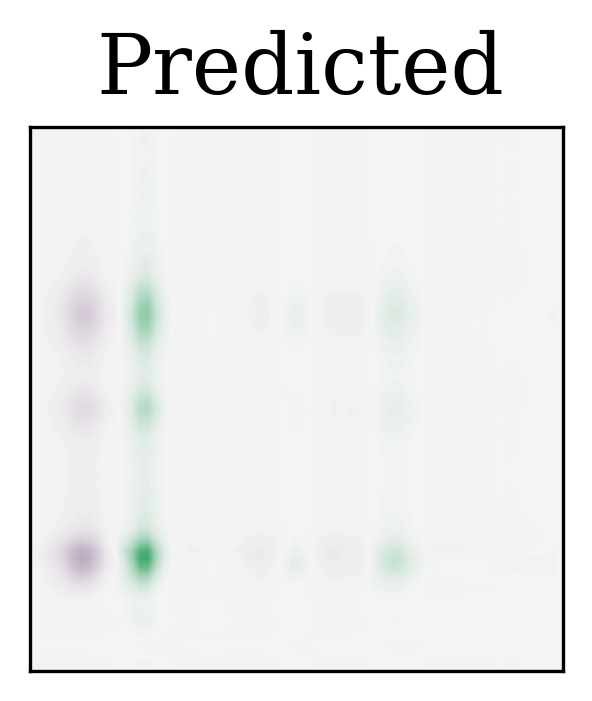}}
           \caption{}
           \label{fig:DCA_prediction}
        \end{subfigure}
    \end{subfigure}
    \hspace{-50pt}
    \begin{subfigure}[b]{0.46\textwidth}
        \resizebox{1.\textwidth}{!}{\includegraphics{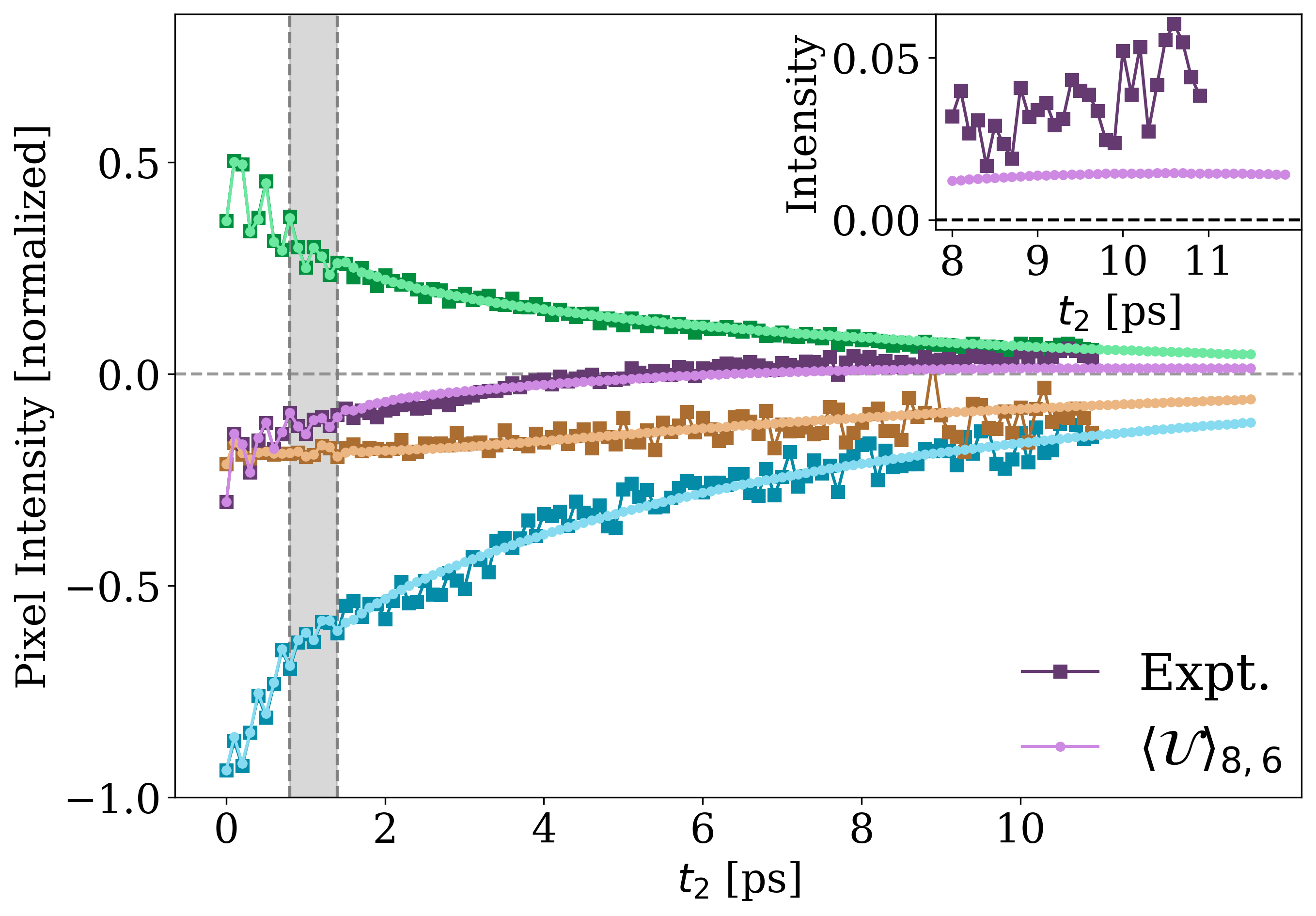}}
        \caption{}
        \label{fig:DCA_pixels}
    \end{subfigure}
    \hspace{-0pt}
    \begin{subfigure}[b]{0.26\textwidth}
        \vspace{0pt}
        \resizebox{1.\textwidth}{!}{\includegraphics{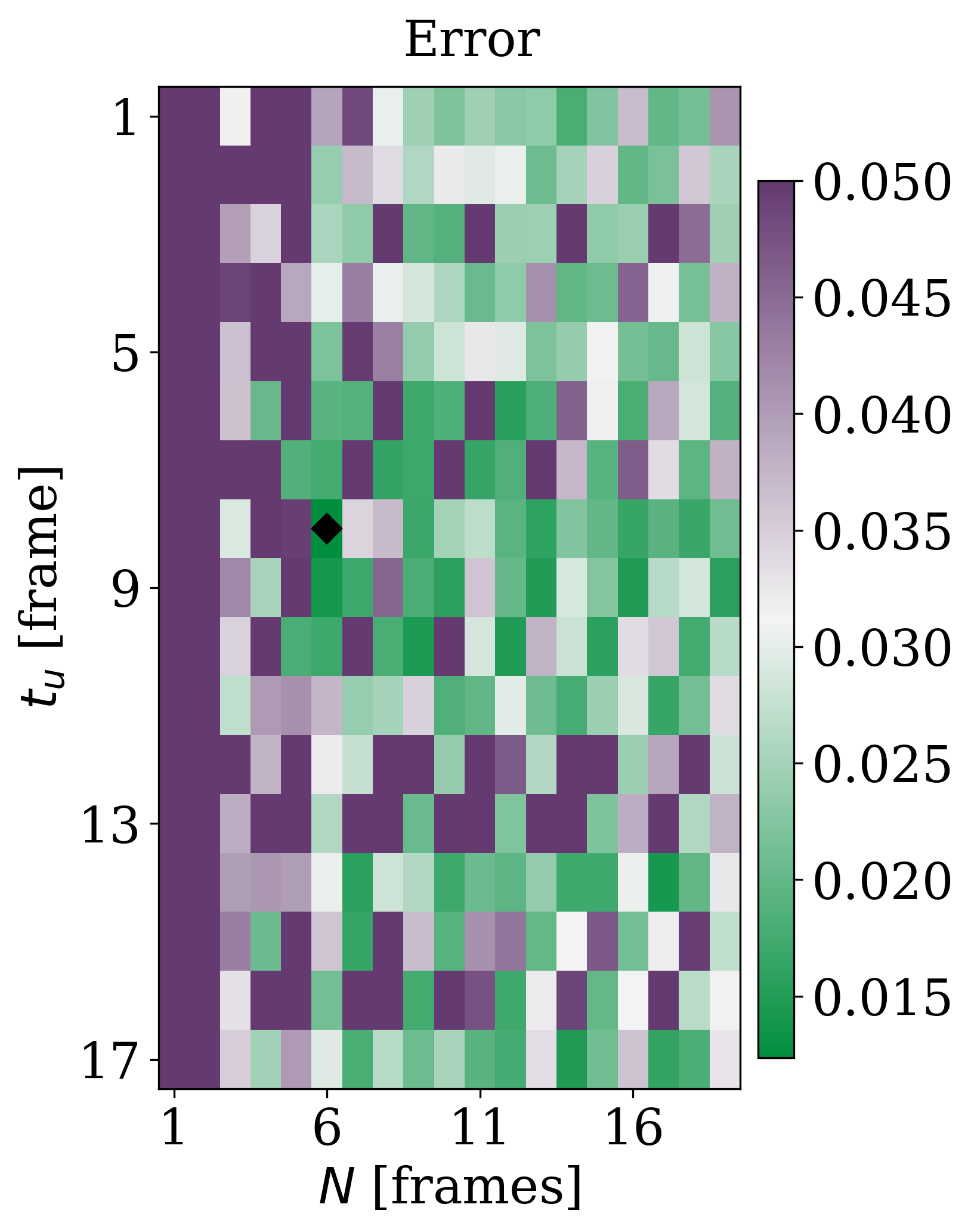}}
        \caption{}
        \label{fig:DCA_error}
    \end{subfigure}
\vspace{+5pt}
\caption{\label{fig:DCA} Applicability of the spectral GME to experimental 2DIR data of the dicyanamide (DCA) vibrational probe in a model ionic liquid electrolyte widely studied in battery science. These 2D spectra span a $64 \times 64$ spectral grid with $\delta \omega_1 = \delta \omega_2 = 3$~cm$^{-1}$ and were sampled at a timestep of $\delta t_2 = 100$~fs. \textbf{(a)} 2DIR spectrum of room temperature ionic liquid mixture [BMIM]$_x$[DCA]$_y$[BF$_4$]$_{1-x-y}$ in the IR at $t_2=0$~ps. We highlight pixels located at $[(5, 13), (9, 13), (13, 13), (13,42)]$ to survey a range of time dependence and intensities. \textbf{(b)} Comparison of 2D spectra at $t_2=11.5$~ps obtained from the experimental benchmark (left) and spectral GME prediction (right). \textbf{(c)} Spectral evolution of the selected pixels along $t_2$. Grey region represents the data used to construct $\langle \bmc{U}_\infty(\delta t_2)\rangle$, the circles denote the spectral GME predictions. The inset shows the dynamics of the bleach signal (purple pixel) that crosses zero. \textbf{(d)} Root mean squared error heatmap showing the sum of only the four tagged pixels at all predicted time points, normalized by the number of predicted time points. The point marked by a black diamond represents the choice in panel~\textbf{(c)} and is the global minimum. Consequently, we choose $t_u=0.8$~ps, $\tU=1.4$~ps, and $\xi=0.01$.}
\end{figure*}

%After appropriately choosing the timestep from the available data, 
We deploy our same workflow that yielded the results in Figure~\ref{fig:theorydata}. Although our pseudoinverse step is successful, the statistical noise in the experimental data prevents finding a stable cutoff, yielding a diverging dynamical system when attempting to evolve Eq.~\eqref{eq:spectralUGME} for any individual choice of $\tU$. This problem arises when constructing GME-type equations from data containing noise,\cite{Cao2020a} and we have previously shown how to solve it.\cite{Dominic2022} Here, we mitigate this noise by adapting the averaging procedure introduced for biomolecular simulations\cite{Dominic2022}, which exploits the plateau of the propagator beyond its Markovian timescale to obtain a stable estimate of $\bmc{U}_\infty(\delta t_2)$. To obtain this stable estimate in practice, one can perform a simple average of $\bmc{U}$ over a window of $N$ frames
\begin{equation}
\label{eq:SGME}
    \ev{\bmc{U}_{\infty}(\delta t_2)} = \frac{1}{N}\sum_{n=u}^{N + u-1} \bmc{U}(t_n),
\end{equation}
and use this averaged value to predict the dynamics at later times
\begin{equation}
\label{eq:AVESGME}
    {\bf S}(\tU + n \delta t_2) = \ev{\bmc{U}_\infty(\delta t_2)}^n {\bf S}(\tU).
\end{equation}
Consistent with Ref.~\onlinecite{Dominic2022}, we replace $\bmc{U}(t_2)$ with $\ev{\bmc{U}_\infty (\delta t_2)}$ at the offset of averaging, $\tU \equiv t_{N+u-1}$, rather than at the onset to maintain `exact' data for longer, resulting in lower error. While our method introduces two new parameters $\{t_u, \tU \}$, we illustrate a simple protocol to automate their selection in SI Sec.~4. Figure~\ref{fig:cy3_error} illustrates how the root mean squared error of the predicted dynamics varies as a function of $\{t_u, \tU \}$. The predicted dynamics shown in~\ref{fig:cy3_pixels} is $(t_u, \tU) = (50, 200)$~ps and have an error of $E = 1.02\%$. We choose the set of $(t_u, \tU)$ in~\ref{fig:cy3_pixels} to minimize the amount of experimental data utilized, while maintaining a low error.

Figure~\ref{fig:cy3_pixels} shows the evolution of representative pixels in the 2D spectrum corresponding to different regions of the spectrum, with different intensities and time-dependence. We find that our predicted spectrum, constructed from $\bm{\mc{U}}(\tau_\mc{U})$ obtained using only the first 200~ps of experimental data, accurately recovers the original experiment over the entire 600~ps timescale over which data were collected, as shown in the comparison of panels in Fig.~\ref{fig:cy3_prediction}. In Fig.~\ref{fig:cy3_pixels}, we extend the evolution to equilibrium, revealing the relaxation of the spectrum over an additional order of magnitude out to 3500~ps. Hence, since statistical noise becomes more experimentally expensive to tame at long $t_2$, the spectral GME can reduce the cost of such a measurement by at least a factor of 20. To confirm the robustness of our protocol and predictions, we repeat the process on two independent replicates of the same experiment, where we again recover remarkable agreement with each data set and globally across all three (see SI Sec.~7).

%%%%%%%%%%%%%%%%%%%%%%%%%%%%%%%%%%%%%%%%%%%%%%%%%%%%
\subsection{Experimental 2DIR Data}
\label{sec:Expt2DIR}

Although we have focused exclusively on 2DES thus far, our spectral GME is broadly applicable to other 2D spectroscopies. Below, we illustrate this breadth by considering 2DIR spectra.

\subsubsection{[BMIM][DCA] Ionic Liquid}
\label{ssec:DCA}

We begin with a model battery electrolyte---an electrochemical system consisting of a mixture of ionic liquids, 1-butyl-3-methylimidazolium tetrafluoroborate (BmimBF4) and 1-butyl-3-methylimidazolium dicyanamide (BmimDCA), at room temperature, between copper electrodes. DCA is an IR-active reporter that is sensitive to water concentration,\cite{Tibbetts2023} thus giving a path to analyzing water migration and solvation dynamics in electrochemical systems, where water content can both enhance or deteriorate cell performance.\cite{Bella2015} 

We have chosen this electrochemical ionic liquid system because it presents significant experimental and theoretical challenges that can be expected to push the limits of applicability of our spectral GME. First, these ionic liquids exhibit slow configurational dynamics characteristic of high viscosity glass formers.\cite{Chang2021} %Bagno2007, Moreno2008, 
Glasses have generally represented a major challenge to GME-based approaches, like mode coupling theory,\cite{BookGoetze} a leading theoretical framework for predicting and understanding the glass transition.
%\cite{Janssen2018} 
This is because GMEs perform best when there is a large separation of timescales between projected and unprojected states. In glassy systems, commonly employed projection operators that probe matter density and current lead to memory kernels with long-time tails---i.e., long-lasting non-Markovianity---that subverts the expected efficiency benefits associated with short-lived non-Markovian generators.\cite{Reichman2005} A critical test of the spectral GME's utility is its ability to offer a short non-Markovian lifetime and thus reduce the measurement burden in characterizing slow solvation dynamics. Second, the combination of slow solvation dynamics and electrochemical degradation occurring during measurement cause 2DIR spectra to exhibit significant noise and spectral background, which increases strongly with increasing $t_2$, to the point where major artifacts appear in the spectra at long $t_2$. Consistent with most efficiency-boosting GMEs,\cite{Shi2003a, Cohen2011a, Montoya-Castillo2016, Mulvihill2021a} the spectral GME utilizes only early-time data to predict the full evolution of the dynamical object, in this case, the 2D spectrum. Hence, one might anticipate that the spectral GME can offer a means to ameliorate the effect of noise and potentially remove the long-$t_2$ artifacts. Thus, the present 2DIR example offers an important test of the spectral GME in its ability to address both slow systems with poor timescale separation and correct for noisy long-$t_2$ spectra subject to spectral artifacts. 

Figure~\ref{fig:DCA_t2-0} shows the 2DIR $t_2=0$ spectrum in the window where, in order of increasing frequency, the DCA $\mathrm{C}\equiv \mathrm{N}$ anti-symmetric stretch, symmetric stretch, and the combination band between $\mathrm{C}-\mathrm{N}$ symmetric and anti-symmetric stretches appear.\cite{Tibbetts2023} Strikingly, the features in this 2DIR spectrum are sharper than those in the 2DES relative to the resolution across both excitation and emission axes, leaving much of the matrix null or, as we discuss below, consisting largely of noise. Figure~\ref{fig:DCA_prediction} shows that, at long $t_2$, scattering of the higher frequency $\omega_3$ signal swamps detection, rendering the cross peak between the combination band and the asymmetric stretch in that region undetectable. Hence, in this protocol, one should not use the total root mean squared error as the convergence metric, because a well-behaved, smooth extrapolation that correctly represents zero signal regions exhibits high error when subtracted from the systematic noise background at long times. We therefore consider only the error of the displayed pixels. The checkerboard pattern of the error heatmap (Fig.~\ref{fig:DCA_error}) represents the frequency of the noise, with choices of $\bm{\mc{U}}(\tau_\mc{U})$ in purple failing to return sensible spectra. Yet, despite this level of noise, the existence of many green regions in the error plot suggests that many sensible approximations to $\bm{\mc{U}}(\tau_\mc{U})$ exist. 

The point of lowest error is marked by a diamond at $t_u=8$~frames ($800$ fs), $N_{\rmm{window}}=6$~frames ($600$ fs). This cutoff is remarkably short given the multi-picosecond timescale of the spectrum, and has two impressive consequences. The first is agreement between the prediction and the measured value of the yellow pixel at long times. This purple pixel is initially within the diagonal bleach, between the diagonal and the anharmonic emission feature to the red. However, a distinct emissive signal grows in at later times from a dark state,\cite{Tibbetts2023} and the pixel signal crosses zero instead of converging to it from below. The spectral GME \textit{predicts} all this, despite seeing only the first 1.4~ps. The second is that the early time data is relatively clean. The result at long times is astonishing: Fig.~\ref{fig:DCA_prediction} shows that our predicted spectra after the cutoff are \textit{noise-free}, uncovering the \textit{real} form of the spectrum in the high $\omega_3$ region at late $t_2$. The ability to remove experimental noise accruing at long times in microscopy data arising directly from experiment indicates that adopting the spectral GME can facilitate significantly more accurate and efficient 2D spectroscopy data collection.

\begin{figure}[!h]
% \vspace{-12pt}
\begin{center} 
    \resizebox{.45\textwidth}{!}{\includegraphics{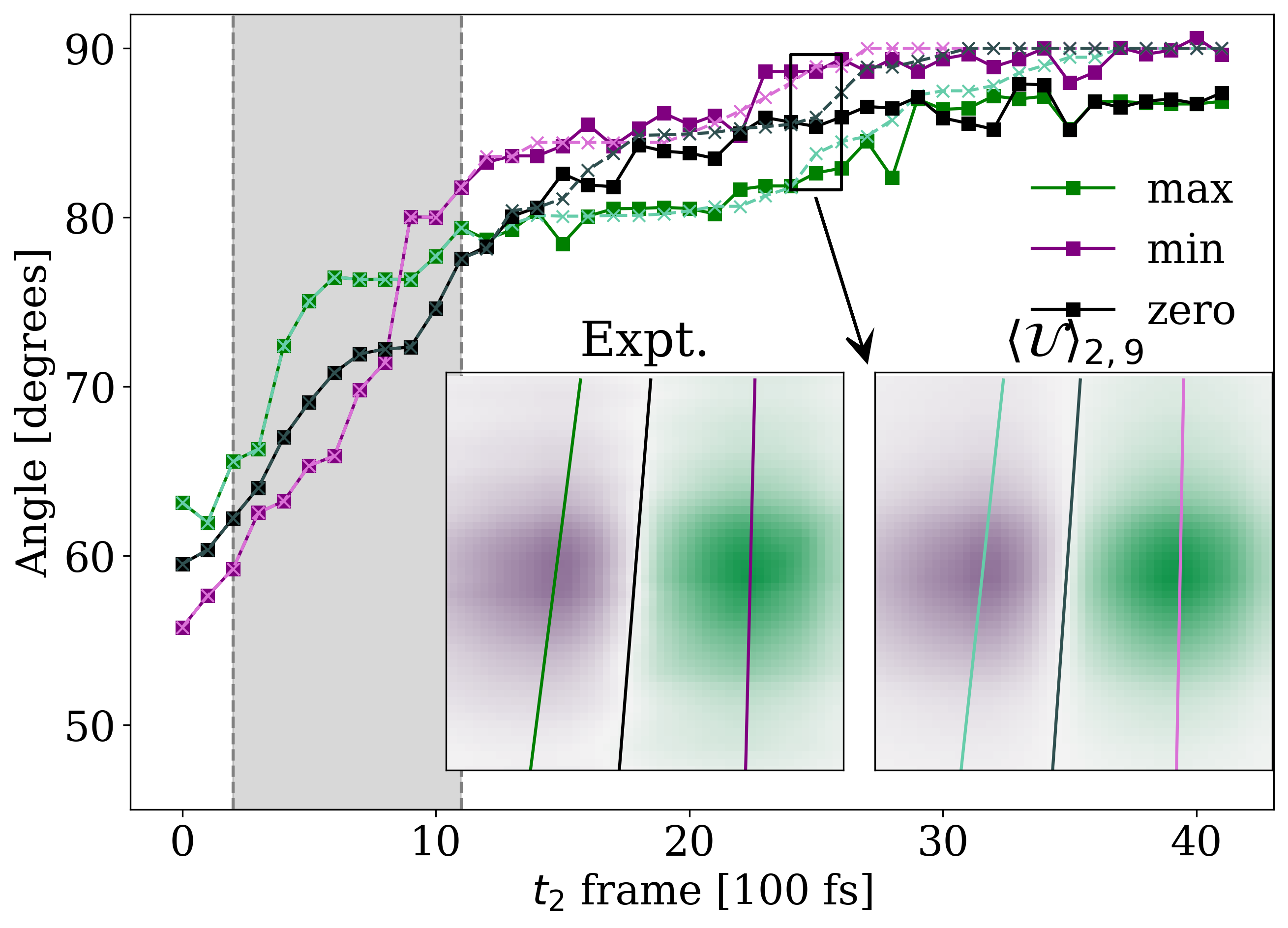}}
\vspace{-2pt}
\caption{\label{fig:lukedata} Applicability of the spectral GME to capture the evolution of the central line slope (CLS) of 2DIR data of the N3 dye undergoing spectral diffusion. We interrogate the evolution of the CLS for two spectral features at $\omega_1=2100$ cm$^{-1}$ and $2125$ cm$^{-1}$, with $\omega_3=2133$ cm$^{-1}$ for both features, corresponding to the symmetric and antisymmetric $\mathrm{C}\equiv\mathrm{N}$ stretches in N3. Insets show both the reference experimental (left) and the spectral GME-predicted (right) CLSs for the selected spectral features. We apply modest linear interpolation to double the number of pixels along each frequency axis, mirroring experimental processing steps for the center line slope. For the result without interpolation, see Fig.~S7.}
\end{center}
\vspace{-18pt}
\end{figure}

%%%%%%%%%%%%%%%%%%%%%%%%%%%%%%%%%%%%%%%%%%%%%%%%%%%%
\subsubsection{Center Line Slope (CLS): A delicate measure}
\label{ssec:CLS}

The full evolution of the 2D spectrum notwithstanding, any tool that aims to remove noise or predict the 2D spectrum based on limited data must also accurately predict the often adopted analysis tools in the field. One such central tool is the center line slope (CLS),\cite{Kwak2007} which is related to (the inverse of) the frequency-frequency correlation function, can reveal the quantitative balance between homogeneous and inhomogeneous contributions to the spectral diffusion, and provides a link between experimental lineshapes and  simulation.\cite{Eaves2005, Corcelli2004, Laage2011, Baiz2020} Predicting the CLS dynamics offers a difficult test for the spectral GME as the low pixel density in the 2DIR features means that small errors can lead to large deviations in the CLS extraction (see Fig.~S7). We illustrate the spectral GME's performance on the commercially available N3 dye, \{[cis-bis(isothiocyanato)bis(2,2'-bipyridyl-4,4'-dicarboxylato ruthenium(II)\}, which is popular for photochemical applications, including in dye sensitized solar cells,\cite{Farah2022} and its ethanoate derivative has been used as a model chromophore for 2DIR studies of vibrational energy propagation pathways and how they differ upon photoexcitation.\cite{Fedoseeva2014} In our spectral range, we probe the isothiocyanate ligands of the fully deprotonated\cite{Farah2021} molecule, N3$^{4-}$. 

Using only raw 2DIR data, it can be challenging to obtain an appropriate SVD cutoff, even when employing the $\boldsymbol{\mathcal{U}}$ averaging procedure. In our example,  it was necessary to truncate the singular values aggressively, incurring minor error before the cutoff---i.e., before any dynamical evolution---to avoid long-time instabilities. While one can still obtain satisfactory agreement (see SI Sec.~8), we found it beneficial to employ data enrichment protocols common in 2DIR to tame this problem and recover excellent performance with the spectral GME. As part of post-processing for kinetic analysis, it is typical to apply spline fits to the data to more faithfully represent the shape of the spectral features and thus track the spectral diffusion mathematically.\cite{Tibbetts2023} Here, we opt for a modest doubling of the number of points along each axis, which again makes it easy to identify the SVD cutoff. 

We present the spectral GME performance on the splined data in Fig.~\ref{fig:lukedata}. Satisfyingly, the predicted CLS is smooth without large jumps and closely matches the experimental (interpolated) data. At the longest $t_2$ times, the predicted data completes diffusion to fully vertical lines, indicating $\omega_1$ and $\omega_3$ are decorrelated, as expected. In contrast, the experimental data yield vertical CLS curves at long $t_2$ only for the negative feature, with the phase line (zero signal) and the companion maximum line both leveling off just before. Such unphysical results are likely a manifestation of error accruing in the long-time spectra, which the spectral GME eliminates. Indeed, these errors are absent in the early time data on which the GME is parameterized, highlighting again the remarkable advantage of correcting long-time errors that the spectral GME provides. 

%%%%%%%%%%%%%%%%%%%%%%%%%%%%%%%%%%%%%%%%%%%%%%%%%%%%
\section{Conclusion}
\label{sec:conclusion}

We have thus developed a dimensionality reduction to \textit{predict} the outcome of 2D spectroscopic experiments using only their early-time data, reducing data collection costs by orders of magnitude and even removing statistical noise and artifacts that prove difficult, even impossible, to remove from experiments directly. We provided rigorous theoretical proof illustrating when and how such a one-time GME can be constructed directly from 2D experiments and tested the applicability of our method using simulated 2D responses from widely invoked energy transfer models in the condensed phase. Beyond such proofs of principle, we demonstrated that our spectral GME works for actual experiments. In particular, we showed that one need not take measurements past a particular waiting time to know the form of the 2D spectrum out to equilibrium in a range of experimental datasets spanning 2DES and 2DIR applied to biologically relevant chromophores exchanging energy and even glassy ionic liquids in model electrochemical cells. We further found that because one can construct the dynamic generator $\bm{\mc{U}}(\tau_\mc{U})$ using only early-time data, where signals are stronger and can be converged more easily, our spectral GME obviates problems associated with poor signal-to-noise ratios or artifacts that become unavoidable at long waiting times, offering a pathway to disentangling otherwise unmeasurable features with no change experimental hardware (and, indeed, less effort). 

To ensure the easy deployability of our spectral GME, we developed a robust data-driven protocol to construct its dynamic generator. It leverages data past the cutoff to identify the optimal onset of Markovianity, therefore requiring at least one detailed study with excess data. In studies where a 2D experiment is performed on a series of system variants (e.g., wild type versus mutants in photosynthesis), or workflows like microscopy where the same 2D experiment must be repeated hundreds or even thousands of times at different locations, temperatures, and compositions, a small number of trial runs to determine $\bm{\mc{U}}(\tau_\mc{U})$ can vastly accelerate full data acquisition, massively cutting down costs. With such savings, one can reallocate time and light pulse budgets to acquire the best, least noisy, highest-resolution data at only early times, and then leverage the spectral GME to propagate these data, with its associated high quality, forward in time at no additional cost. Thus, signals and data quality that was previously unaffordable is now within reach.

\section{Methods}

\subsection{Preprocessing}
\label{ssec:preproc}

Before employing Eq.~\eqref{eq:SGME} to extract the spectral propagator, we first preprocess the $L$ frames of experimental 2D spectra $\{ {\bf S}(t_n) \}_{n=0}^{L-1}$. We specifically ensure uniform sampling along the $\omega_1$ and $\omega_3$ axes and a constant timestep along the $t_2$-axis. We accomplish uniform sampling by applying linear interpolation to ensure both axes have the same density of points. We ensure the constant timestep in $t_2$ by identifying the least common multiple of the available $t_2$ time steps. 

In the case of the 2DES data on the Cy3-Cy5 dimer (see Fig.~\ref{fig:cy3}), which has $t_2$ timesteps of $[1, 2.5, 5, 25, 50]$~ps, we utilize a $25$~ps timestep. This larger timestep enables us to sparsely sample early-time dynamics using timesteps separated by factors of $25$, while still averaging $\bmc{U}(t_2)$ over later times. Such averaging remains essential because we analyze data up to $200$~ps within the $\delta t_2=25$~ps region. To maintain compatibility with experimental datasets, the timestep grid must therefore satisfy a simple constraint: \textit{later timesteps must be integer multiples of earlier ones}. Exponentially spaced timesteps do not satisfy this requirement and therefore are incompatible with our approach.

After preprocessing, each spectrum is a square $M \times M$ matrix with elements ${\bf S}_{ij}(t)$ corresponding to excitation ($\omega_1$) and emission ($\omega_3$) frequencies. Additionally, we normalize each spectrum by the intensity of the most intense pixel in the spectrum, which typically occurs at time zero.

\subsection{Building the Spectral GME}
\label{ssec:BuildingU}

With the preprocessed frames $\{ {\bf S}(t_n) \}_{n=0}^T$ in hand, one extracts the corresponding propagator time series $\{ \bmc{U}(t_n) \}$ by inverting Eq.~\eqref{eq:spectralUGME}:
\begin{equation}
\label{eq:U-constructor}
    \bm{\mc{U}}(t) = {\bf S}(t+\delta t){\bf S}^{-1}(t).
\end{equation}
Hereafter, we use $t$ to denote time rather than $t_2$. Because the spectral frames are sparse and generally singular, one must replace the inverse in Eq.~\eqref{eq:U-constructor} with a numerically stable pseudoinverse.

To achieve this, we employ singular value decomposition (SVD) to characterize the spectral decomposition of the frames and use this analysis to inform our pseudoinverse. The SVD of the $n^{\rm th}$ frame is given by 
\begin{equation}
\label{eq:SVDecomp}
    {\bf S}(t_n) = {\bf{U \Sigma V}}^\dagger = \sum_{k=1}^M \sigma_k \vec{{\bf u}}_k \vec{\bf v}^{\dagger}_k .
\end{equation}
In Eq.~\eqref{eq:SVDecomp}, ${\bf U}$ and ${\bf V}$ are unitary matrices whose columns contain the left and right singular vectors, respectively, and ${\bf \Sigma}$ is a diagonal matrix with ordered singular values $\sigma_{1} \geq \sigma_2 \geq \cdots \geq \sigma_M \geq 0$. The $k^{\rm th}$ left ${\bf u}_k$ and right ${\bf v}_k$ singular vectors identify dominant emission and excitation-frequency patterns in the data, while their singular values $\sigma_k$ indicate how much this excitation-emission pair contributes to the overall spectrum. 

We construct a stable pseudoinverse of ${\bf S}(t_n)$ by truncating singular values below a user-defined threshold $\xi$ such that all $\sigma_k < \xi$ are set to zero. We implement this protocol in Python using the \texttt{numpy.linalg} package. This truncation suppresses directions in the spectral data that are poorly resolved, redundant, or dominated by experimental noise, thereby stabilizing the subsequent construction of the spectral propagator. We analyze our choice of $\xi$ using the error metric defined in the next section, and provide additional insight into our SVD implementation and $\xi$ in SI Sec.~5.

\subsection{Error Metric}
\label{ssec:errormetric}

To extract the spectral propagator $\ev{\bmc{U}_\infty(\delta t)}$ in Eq.~\eqref{eq:SGME} from the $T$ experimental frames, one must determine three parameters: the onset of averaging $t_u$, the offset of averaging $\tU$, and the pseudoinverse threshold $\xi$. We determine the optimal values of $\{ t_u, \tU, \xi \}$ as those requiring the minimal number of frames while satisfying two criteria:
\begin{enumerate}
    \item \textbf{Short-time accuracy.} The propagator reproduces the experimentally measured dynamics.
    \item \textbf{Long-time stability.} The propagator predicts stable dynamics when propagated beyond the experimentally accessible timescales.
\end{enumerate}

To quantify the extent to which these choices satisfy our criteria, we introduce the root mean squared error metric:
\begin{equation}
\label{eq:error}
     \! E(\tU, \xi) \! =
     \left\{ 
     \sum_{i=\tU+1}^{T}\!
     \sum_{j,k}^M
     \frac{[{\bf S}_{jk}^{\rm Ref}(i) - {\bf S}_{jk}(i;\tU, \xi)]^2}{M^2(L-\tU)} \! \right\}^{1/2} \!.
 \end{equation}
This measures the root mean squared error between the experimental reference spectra ${\bf S}_{jk}^{\rm Ref}(i)$ and the spectral GME prediction obtained using trial values $\{ t_u, \tU, \xi \}$, averaged over all spectral pixels $(j, k)$ and timesteps $i > \tU$. Because we replace $\bmc{U}(t)$ with $\ev{\bmc{U}_\infty (\delta t)}$ after the offset of averaging, the first summation in Eq.~\eqref{eq:error} only includes timesteps $i> \tU$. Frames before $\tU$ do not contribute to the error as our method reproduces these exactly by construction. In principle, these error curves should monotonically approach zero. In practice, however, experimental data are finite and noisy. We therefore consider the error on a case-by-case basis and refer the reader to SI Sec.~4 and Sec.~6 for more details. We provide pseudocode for our algorithm in SI Sec.~4.

\subsection{2DIR Experimental Method: [BMIM][DCA]}

We collected the 2DIR spectra from a home-built 2DIR spectrometer (see Refs.~\onlinecite{Luther2016, Luther2025} for full details on the workflow). In these experiments, we used a pump-probe beam geometry with pulses centered at 4700~nm and a duration of 100~fs. We set the average power before the sample to 7~mW in the probe line and 14~mW in the pump line, corresponding to 70~nJ per pulse at a repetition rate of 100 kHz. 

We used a mid-IR pulse shaper to control the phase and time delays between the two pump pulses.  The $t_1$ delay time is scanned from 0 ps to 4~ps in steps of 10~fs with a total of 401 delay steps. We zero-padded the data along the pump axis to 802 points and Fourier transformed to generate the $\omega_1$ frequency axis.  The third-order signal is sent to a monochromator and collected on a 64-element MCT array detector. The spectra were collected using a four-frame phase cycle with a rotating frame of 1900 cm$^{-1}$. These experimental conditions produce a spectral resolution of 4.2 cm$^{-1}$ for $\omega_1$ and 3.4 cm$^{-1}$ for $\omega_3$ in the 2D IR spectra.  Each 2DIR spectrum collected for each $t_2$ step consists of 300 2D IR spectra averaged together. A polarizer is used to split the probe line to implement a noise reduction scheme using two-detectors.\cite{Feng2019} The waiting time ($t_2$) between the second pump pulse and the probe pulse was scanned from 0 to 10.9~ps in 100~fs steps.

\section{Acknowledgments}
This work was partially supported (A.M.C.) by an Early Career Award in CPIMS program in the Chemical Sciences, Geosciences, and Biosciences Division of the Office of Basic Energy Sciences of the U.S.~Department of Energy under Award DE-SC0024154. A.M.C.~also acknowledges the support from a David and Lucile Packard Fellowship for Science and Engineering. T.S.~is the recipient of an Early Career Fellowship from the Leverhulme Trust.
%\end{acknowledgments}

 \section*{AUTHOR DECLARATIONS}
% \vspace{-12pt}
 \section*{Conflict of Interest}
% \vspace{-8pt}
 The authors have no conflicts to disclose.
% \vspace{-12pt}
% \section*{Author Contributions}
% \vspace{-8pt}
% Thomas Sayer: Formal analysis (lead); Investigation (lead);
% Writing – original draft (lead); Writing – review \& editing (equal).
% Andrés Montoya-Castillo: Conceptualization (lead); Supervision
% (lead); Writing – review \& editing. Other people: Experimental data; hopefully they/you review and edit!
% (equal).

\section*{DATA AVAILABILITY}
\vspace{-8pt}
The data that support the findings of this study are available
from the corresponding author upon reasonable request.

\subsection*{References}
\vspace{-14pt}
\bibliography{sgme-refs}

\end{document}

\title{Supplementary information for "Short-lived memory in multidimensional spectra encodes full signal evolution"}
\author{Thomas Sayer}
\thanks{These two authors contributed equally}
\affiliation{Department of Chemistry, University of Colorado Boulder, Boulder, CO 80309, USA\looseness=-1} 
\affiliation{Department of Chemistry, Durham University, South Road, Durham, DH1 3LE, United Kingdom\looseness=-1} 

\author{Ethan H. Fink}
\thanks{These two authors contributed equally}
\affiliation{Department of Chemistry, University of Colorado Boulder, Boulder, CO 80309, USA\looseness=-1} 

\author{Zachary R. Wiethorn}
\affiliation{Department of Chemistry, University of Colorado Boulder, Boulder, CO 80309, USA\looseness=-1} 
\author{Devin R. Williams}
\affiliation{Department of Chemistry, Colorado State University, Fort Collins, Colorado 80523-1872, USA\looseness=-1} 
\author{Anthony J.~Dominic III}
\affiliation{Department of Chemistry, University of Colorado Boulder, Boulder, CO 80309, USA\looseness=-1} 
\author{Luke Guerrieri}
\affiliation{Department of Chemistry, Colorado State University, Fort Collins, Colorado 80523-1872, USA\looseness=-1} 
\author{Yi Ji}
\affiliation{Department of Chemistry, Massachusetts Institute of Technology, Cambridge, Massachusetts 02139, USA\looseness=-1} 
\author{Veronica Policht}
\affiliation{Department of Physics, Colorado School of Mines, Golden, Colorado 80401, USA\looseness=-1} 
\author{Jennifer Ogilvie}
\affiliation{Department of Physics, University of Ottawa, 150 Louis-Pasteur Pvt., Ottawa, ON K1N 6N5, Canada\looseness=-1} 
\author{Gabriela Schlau-Cohen}
\affiliation{Department of Chemistry, Massachusetts Institute of Technology, Cambridge, Massachusetts 02139, USA\looseness=-1} 
\author{Amber Krummel}
\affiliation{Department of Chemistry, Colorado State University, Fort Collins, Colorado 80523-1872, USA\looseness=-1} 

\author{Andr\'{e}s Montoya-Castillo}
\homepage{Andres.MontoyaCastillo@colorado.edu}
\affiliation{Department of Chemistry, University of Colorado Boulder, Boulder, CO 80309, USA\looseness=-1}

\maketitle

\tableofcontents
\newpage

\setcounter{page}{1}
\renewcommand{\thepage}{S\arabic{page}}
\setcounter{figure}{0}
\renewcommand{\thefigure}{S\arabic{figure}}
\setcounter{enumiv}{0}
\renewcommand{\theenumiv}{S\arabic{enumiv}}

\section{Dimensionality Reduction}\label{appendix:dimensionality_reduction}
GMEs aid interpretability by reducing the dimensions one tracks in time from macroscopically many to only a few of interest. This number of dimensions equals the number of components in the time-dependent correlation matrix, $\mc{C}(t)$, where the rows and columns encode the few observables one wishes to correlate at the initial and final times. In the case of, for example, the side-side correlation function description of protein folding kinetics,\cite{Dominic2022, Husic2018, Swope2004, Hummer2015} the number of dimensions reduces from scaling with the system size to being equal to the number of metastable states one intends to track, e.g., the properly folded, unfolded, and a few misfolded states. These states, then, facilitate the interpretation of observed phenomena.

The reduction in the cost of encoding and predicting the observable dynamics depends on the separation of timescales between the explicitly tracked and implicitly encoded states in the GME. The generator of the GME dynamics (i.e., memory kernel, time-dependent rate matrix, or transfer operator, depending on the type of GME) has a finite lifetime and, depending on the balance of tracked and untracked states, this lifetime can be shorter than the lifetime of $\mc{C}(t)$ itself. This can greatly enhance efficiency in prediction and storage. Recent algorithmic advances have enabled researchers to employ short-time reference $\mc{C}(t)$ dynamics to predict full dynamics exactly and cheaply,\cite{Shi2003a, Cohen2011a, Montoya-Castillo2016, Mulvihill2021a} offering multiple order-of-magnitude speedups in simulating, for example, protein folding,\cite{Dominic2022, Dominic2023} polaron formation and transport in solid-state semiconductors,\cite{Bhattacharyya2024b, Bhattacharyya2025b} and exciton transfer pathways in light-harvesting complexes.\cite{Sayer2023,  Mulvihill2021a, Pfalzgraff2019a} Critically, for a 2D spectroscopic experiment, being able to employ the GME framework could reduce the time between experiments and circumvent resource-intensive convergence at long waiting times where experimental error accrues most dramatically, since signals generally assume their smallest values compared to the background. Employing a GME directly on spectroscopic data in the post-processing phase might thus offer a system-agnostic means to reduce the effort needed to obtain 2D spectra. 

Why does combining the GME framework with experimentally measured spectroscopic data present a significant challenge to theory? Consider the simpler case of linear spectroscopy, where the spectrum takes the form,
\begin{equation}\label{appendix-eq:linear-spec}
    I(\omega) \propto \int \dd{t} e^{i\omega t} R^{(1)}(t),
\end{equation}
with the first-order response function, 
\begin{equation}\label{appendix-eq:R1}
    R^{(1)}(t) = \mathrm{Tr}\{\tilde{\bm{\mu}}(0) \hat{\bm{\mu}}(t) \hat{\rho}_{0}\}.
\end{equation}
Here, $\tilde{\bm{\mu}}(t)\ \cdot\ \equiv [\hat{\bm{\mu}}(t),\ \cdot\ ]$ is the commutator with the transition dipole, $\hat{\bm{\mu}} = \sum_{i\neq j}\bm{\mu}_{i,j}(\mathbf{q})(\ket{i}\bra{j} + \ket{j}\bra{i})$ is the transition dipole connecting discrete states, $\bm{\mu}_{i,j}(\mathbf{q})$ is the matrix element of the transition dipole in the basis whose transitions one is interested in (which generally depends on the environmental conformation encoded by atomic positions $\mathbf{q}$), and $\rho_{0} = e^{-\beta \hat{H}}/{\rm Tr}\{ e^{-\beta \hat{H}}\}$ is the canonical density of the entire system. Time-dependent operators are given by the action of the propagator, $\hat{O}(t) = e^{i\mathcal{L}t}\hat{O}$, where $\mathcal{L} = [\hat{H}, ...]$ is the Liouvillian.

The linear response function, $R^{(1)}(t)$, is reminiscent of the type of correlation function for which one could design a GME with a short-lived generator. For example, one could pursue the family of equilibrium correlation functions obtained from the outer product of all discrete states probed by the spectroscopy, $\mc{C}_{ij,kl}(t) = \mathrm{Tr}\{\rho_{eq}\ket{i}\bra{j}e^{i\mathcal{L}t}\ket{k}\bra{l}\}$.\cite{Montoya-Castillo2017} When the excitation energy is larger than the thermal energy, one can even adopt a simpler nonequilibrium formulation that also probes the spectroscopically active states and has formed the basis of many GME-based approaches to the dynamics of open quantum systems.\cite{Shi2004b, Cohen2011a, Cohen2013, Kidon2018a, Kelly2013a, Montoya-Castillo2016, Mulvihill2021a} All types of correlation functions have been shown to have generally short-lived generators. Crucially, one can express $R^{(1)}(t)$ as a linear combination of $\mc{C}_{ij,kl}(t)$. \textit{Yet, the experimentally accessible linear spectrum does not give access to these terms individually, preventing one from directly employing experimental data to build such a GME.}

The problem worsens in 2D spectra. Specifically, within linear response theory, the 2D spectrum arises from a double Fourier transform of the 3-time response function
\begin{equation}\label{appendix-eq:2d-spectrum}
    S(\omega_1, t_2, \omega_3) = \sum_{\pm} \iint \dd{t_1} \dd{t_3}\ e^{i\omega_3 t_3 \pm i\omega_1 t_1} R^{(3)},
\end{equation}
where
\begin{equation}\label{appendix-eq:third-order-optical-response}
    R^{(3)} = {\rm Tr} \Big\{ \tilde{\bm{\mu}}(t_3+t_2+t_1) \tilde{\bm{\mu}}(t_1+t_2) \tilde{\bm{\mu}}(t_1) \hat{\bm{\mu}}(0)\hat{\rho}_0\Big\},
\end{equation}
where the sum over rephasing ($-$) and nonrephasing ($+$) contributions yield the absorptive spectrum.\cite{BookMukamel} In simulations of the spectral response two approximations are commonly invoked: (1) the impulsive limit, in which the spectral bandwidth of pump and probe pulses are treated as constant over the range of the 2D spectrum; and (2) the rotating wave approximation, which allows one to decompose the total response into rephasing and nonrephasing contributions by expanding the commutators of Eq.~1 and neglecting rapidly oscillating terms. In Eq.~2 we have adopted the impulsive limit and justify this choice in SI Sec.~2, which rigorously derives Eq.~2. While we\cite{Sayer2024} have built on earlier work\cite{Ivanov2015} to develop a multitime GME that reduces the computational cost of simulating 2D spectra---and even offered a new way of quantifying $n$-photon correlations in the spectra---the resulting multitime GME requires all 1-, \mbox{2-,} and 3-photon correlators separately to construct the 2D spectrum, and in the same basis of outer product states, $\ket{i}\bra{j}$, which are not accessible individually in the 2D spectrum. That is, upon measuring the 2D spectrum, one loses the information required to construct a multitime GME. 

%-------------------------------------------------------------------------
%-------------------------------------------------------------------------
\section{Derivation for the 2D-spectrum with minimal approximations}
\label{appendix:derivation-for-the-impulsive-spectrum}

To simulate absorptive 2D spectra requires one to sum the rephasing and non-rephasing contributions. These two contributions are obtained by Fourier transforms of kernels along the $t_3$ axis ($e^{i \omega_3 t_3}$) and conjugate kernels along the $t_1$ axis ($e^{-i \omega_1 t_1}$ and $e^{+i \omega_1 t_1}$ for rephasing and non-rephasing, respectively). This choice is commonly justified following multiple approximations to the third-order polarization function which encodes the third-order optical response function (Eq.~1). 

In contrast to the conventional approach \cite{BookMukamel}, our expression for the 2D-spectrum in Eq.~2 does not partition the total response into separate contributions while maintaining the conventional definition of rephasing and non-rephasing Fourier transformations. Beginning with the formal definition of the measured third-order polarization, we show that the conjugate kernels over $t_1$ arise naturally from a spatio-temporal relationship arising from the phase matching conditions prescribed by the collective wave vector of the measured third-order polarization field. Once established, we arrive at Eq.~2 by invoking the impulsive limit which constitutes the only approximation made to our expression for the 2D-spectrum. Lastly, we briefly discuss the RWA to show how rephasing and non-rephasing response functions can be obtained from Eq.~2 of the main text, and spectra obtained in and out of the RWA.

%-------------------------------------------------------------------------
\subsection{The third-order polarization envelope}

In this section, we illustrate how we arrive at the expression for the complex 2D spectrum in Eq.~2. To do this, we begin with a general expression for the third-order macroscopic polarization,
\begin{equation}
\begin{split}\label{appendix-eq:initial-third-order-polarization}
    \bm{P}^{(3)} (t, T, \tau) = &\iiint_{- \infty}^{\infty}  \dd{t_3} \dd{t_2} \dd{t_1}  S^{(3)} (t_3, t_2, t_1) \\ & \times \bm{E}^{(3)} (t-t_3, T-t_2, \tau-t_1).
\end{split}
\end{equation}
Here, $S^{(3)}$ is the impulse response containing the third-order optical response function $R^{(3)}$ (given by Eq.~1),
\begin{equation}\label{appendix-eq:third-order-impulse-response}
    S^{(3)}(t_3, t_2, t_1) = i^3 \Theta(t_3)\Theta(t_2)\Theta(t_1) R^{(3)}(t_3, t_2, t_1).
\end{equation}
The third-order electric field in Eq.~\eqref{appendix-eq:initial-third-order-polarization} is given by, 
\begin{equation}\label{appendix-eq:third-order-electric-field-expanded}
\begin{split}
    &\bm{E}^{(3)} (t-t_3, T-t_2, \tau-t_1) =  \bm{E}_3 (t - t_3)  \\&\times\bm{E}_2 (t + T - t_2 - t_3)  \bm{E}_1 (t + T + \tau - t_1 - t_2 - t_3).
\end{split}
\end{equation}
Here, we see that Eq.~\eqref{appendix-eq:third-order-electric-field-expanded} contains combinations of various time arguments with $t$ denoting an excitation timescale and $\tau$ and $T$ denoting the optical time delays between the two pump-pulses, $\bm{E}_1$ and $\bm{E}_2$, and the second pump- and probe-pulse, $\bm{E}_2$ and $\bm{E}_3$, respectively. We now define a polarization envelope by first expressing each electric field as a sum over conjugate pairs, 
\begin{equation}\label{appendix-eq:complex-field-general}
    \bm{E}_i (t) = \sum_{\sigma_i} e^{i \bm{k}_{\sigma_i} \cdot \bm{r}} \int_{-\infty}^{\infty} {\rm d} \omega_i  \bm{E}_i (\omega_i) e^{- i \omega_{\sigma_i} t}.
\end{equation} 
Here, $\bm{k}_{\sigma_i}$, $\omega_{\sigma_i}$ and $\bm{E}_i(\omega_{i})$ denote the phase wave vector, phase frequency and complex frequency distribution of the field with center frequency $\Omega_i$ ($\omega_i = \omega_{\sigma_i} - \Omega_i$), respectively. Phases of these quantities are determined by $\sigma_i \equiv \{-,+\}$ with the following definitions:
\begin{subequations}
\begin{align}
    \bm{k}_{\sigma_i} &\equiv \sigma_i \bm{k}_{i} = \pm \bm{k}_{i}\label{eq:wavevector_sigma_deff}, \\ 
    \omega_{\sigma_i} &\equiv \sigma_i \omega_{i} = \pm \omega_i \label{eq:omega_sigma_deff}. 
\end{align}
\end{subequations}
Lastly we note 
\begin{equation*}
    \bm{E}_i (t) \equiv \bm{E}_{+_i} (t) + \bm{E}_{-_i} (t),
\end{equation*}
where 
\begin{subequations}
\begin{align*}
    \bm{E}_{+_i} (t) &= \bm{E}_{i} (t), \\
    \bm{E}_{-_i} (t) & = \bm{E}^*_{i} (t).
\end{align*}
\end{subequations}
Now that we have established our notation, we derive the closed form for the 2D spectrum. 

To begin, we note that the \textit{measured signal polarization} is proportional to Eq.~\eqref{appendix-eq:initial-third-order-polarization} by a phase: $\bm{P}^{(3)}_s \propto i \bm{P}^{(3)}$. We isolate the temporal component of $\bm{P}^{(3)}_s$ from its spatial component through a corresponding polarization envelope, $\Pi^{(3)}_{\{\sigma\}}$, that contains all dynamical information. Doing this and substituting Eq.~\eqref{appendix-eq:complex-field-general} into Eq.~\eqref{appendix-eq:initial-third-order-polarization}, we write $\bm{P}_s^{(3)}$ as,
\begin{equation}\label{appendix-eq:signal-polorization}
    \bm{P}^{(3)}_s (t, T, \tau) \propto \sum_{\{\sigma\}} e^{i \bm{k}_{\{\sigma\}} \cdot \bm{r}} \Pi^{(3)}_{\{\sigma\}} (t, T, \tau).
\end{equation}
Here, $\bm{k}_{\{\sigma\}} = \bm{k}_{\sigma_1} + \bm{k}_{\sigma_2} + \bm{k}_{\sigma_3}$ represents the \textit{signal wavevector} distinguishing specific third-order signals such as rephasing ($\bm{k}^{\rm rp}_{\{\sigma\}} = \bm{k}_{-,+,+} =  -\bm{k}_{1} + \bm{k}_{2} + \bm{k}_{3}$) and non-rephasing ($\bm{k}^{\rm nr}_{\{\sigma\}} = \bm{k}_{+,-,+} =\bm{k}_{1} - \bm{k}_{2} + \bm{k}_{3}$) signals
that can be isolated and measured separately in experiment. The signal envelope is now the quantity of interest, 
\begin{equation}\label{appendix-eq:signal-envelope}
\begin{split}
    &\Pi^{(3)}_{\{\sigma\}} (t, T, \tau) = \\ i\iiint_{- \infty}^{\infty}& \dd{\omega_3} \dd{\omega_2} \dd{\omega_1} \bm{E}_3(\omega_3) \bm{E}_2(\omega_2) \bm{E}_1(\omega_1)
     \\  \times \iiint_{- \infty}^{\infty} &  \dd{t_3} \dd{t_2} \dd{t_1} S^{(3)}(t_3, t_2, t_1) e^{-i (\omega_{\sigma_3} + \omega_{\sigma_2} + \omega_{\sigma_1})(t-t_3) } \\&\times e^{-i (\omega_{\sigma_2} + \omega_{\sigma_1})(T-t_2) } e^{-i \omega_{\sigma_1} (\tau-t_1) } .
\end{split}
\end{equation}
Finally, the complex 2D spectrum is given by Fourier transforming against the detection $t$ and excitation $\tau$ time-intervals, 
\begin{equation}\label{appendix-eq:signal-envelope-frequency-domain}
    \Pi^{(3)}_{\{\sigma\}} (\omega_t, T, \omega_\tau) \equiv \iint_{- \infty}^{\infty} \dd{t}\dd{\tau} e^{i(\omega_t t + \omega_{\tau} \tau)} \Pi^{(3)}_{\{\sigma\}} (t, T, \tau).
\end{equation}
Up to this point, we have yet to establish a relationship between the phase of the Fourier kernel chosen when transforming with respect to the excitation ($t$) and detection ($\tau$) timescales (such as Eq.~2 with $t\equiv t_3$ and $\tau \equiv t_1$). To establish this relationship, we first obtain a closed form for Eq.~\eqref{appendix-eq:signal-envelope-frequency-domain}. We then connect our result with the phase-matching prescribed by the measured signal polarization in Eq.~\eqref{appendix-eq:signal-polorization} to show how this spatio-temporal relationship emerges from the mathematics. 

We obtain a closed form of the 2D spectrum by first relating a new set of frequencies $\{ \nu_{\sigma_i}\}$ to the former $\{ \omega_{\sigma_i} \}$ using $\nu_{\sigma_3} = \omega_{\sigma_3} + \omega_{\sigma_2} + \omega_{\sigma_1}$, $\nu_{\sigma_2} = \omega_{\sigma_3} + \omega_{\sigma_2} $, and $\nu_{\sigma_1}~=~\omega_{\sigma_1}$ allowing us to write, 
\begin{subequations}\label{appendix-eq:variable-changes}
\begin{gather}
    \omega_3 = \nu_3 - \sigma_3 \nu_2 \nonumber, \\ \omega_2 = \nu_2 - \sigma_2 \nu_{\sigma_1} \nonumber, \\ \omega_1 = \nu_{1} \nonumber.
\end{gather}
\end{subequations}
We return to Eq.~\eqref{appendix-eq:signal-envelope} and perform the Fourier transform with respect to $t_3$ and $t_1$, 
\begin{equation}\label{appendix-eq:signal-envelope-time-domain-final}
\begin{split}
    &\Pi^{(3)}_{\{\sigma\}} (t, T, \tau) = \\ & i\iiiint_{- \infty}^{\infty} \dd{t_2} \dd{\nu_3} \dd{\nu_2} \dd{\nu_1} S^{(3)}(\nu_3, t_2, \nu_1) e^{-i(\nu_{\sigma_3}t + \nu_{\sigma_1}\tau)} \\ \times&   e^{-i\nu_{\sigma_2}(T-t_2)} \bm{E}_3 (\nu_3 - \sigma_3 \nu_2) \bm{E}_2 (\nu_2 - \sigma_2 \nu_1) \bm{E}_{1} (\nu_1).
\end{split}
\end{equation}
Finally, by substituting Eq.~\eqref{appendix-eq:signal-envelope-time-domain-final} into Eq.~\eqref{appendix-eq:signal-envelope-frequency-domain} and first focusing solely on the integrations over $t$ and $\tau$, we write,    
\begin{equation}
\begin{split}
    &\iint_{- \infty}^{\infty} \dd{t} \dd{\tau} S^{(3)}(\nu_{\sigma_3}, t_2, \nu_{\sigma_1}) e^{i(\omega_t t + \omega_{\tau} \tau)} e^{-i(\nu_{\sigma_3}t + \nu_{\sigma_1}\tau)} 
    \\ & =
    S^{(3)}(\nu_{\sigma_3}, t_2, \nu_{\sigma_1}) \delta(\nu_{\sigma_3}-\omega_t) \delta(\nu_{\sigma_1}-\omega_\tau)
    \\ & =
    S^{(3)}(\nu_{\sigma_3}, t_2, \nu_{\sigma_1}) \delta(\nu_3-\sigma_3\omega_t) \delta(\nu_1-\sigma_1\omega_\tau).
\end{split}
\end{equation}
Thus, the spectrum of the signal polarization becomes,
\begin{equation}\label{appendix-eq:polarization-spectrum}
\begin{split}
    &\bm{P}_s^{(3)} (\omega_t, T, \omega_\tau) = \sum_{\{\sigma\}} e^{i \bm{k}_{\sigma} \cdot \bm{r} } \bm{E}_1(\sigma_1 \omega_\tau) \\ \times \iint_{- \infty}^{\infty}  &  \dd{t_2} \dd{\nu_2} \Bigg\{  e^{-i \nu_{\sigma_2} (T-t_2)} \Pi^{(3)}_{\{\sigma\}} (\sigma_3 \omega_t, T, \sigma_1\omega_\tau) \\ & \hspace{1cm } \times \bm{E}_3 (\sigma_3[\omega_t - \nu_2]) \bm{E}_2 (\nu_2 - \sigma_2 \sigma_1 \omega_\tau) \Bigg\}.
\end{split} 
\end{equation}
Although this result appears rather complex, one can easily write the rephasing (-) and non-rephasing (+) 2D spectral profiles by considering the cases $\{\sigma_1, \sigma_2, \sigma_3\} = \{-, +, +\}$ and $\{\sigma_1, \sigma_2, \sigma_3\} = \{+, -, +\}$, respectively,
\begin{equation}\label{appendix-eq:rephasing-and-nonrephasing-polarization-spectra}
\begin{split}
    &\bm{P}_{\mp}^{(3)}(\omega_t, T, \omega_\tau) = e^{i(\bm{k}_3 \pm \bm{k}_2 \mp \bm{k}_1)\cdot \bm{r}} \Pi^{(3)}(\omega_t, T, \mp\omega_\tau). 
\end{split}
\end{equation}
While this result illustrates why one must consider conjugate phases in the Fourier kernels when transforming over the excitation ($\tau$) and detection timescales ($t$), we have yet to establish how Eq.~\eqref{appendix-eq:rephasing-and-nonrephasing-polarization-spectra} reduces to Eq.~2, which does not contain any of the electric fields. To make this connection, one must consider a commonly invoked approximation to calculate 2D spectra that considers light pulses with extremely short (broad) temporal (frequency) envelopes.

%-------------------------------------------------------------------------
\subsection{Rephasing and non-rephasing impulse spectra }

We make only one approximation to Eq.~\eqref{appendix-eq:rephasing-and-nonrephasing-polarization-spectra} to arrive at Eq.~2 of the main text. Here we invoke the ``impulsive limit'' in which the ultra-fast laser pulses are expected to produce $\bm{E}_i (\omega)$ with frequency envelopes broad enough that they be considered constant over the spectral window of interest. When this is the case, one can simply take $\bm{E}_i (\omega) = \bm{E}_i$. Taking $\bm{E}_i = \bm{1}$ and defining $\Pi^{(3)}(\omega_t, T, \mp\omega_\tau) \equiv \Pi^{(3)}_{\mp} (\omega_t, T, \omega_\tau)$, in the impulsive limit and using Eqs.~\eqref{appendix-eq:signal-envelope}, \eqref{appendix-eq:variable-changes}, and \eqref{appendix-eq:rephasing-and-nonrephasing-polarization-spectra}, we write $\tilde{\Pi}^{(3)}_{\mp}$ as,  
\begin{equation}\label{appendix-eq:impulsive-rephasing-and-nonrephasing-spectral-envelope-first-step}
\begin{split}
    &\tilde{\Pi}^{(3)}_{\mp}(\omega_t, T, \omega_\tau) = 
    \\& i\iint_{-\infty}^{\infty}\dd{t} \dd{\tau} \!\!\iiint_{- \infty}^{\infty}  \dd{t_3} \dd{t_2} \dd{t_1} e^{i(\omega_t t \mp \omega_\tau \tau)} S^{(3)}(t_3, t_2, t_1) \\&\times \iiint_{- \infty}^{\infty} \dd{\nu_3} \dd{\nu_2} \dd{\nu_1} e^{-i\nu_{\sigma_3}(t-t_3)} e^{-i\nu_{\sigma_2}(T-t_2)} e^{-i\nu_{\sigma_1}(\tau-t_1)}.
\end{split}
\end{equation}
By invoking the identity Fourier definition of the Dirac delta function, $\delta(t-t') = \int \dd{\omega} e^{\pm i \omega (t-t')}$, for the integrals over $\{\nu\}$, Eq.~\eqref{appendix-eq:impulsive-rephasing-and-nonrephasing-spectral-envelope-first-step} reveals that in the impulsive limit, where pump and probe pulses can be treated as having infinite spectral bandwidth, spectroscopic and molecular time variables are interchangeable: $\{t,T,\tau\} \iff \{t_3,t_2,t_1\}$. Relabeling the variables, we finally arrive at Eq.~2 of the main text,

\begin{equation}\label{appendix-eq:final-impulsive-rephasing-and-nonrephasing-spectral-envelope}
\begin{split}
    \tilde{\Pi}^{(3)}_{\mp}(\omega_3, t_2, \omega_1) =& \iint_{0}^{\infty} \dd{t_3} \dd{t_1} e^{i(\omega_3 t_3 \mp \omega_1 t_1)} \, R^{(3)}(t_3, t_2, t_1) \\ \equiv& S^{(3)}_{\mp}(\omega_3, t_2, \omega_1).
\end{split}
\end{equation}
We note that this result is defined for $t_2 > 0$. 

\begin{figure}[h!]
    \centering
    \includegraphics[trim={0, 0, 0, 0},clip,width=.45\columnwidth]{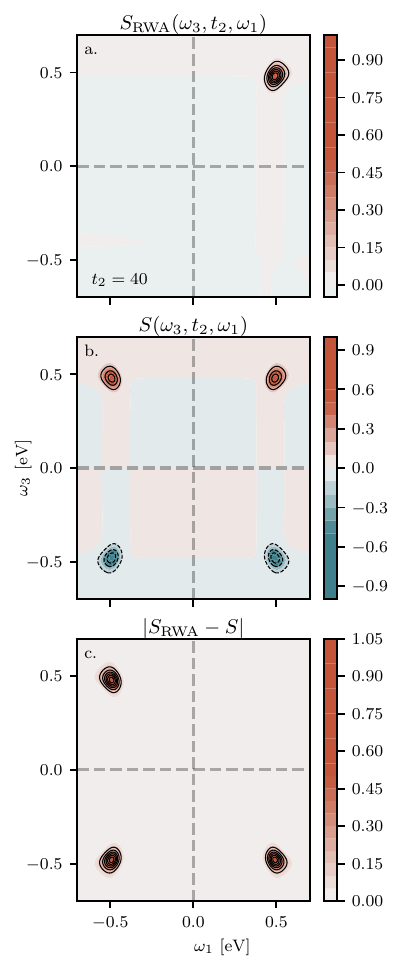}
    \caption{2D spectra for a 2LS from invoking the RWA (a) and without invoking the RWA (b). Here, one observes that in quadrants along negative $\omega_3$ and $\omega_1$, in contrast to $S$, $S_{\rm RWA}$ bears no spectral weight whereas in the first quadrant the two lineshapes are identical. We quantify in panel (c) showing the absolute difference of panels (a) and (b).}
    \label{fig:full-and-rwa-spectra-2ls}
\end{figure}

%----------------------------------------------------------------------------
\subsection{Rotating wave approximation}

Finally, we briefly discuss the rotating wave approximation and how invoking this approximation to Eq.~1 allows one to recover the common form of the rephasing and non-rephasing spectra. To begin, we note that expanding out the commutators in Eq.~1 reveals that one can write $R^{(3)}(t_3, t_2, t_1) = \sum_{\alpha = 1}^4 [R^{(3)}_{\alpha} - R^{(3)}_{-\alpha}]$, where $R^{(3)}_{-\alpha} \equiv [R^{(3)}_{\alpha}]^*$. Next, by assuming perfect phase matching, the signal field must yield carrier frequencies with phases commensurate with the incident fields. One can thus posit, 
\begin{equation}
    R^{(3)}_{\alpha} (t_3, t_2, t_1) \equiv e^{i(\pm \omega_3^{\alpha} t_3 \pm \omega_2^\alpha t_2 \pm \omega_1^\alpha t_1)} \mathcal{R}^{(3)}_{\alpha} (t_3, t_2, t_1),
\end{equation}
where $\omega^\alpha_i$ represents the induced coherence between two states during the time intervals $t_i$. By substituting this expression into Eq.~\eqref{appendix-eq:final-impulsive-rephasing-and-nonrephasing-spectral-envelope}, one finds that there are certain phases of $\omega_i^{\alpha}$ that are ``on resonance" with the rephasing and non-rephasing kernels. For the rephasing spectra, response functions with coherence phases $-i\omega_3^\alpha$ and  $i\omega_1^\alpha$ are resonantly enhanced, whereas for the non-rephasing spectra, response functions with coherence phases $-i\omega_3^\alpha$ and  $i\omega_1^\alpha$ are resonantly enhanced. Employing diagramatic perturbation theory, which we do not discuss here but refer the reader to Ref.~\onlinecite{BookMukamel} for more details, one finds that for a multi-state system, 
\begin{equation}\label{appendix-eq:rwa-impulsive-rephasing-and-nonrephasing-spectral-envelope}
\begin{split}
    R^{(3)}_{\mp}&(\omega_3, t_2, \omega_1)  \sim \iint_{0}^{\infty} \dd{t_3} \dd{t_1}  e^{i(\omega_3 t_3 \mp \omega_1 t_1)} \\&\times \left[ R_{2/1}^{(3)}  + R_{3/4}^{(3)} - R_{-1'/-2'}^{(3)} \right](t_3, t_2, t_1)
\end{split}
\end{equation}
Here, we mark $R_{-1'}^{(3)}$ and  $R_{-2'}^{(3)}$ with primed indices to indicate that they contribute excited state absorption features to the the rephasing and non-rephasing spectra, respectively. These two contributions carry signal frequencies that can be deduced from Feynman diagrams but cannot be determined simply by taking the complex conjugate of each $R_1^{(3)}$ and $R_2^{(3)}$ as for a two-level system $R^{(3)}_\mp \sim R_{2/1}^{(3)} + R_{3/4}^{(3)}$.

We note that the rotating wave approximation is valid when the coherence frequencies $\omega^{\alpha}_i$ are large, as the off-resonant terms can lie far outside of the spectral window of interest and Eqs.~\eqref{appendix-eq:final-impulsive-rephasing-and-nonrephasing-spectral-envelope} and \eqref{appendix-eq:final-impulsive-rephasing-and-nonrephasing-spectral-envelope} are nearly equivalent. We show this in Fig.~\ref{fig:full-and-rwa-spectra-2ls}, which displays spectra generated from the RWA response function (a), the full response function (b), and the absolute difference between the latter spectra normalized to respective intensity maxima. Here, one observes that while $S_{\rm RWA}$ only has spectral weight in the first quadrant ($\omega_1>0$, $+\omega_3>0$), $S$ displays positive and negative peaks for $\omega_1>0$ and $\omega_1<0$, respectively. While the spectra generated within and beyond the RWA differ, Fig.~\ref{fig:full-and-rwa-spectra-2ls}(c) confirms that the two generate identical lineshapes in when $\omega_1>0$ and $\omega_3>0$ thus rendering the two equivalent within the context of conventional analysis of the 2D spectrum. Because 2D spectra are measured over positive frequencies, and because our tests already demonstrate the spectral GME's applicability over the doubly positive quadrant, we can expect the spectral GME to be indifferent to the presence or absence of the RWA in computationally generated data. 

%-------------------------------------------------------------------------
%-------------------------------------------------------------------------
\section{The 2D response obeys a one-time GME.}\label{app:ethan-ORF-derivation}
      
Here, we demonstrate that the third order response function in Eq.~\eqref{appendix-eq:third-order-optical-response} obeys a one-time GME. To achieve this, we perform a series of manipulations, ultimately resulting in Eq.~4. We then offer a brief discussion of these equations in the context of a single-mode Brownian oscillator model to provide insight into Eq.~6. 

We begin by reordering the commutators of transition dipoles in Eq.~\eqref{appendix-eq:third-order-optical-response} to write, 
\begin{equation}\label{appendix-eq:R3-deriv}
    R^{(3)} = -{\rm Tr} \Big\{ \hat{\rho}_0 \tilde{\bm{\mu}}(0) \tilde{\bm{\mu}}(t_1) \tilde{\bm{\mu}}(t_1+t_2) \hat{\bm{\mu}}(t_3+t_2+t_1)\Big\},
\end{equation}
where $\tilde{\bm{\mu}}(0) \equiv  \comm{\hat{\bm{\mu}}(0), \cdot}$ and $\tilde{\bm{\mu}}(t_1) = e^{i\mc{L}t_1}\tilde{\bm{\mu}}(0)e^{-i\mc{L}t_1}$ are the Liouville space transition dipole superoperator and its time evolved form, respectively. We emphasize the difference between $\tilde{\bm{\mu}}(0)$ and $\hat{\bm{\mu}}(0)$, with the former denoting a superoperator and the latter denoting a vectorized operator which evolves in time as $\hat{\bm{\mu}}(t_1) = e^{i\mc{L}t_1}\hat{\bm{\mu}}(0)$. We isolate the $t_2$ propagator from all other time evolution, expand Eq.~\eqref{appendix-eq:R3-deriv} explicitly to cancel several time propagators, and leverage the cyclic invariance of the trace to obtain the equivalent expressions,
\begin{equation}\label{appendix-eq:expanded-props}
\begin{split}
    R^{(3)} &= -{\rm Tr}\Big\{\hat{\rho}_0\tilde{\bm{\mu}}(0) e^{i\mc{L}t_1}\tilde{\bm{\mu}}(0)e^{i\mc{L}t_2} \tilde{\bm{\mu}}(0) e^{i\mc{L}t_3}\hat{\bm{\mu}}(0)\Big\}\\
    &= {\rm Tr}\Big\{[\tilde{\bm{\mu}}(0)\hat{\rho}_0]e^{i\mc{L}t_1}\tilde{\bm{\mu}}(0)e^{i\mc{L}t_2} \tilde{\bm{\mu}}(0) e^{i\mc{L}t_3}\hat{\bm{\mu}}(0)\Big\}\\
    &= -{\rm Tr}\Big\{[\tilde{\bm{\mu}}(0)e^{-i\mc{L}t_1}\tilde{\bm{\mu}}(0)\hat{\rho}_0] e^{i\mc{L}t_2} \tilde{\bm{\mu}}(0) e^{i\mc{L}t_3}\hat{\bm{\mu}}(0)\Big\} \\
    &={\rm Tr}\Big\{[\tilde{\bm{\mu}}(0)e^{i\mc{L}t_3}\hat{\bm{\mu}}(0)]^{\dagger} e^{-i\mc{L}t_2}[\tilde{\bm{\mu}}(0)e^{-i\mc{L}t_1}\tilde{\bm{\mu}}(0)\hat{\rho}_0]\Big\}.
\end{split}
\end{equation}

\begin{figure*}[t!]
    \centering
    \begin{subfigure}{0.32\linewidth}
        \centering
        \includegraphics[width=\linewidth]{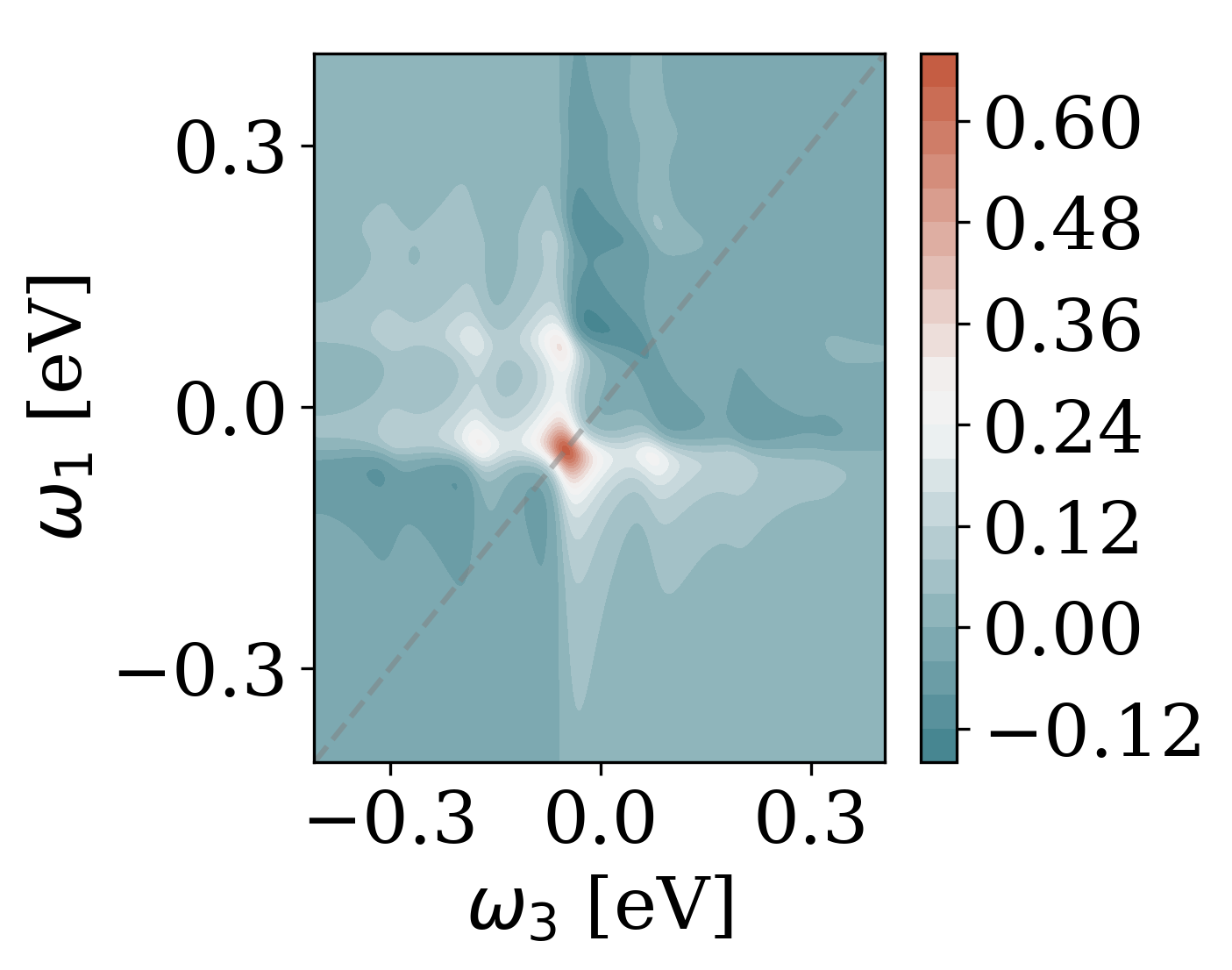}
        \caption{}
    \end{subfigure}
    \hfill
    \begin{subfigure}{0.32\linewidth}
        \centering
        \includegraphics[width=\linewidth]{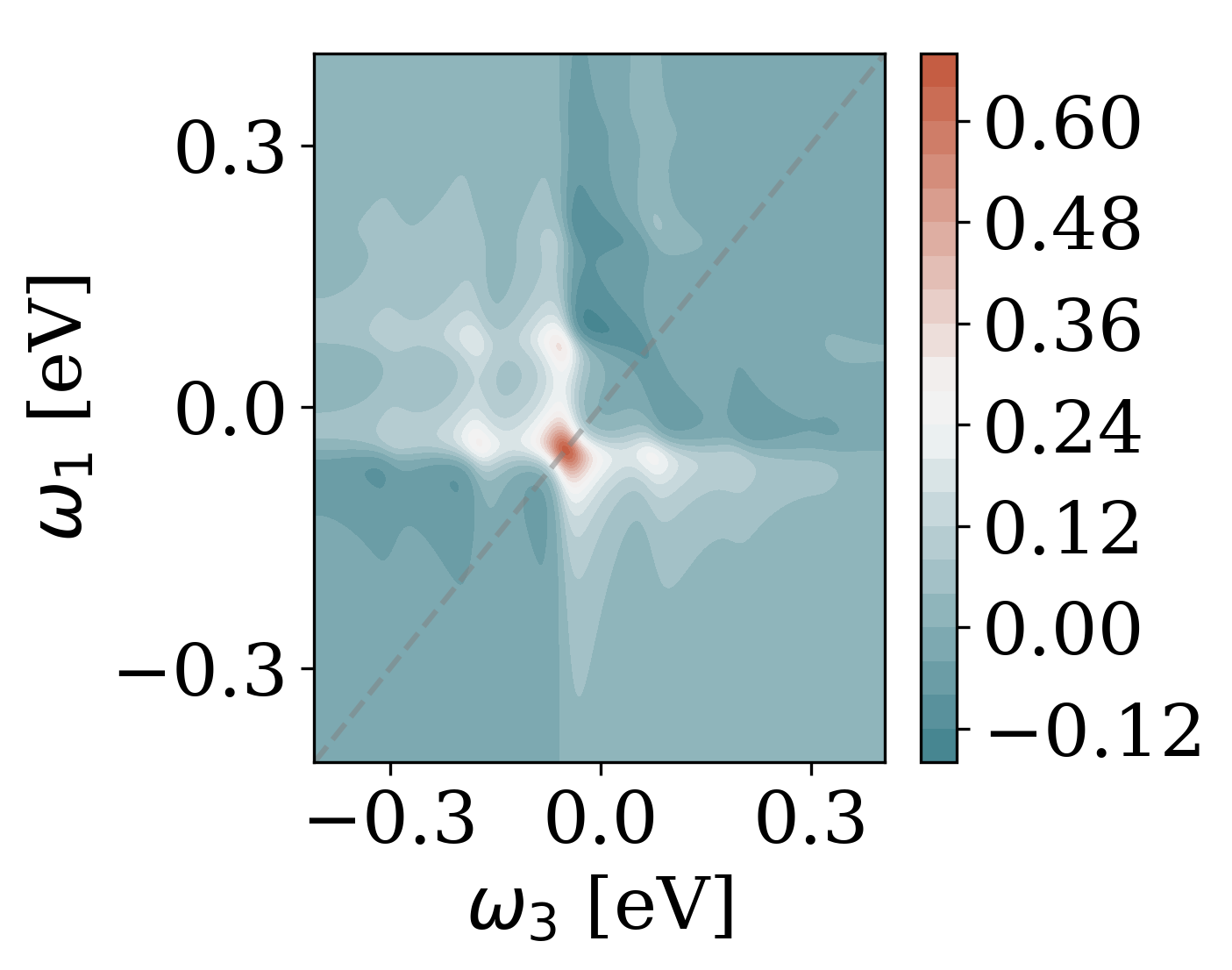}
        \caption{}
    \end{subfigure}
    \hfill
    \begin{subfigure}{0.32\linewidth}
        \centering
        \includegraphics[width=\linewidth]{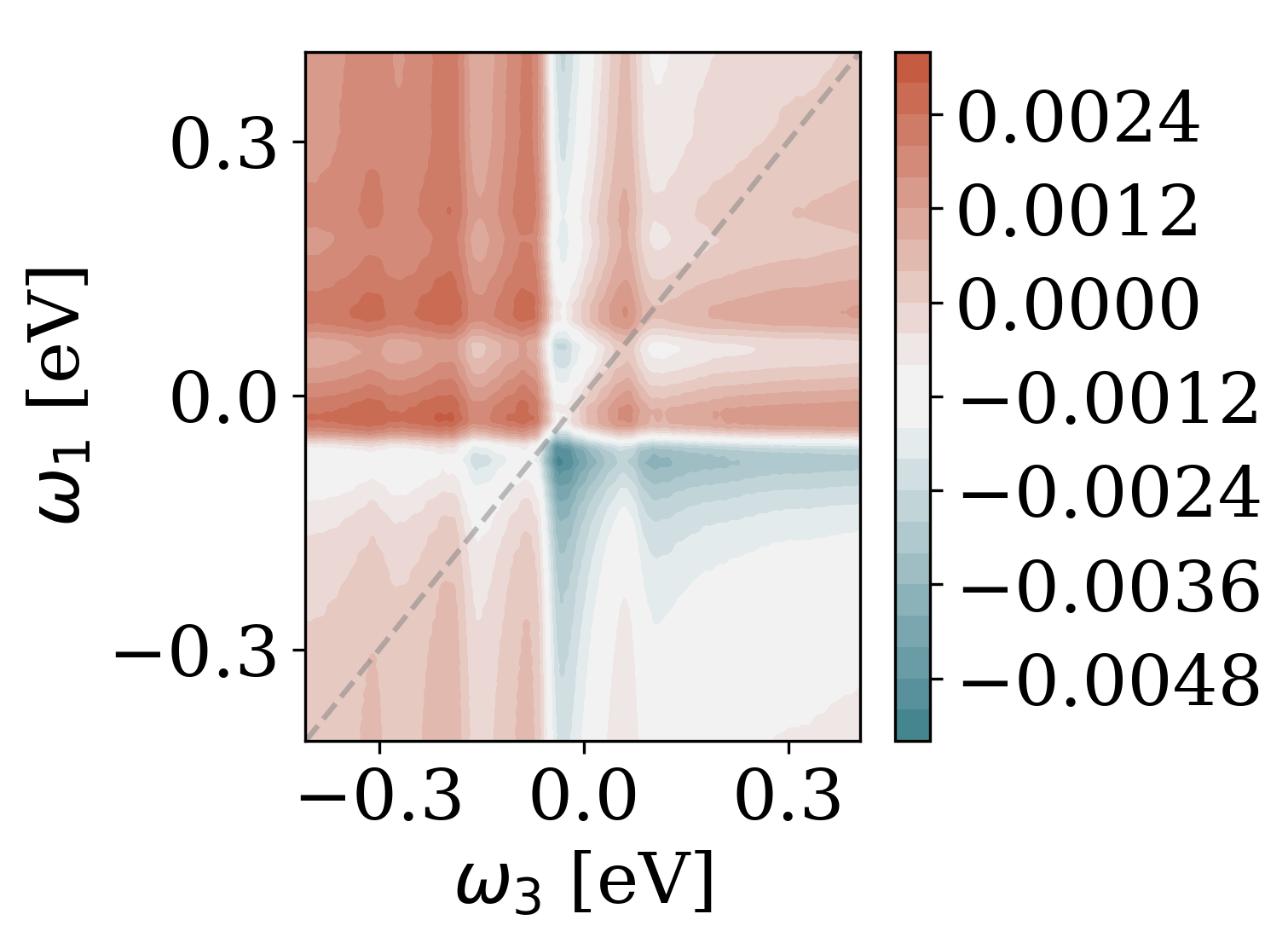}
        \caption{}
    \end{subfigure}
    
    \caption{Equivalence of the resolvent formulation. Representative frame of a spectrum simulated with a numerical 2D Fourier-Laplace transform of dynamics generated from our minimal model in \textbf{(a)}, and calculated directly with the analytically exact resolvent (Eqs.~\ref{appendix-eq:modified_liouville_S2} and~\ref{appendix-eq:resolvent}) in \textbf{(b)}. \textbf{(c)} Signed difference when panel (a) is subtracted from panel (b).}
    \label{fig:compare-resolvent-n-ft}
\end{figure*}

We can substitute Eq.~\eqref{appendix-eq:expanded-props} into Eq.~\eqref{appendix-eq:final-impulsive-rephasing-and-nonrephasing-spectral-envelope} and perform the numerical Fourier transform to obtain the two-dimensional spectrum. 

To solidify the connection between Eq.~\eqref{appendix-eq:expanded-props} and a one-time GME we must interrogate the analytical form of the Fourier transform. To illustrate this connection, we employ a single-mode Brownian Oscillator model\cite{BookMukamel}. This model provides a physically intuitive understanding of the Fourier integral's convergence parameter, $\epsilon$, in Eq.~6 of the main text. The Hamiltonian for this model is, 
\begin{subequations}
\begin{align}
    {\rm H_g} &= \frac{1}{2}[\hat{p}^2+\omega^2\hat{q}^2] \\
    {\rm H_e} &= \hbar\omega_{eg} + \frac{1}{2}[\hat{p}^2+\omega^2(\hat{q}-d)^2]
\end{align}
\end{subequations}
where $\hat{p} = \hat{P}/\sqrt{m}$ and $\hat{q} = \sqrt{m}\hat{Q}$ are the mass-weighted momentum and position, and $d$ denotes the shift in oscillator minima upon excitation. Direct simulation of this Hamiltonian yields only Rabi oscillations, so we introduce simple exponential damping along the $t_1$ and $t_3$ axes to obtain a dissipative spectrum. This results in a modified version of Eq.~\eqref{appendix-eq:expanded-props}. 
\begin{align}\label{appendix-eq:modified_liouville_S2}
\notag
    R^{(3)} (t_3, t_2, t_1) =  {\rm Tr}\Big\{&[\tilde{\boldsymbol{\mu}}(0)e^{(i\mc{L}-\epsilon)t_3}\hat{\boldsymbol{\mu}}(0)]^{\dagger}e^{-i\mc{L}t_2} 
    \\ 
    &[\tilde{\boldsymbol{\mu}}(0)e^{(-i\mc{L}-\epsilon)t_1}\tilde{\boldsymbol{\mu}}(0)\hat{\rho}_0]\Big\}
\end{align}
Upon applying the double Fourier-Laplace transform, we can analytically express the resolvents as, 
\begin{align}\label{appendix-eq:resolvent}
    &\tilde{\mathbf{R}}^{\mp}(\pm \omega) = \int^{\infty}_{0}\dd{t}e^{i(\pm \omega + i\epsilon)t}e^{\mp i{\mc L}t} = \frac{i}{\pm \omega \mp{\mc L}+i\epsilon}
\end{align}
with the distinction from Eq.~6 that instead of considering $\lim_{\epsilon \to 0}$ in the context of this model $\epsilon$ takes on physical meaning, as opposed to the general case where $\epsilon$ is a convergence criterion of the Fourier-Laplace transform. We validate the equivalence of the Resolvent and numerical Fourier transform approaches through simulation, as shown in Fig.~\ref{fig:compare-resolvent-n-ft}. We emphasize that while there is numerical error (panel c) between the 2D Fourier-Laplace transform (panel a) and the resolvent formulation (panel b), this error converges as the number of simulated points across the $t_3$ and $t_1$ axes increase. With the set of parameters shown, the largest error is less than $0.5\%$ in terms of pixel intensity.  

%-------------------------------------------------------------------------
%-------------------------------------------------------------------------
\section{A guide to applying the spectral GME}
\label{app:ethan-workflow}

\begin{figure*}[t!]
    \centering
    \begin{subfigure}{0.24\linewidth}
        \centering
        \includegraphics[width=\linewidth]{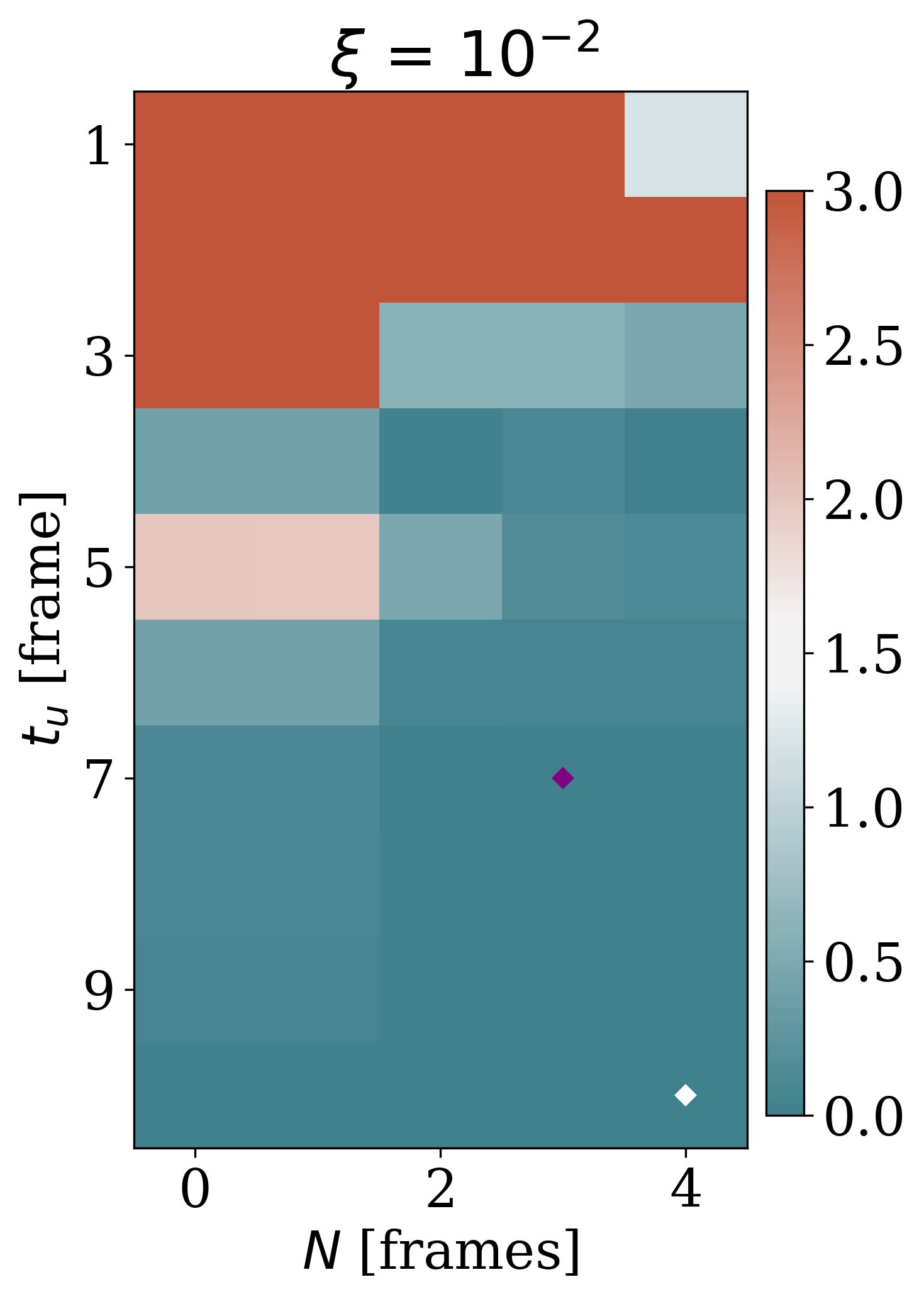}
        \caption{}
    \end{subfigure}
    \hfill
    \begin{subfigure}{0.24\linewidth}
        \centering
        \includegraphics[width=\linewidth]{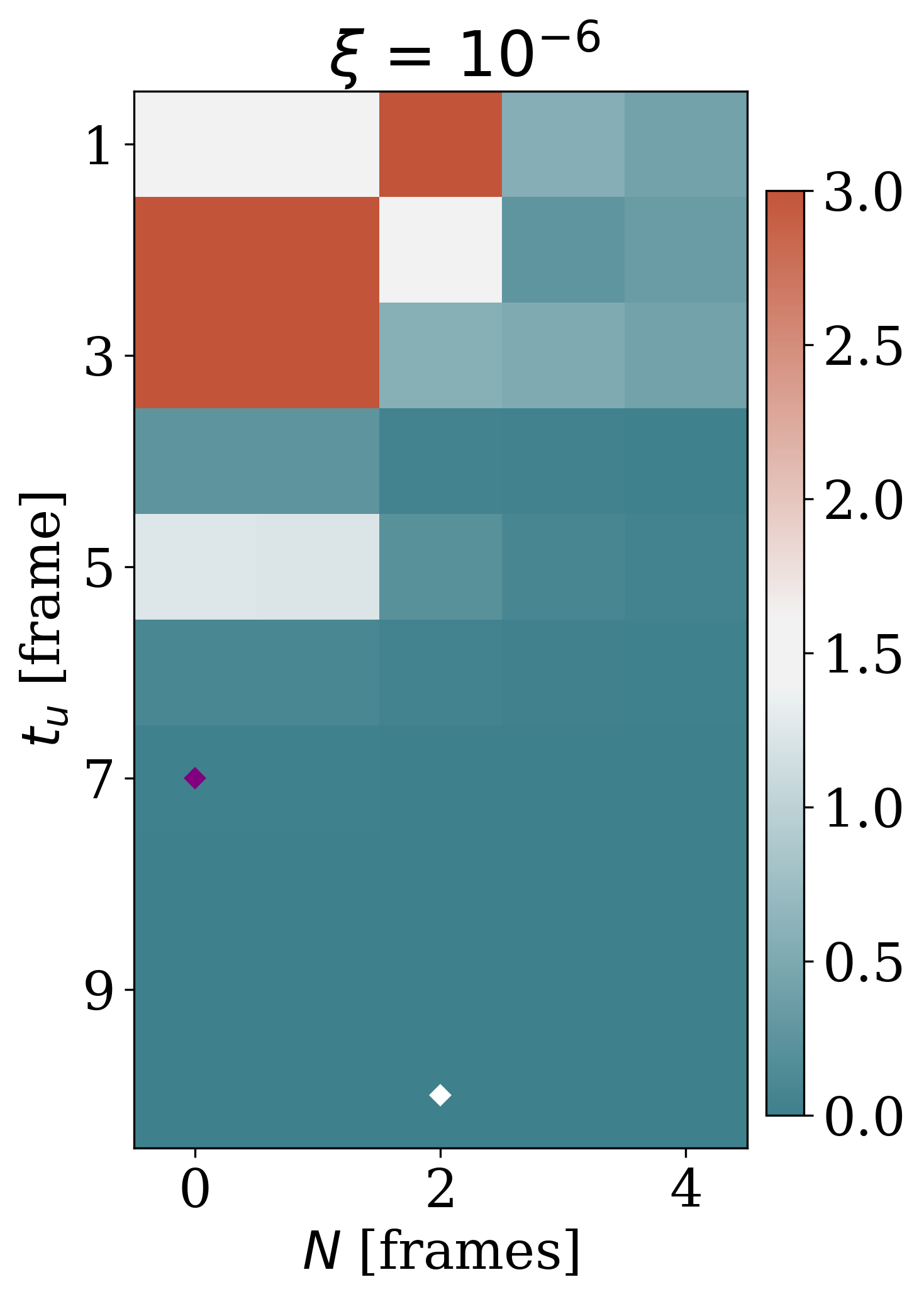}
        \caption{}
    \end{subfigure}
    \hfill
    \begin{subfigure}{0.24\linewidth}
        \centering
        \includegraphics[width=\linewidth]{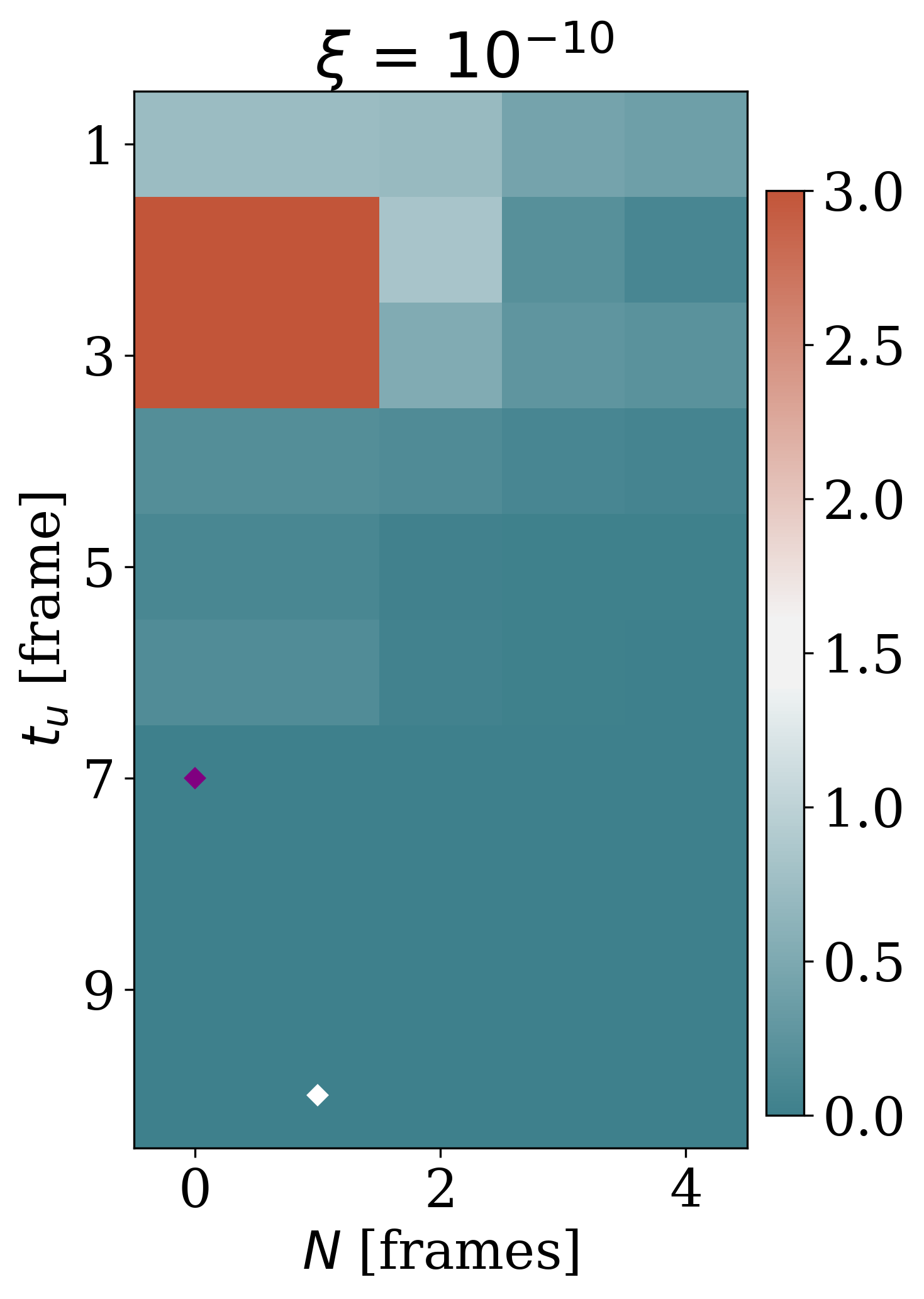}
        \caption{}
    \end{subfigure}
    \hfill
    \begin{subfigure}{0.24\linewidth}
        \centering
        \includegraphics[width=\linewidth]{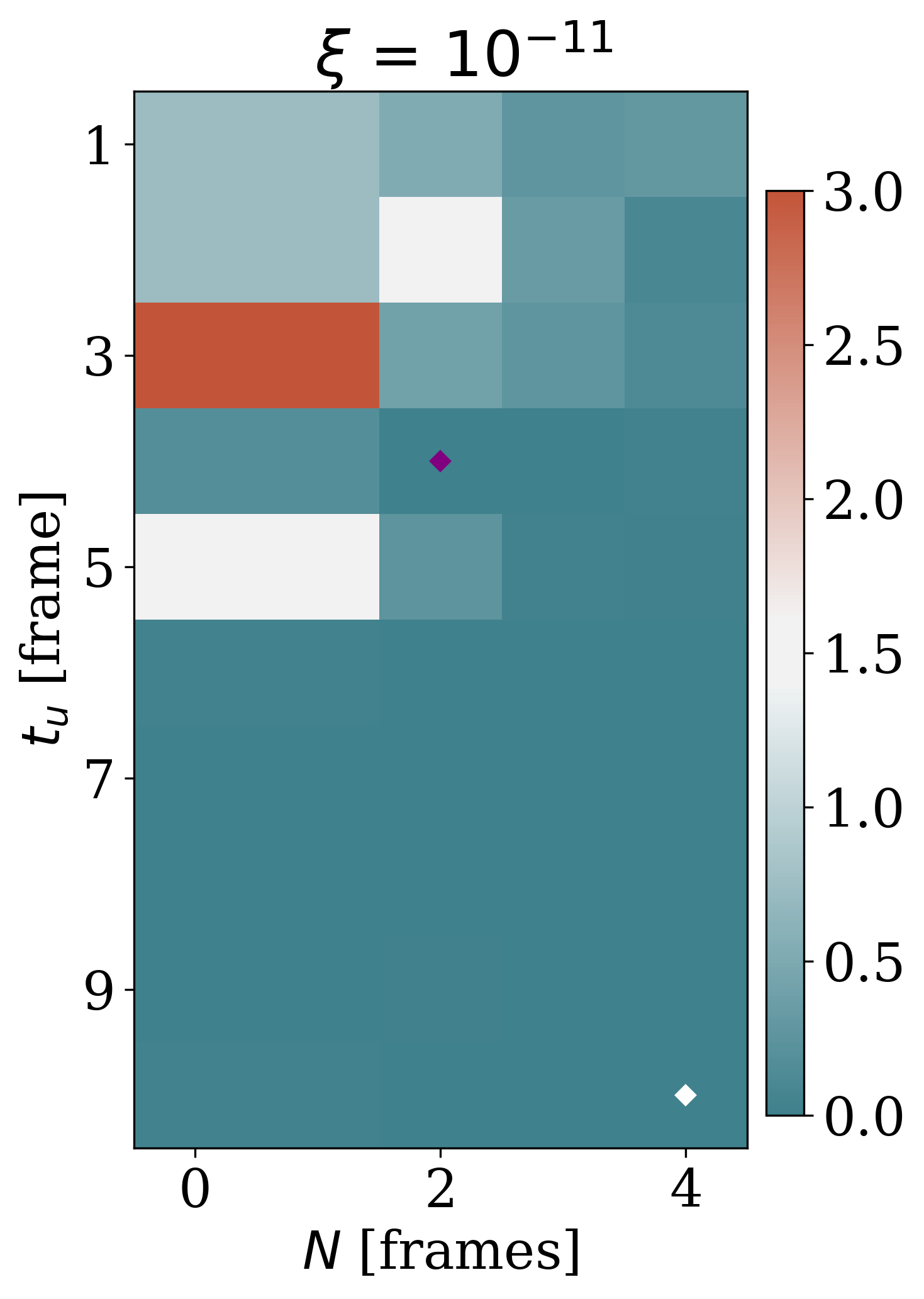}
        \caption{}
    \end{subfigure}
    \caption{Root-mean squared error maps (see Eq.~12 of the main text) for $\xi$ ranging from $10^{-2}$ to $10^{-11}$ using the excitation energy transfer dimer of Fig.~2 of the main text. The x-axis is the number of frames averaged ($N$), which is zero in this case, and the y-axis is the cutoff frame in $t_2$ ($t_u$). The white diamond on each heatmap represents the minimum error combination of cutoff and window parameters. The purple diamond denotes the set of $t_u$ and $N$ that yields an error less than $2.5\%$}
    \label{fig:SVD-boundary-convergence}
\end{figure*}

To apply the spectral GME in Eqs.~8~and~9, one must determine three parameters: the onset of averaging $t_u$, the offset of averaging $\tau_\mc{U} \equiv t_{N+u-1}$, and the pseudoinverse threshold $\xi$. In this appendix, we leverage the HEOM data for the electronic energy transfer dimer system discussed in Sec.~III of the main text to illustrate our protocol for systematically identifying these parameters.

\textbf{Choosing $t_u$ and $\boldsymbol{\tau}_{\mc{U}}$:} Before optimizing $\xi$, we first select a range of values for $t_u$ and $\tau_\mc{U}$ for our initial tests. If during the refinement of $\xi$, these initial choices do not yield a sufficiently accurate and stable construction of the spectral GME, we adjust the ranges of $t_u$ and $\tU$ and repeat the refinement of $\xi$ that we detail below. We recommend initially exploring a wide range of $t_u$ and $\tU$ values while ensuring that $\tU$ does not reach the end of the available data. Because the HEOM dataset contains 16 frames across $t_2$, we test values in the range $1\leq t_u \leq 10$ frames and $0 \leq N \leq 4$. If the initial choices of $t_u$ and $\tU$ still fail to produce a suitable spectral GME parameterization, we adjust the ranges and repeat the procedure in Algorithm~1.

\begin{algorithm}
\caption{Spectral GME Workflow}\label{alg:alg1}

Load experimental spectra $\{\mathbf{S}(t_k)\}_{n=0}^{L-1}$ as a function of $t_2$\;

Preprocess the spectra (see Sec.~V of the main text) so that $\mathbf{S}$ has\;
\quad Uniform density across $\omega_1$ and $\omega_3$\;
\quad Regular timestep spacing in $t_2$\;

Initialize onset and offset averaging parameters $\{t_u,\; \tU\}$\;

Choose a range of SVD thresholds $\xi = [\xi_1,\xi_2,...,\xi_k]$\;

\For{$\{t_u,\; \tU,\; \xi\}$}{
    Compute the averaged propagator
    $\langle\bmc{U}_{\infty}(\delta t)\rangle = \frac{1}{N}\sum_{n=u}^{N+u-1}\bmc{U}(t_n)$\;
    
    Predict spectral dynamics using
    ${\bf S}(\tU + n\delta t) = \langle\bmc{U}_\infty(\delta t)\rangle^n{\bf S}(\tU)$\;
    
    Compute the error (Eq.~12)\;
}

Identify optimal parameters based on short-time accuracy (low error) and long-time stability\;

\end{algorithm}

\textbf{Identifying an optimal ${\boldsymbol{\xi}}$:} We now seek to locate an optimal value for ${\xi}$ by varying $\xi$ and examining the error (Eq.~12) associated with every combination of our chosen $t_u$ and $\tU$. We recommend varying ${\xi}$ by orders of magnitude spanning the full range of singular values. For the present example, the range encompasses $\sim10^1$ to $10^{-16}$. Figure~\ref{fig:SVD-boundary-convergence} shows error maps (as a function of $t_u$ on the y-axis and $N$ on the x-axis) for several choices of $\xi$, with the error calculated using Eq.~12. The white diamonds across the panels of Figure~\ref{fig:SVD-boundary-convergence} correspond to the minimum error choice of $t_u$ and $\tU$ for a given $\xi$. The purple diamond denotes the earliest choice of parameters in $t_2$ that reproduces the spectrum within $2.5\%$ error. We note that when $\xi = 10^{-2}$ and $\xi = 10^{-11}$ (Fig.~\ref{fig:SVD-boundary-convergence}b and ~\ref{fig:SVD-boundary-convergence}d), the first set of parameters below $2.5\%$ error requires averaging over several frames. This may appear surprising since the HEOM data we use to illustrate this protocol are exact and free of statistical noise, leading one to expect that one should not require any averaging over any frames. We interrogate the cause for this in SI Sec.~5. The $\xi=10^{-6}$ and $\xi=10^{-10}$ error maps satisfy our error threshold without averaging. Given that these data are free from experimental noise, we expect that $t_u = \tU$, meaning both $\xi=10^{-6}$ and $\xi=10^{-10}$ are reasonable choices for $\xi$. We select $\xi=10^{-10}$ to minimize the amount of spectral data truncated by SVD while still maintaining short-time accuracy (low error). We provide an intuitive framework to help understand the different behaviors of Fig.~\ref{fig:SVD-boundary-convergence} in SI Sec.~5.

\section{Understanding the influence of $\xi$ in noise-free spectra}
\label{app:role-of-xi}

Here, we provide an intuitive framework to characterize the impact of our threshold choice $\xi$ on the spectral GME. Building on the results of SI Sec.~4, we consider HEOM spectra free from experimental noise and characterize the behaviors shown by the purple diamonds in Fig.~\ref{fig:SVD-boundary-convergence}. These purple diamonds denote the earliest set of parameters (in $t_2$) that reproduce the spectrum within $2.5\%$ error. In this analysis we address two questions:
\begin{enumerate}
    \item[Q1:] Why do some thresholds $\xi$ require averaging, while others do not?
    \item[Q2:] How does $\xi$ impact the obtained memory timescales?
\end{enumerate}

\begin{figure}[h!]
    \centering
    \includegraphics[clip,width=.3\columnwidth]{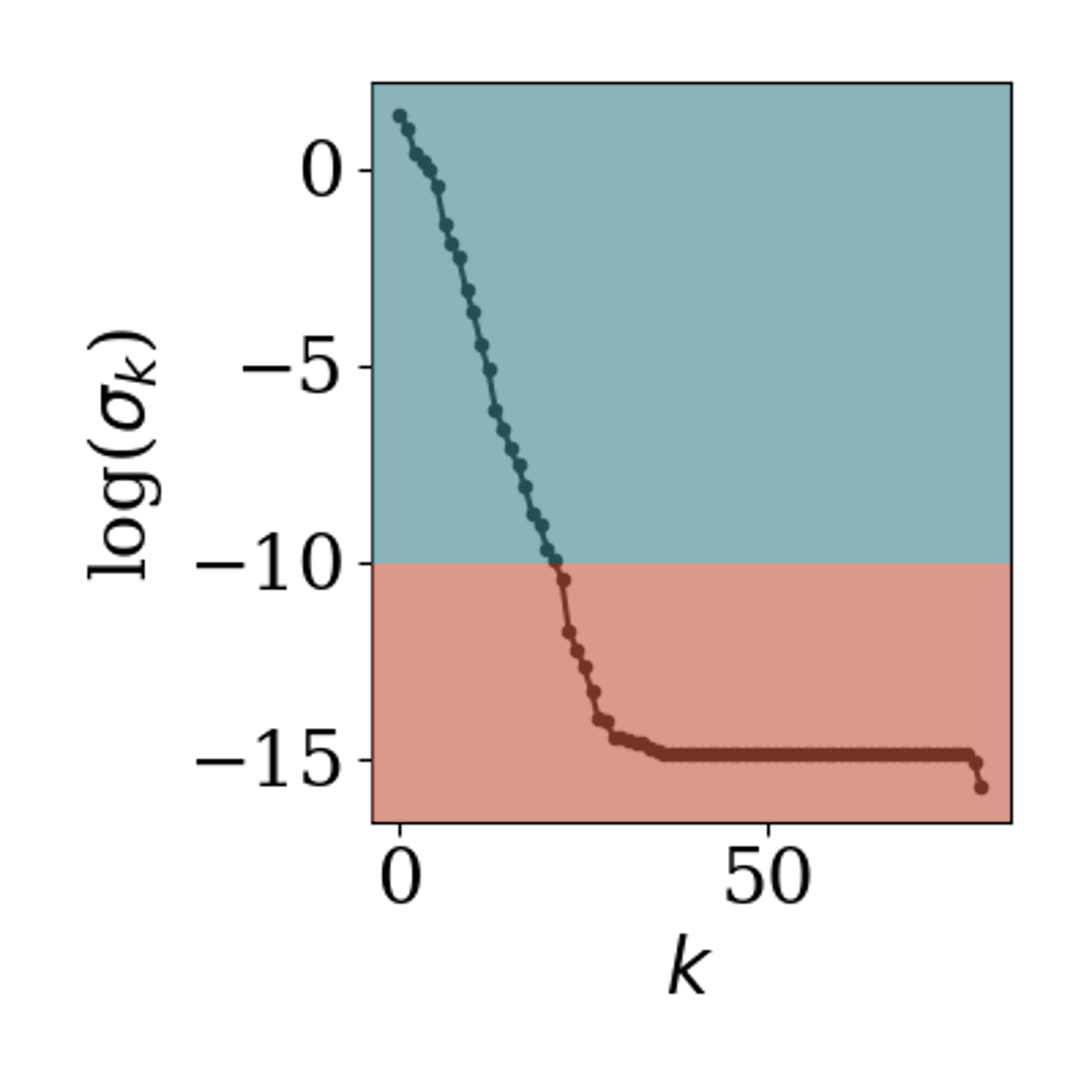}
    \caption{The $\log_{10}$-scaled singular values $\{ \sigma_k \}$ at $t_2 = 0$~fs for the excitation energy transfer dimer of Fig.~2 of the main text as a function of index, $k$. The boundary between the blue and red shaded regions corresponds to the optimal value of $\xi$. The modes in the blue region $\sigma_k > \xi$ are the true dynamical modes of the system, while those in the red shaded region $\sigma_k < \xi$ correspond to modes dominated by numerical noise. }
    \label{fig:singular-val-spectra}
\end{figure}

We answer Q1 and Q2 by characterizing the $M$ singular values of ${\bf S}$ and the impact of choices of $\xi$ (see Sec.~V~B of the main text for details). The answers to both questions lies in understanding how our low-rank approximation of ${\bf S}^{-1}(t_2)$ varies with $\xi$. Specifically, we disentangle the influence of $\xi$ by partitioning the singular value spectrum into two shaded regions, where the boundary marks the optimal cutoff of $\xi$ that separates the blue region containing the true dynamical modes from the red region containing noise (see Fig.~\ref{fig:singular-val-spectra}). When we approach this boundary from above, we incorporate dynamical modes into ${\bf S}^{-1}(t_2)$ until we recover all true modes at the optimal boundary, which may contain few or zero numerical noise modes. Conversely, when we approach from below, we progressively remove noise modes until only the true dynamical modes remain. With this perspective, we can address Q1 and Q2.

\textbf{Answer to Q1:} Finite averaging in this \textit{overdetermined, noise-free} system arises because some values of $\xi$ either over-truncate or under-truncate the pseudoinverse. In the $\xi=10^{-11}$ case (Fig.~\ref{fig:SVD-boundary-convergence}d), we under-truncate the spectrum, and thus include too much noise. As a result, we must average to suppress the influence of these numerical noise modes. In contrast, in the $\xi=10^{-2}$ case (Fig.~\ref{fig:SVD-boundary-convergence}b), we over-truncate the spectrum and discard true dynamical modes. In this example, we can recover a stable propagator $\bmc{U}$ by implementing our averaging procedure in Eq.~9.

\textbf{Answer to Q2:} Examining the minimum error parameterizations (white diamonds) in Fig.~\ref{fig:SVD-boundary-convergence}, we observe that $\xi=10^{-10}$ has a notably shorter memory timescale than the other choices of $\xi$. This behavior reflects the fact that removing dynamical modes requires more dynamical memory to accurately track the remaining dynamics. From a numerical perspective, the optimal threshold $\xi$ for approximating ${\bf S}^{-1}(t_2)$ increases with $t_2$ as the system approaches equilibrium. This example highlights that the memory decay timescale $\tU$ depends on the choice of $\xi$ and that \textit{careful, principled optimization of $\xi$ leads to reduced data requirements and more accurate results}.

%-------------------------------------------------------------------------
%-------------------------------------------------------------------------
\section{The effects of white noise on the spectral GME}
\label{app:ethan-noisy-HEOM}

Here, we test the robustness of the spectral GME workflow (see Sec.~V~B and SI Sec.~4), by applying it to HEOM-generated 2D spectra for an EET dimer with additive noise. Specifically, we add Gaussian (independently and identically distributed) white noise with a mean of $0$ and standard deviation, $10^{-4}$ to each pixel in each frame. By considering additive noise, we demonstrate how singular value truncation complements the spectral GME framework in suppressing statistical noise and provide guidance for applying the method to noisy datasets. 

\begin{figure*}[h!]
    \begin{subfigure}{0.6\linewidth}
        \centering
        \includegraphics[width=\linewidth]{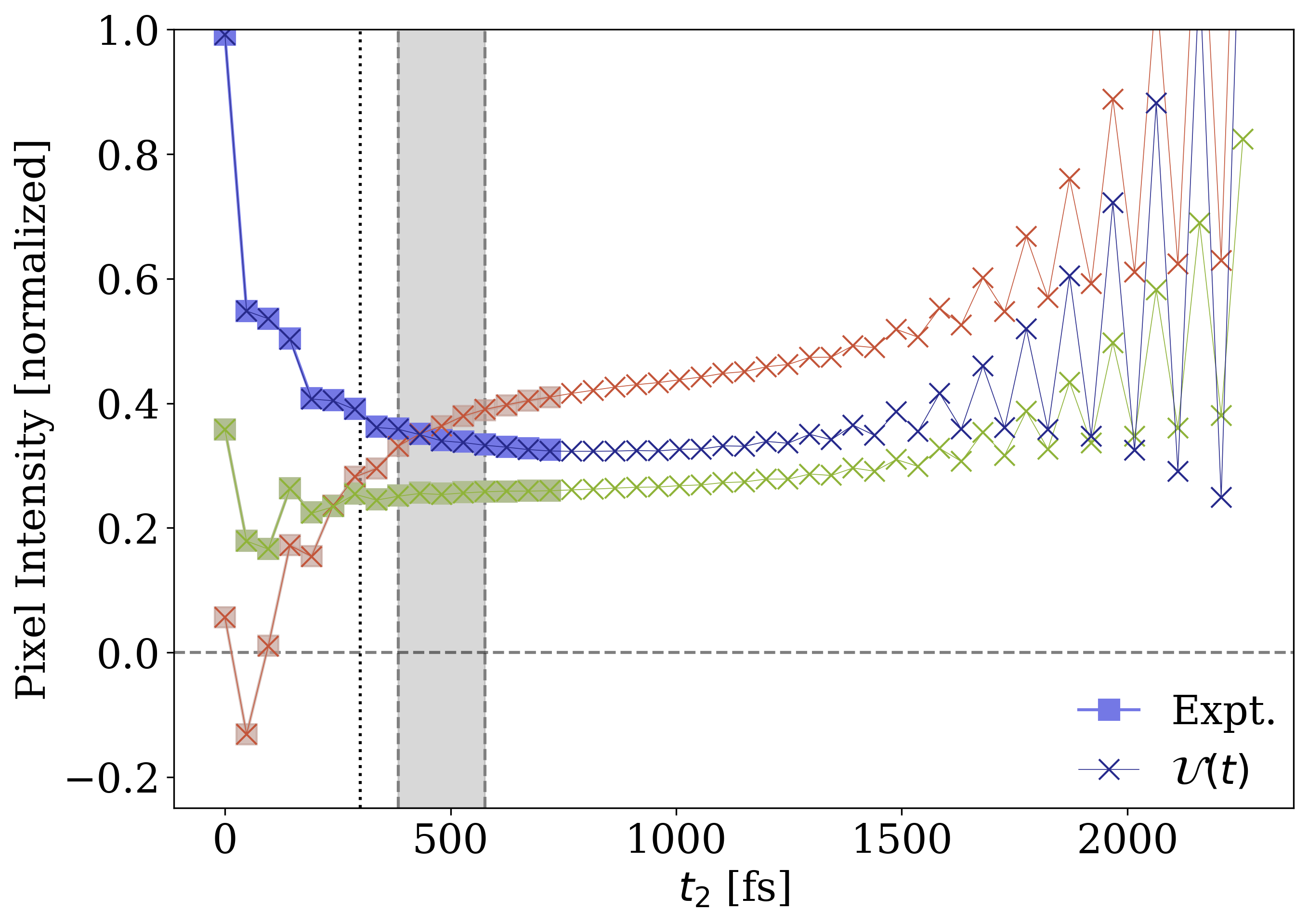}
        \caption{}
    \end{subfigure}
 
    \begin{subfigure}{0.6\linewidth}
        \centering
        \includegraphics[width=\linewidth]{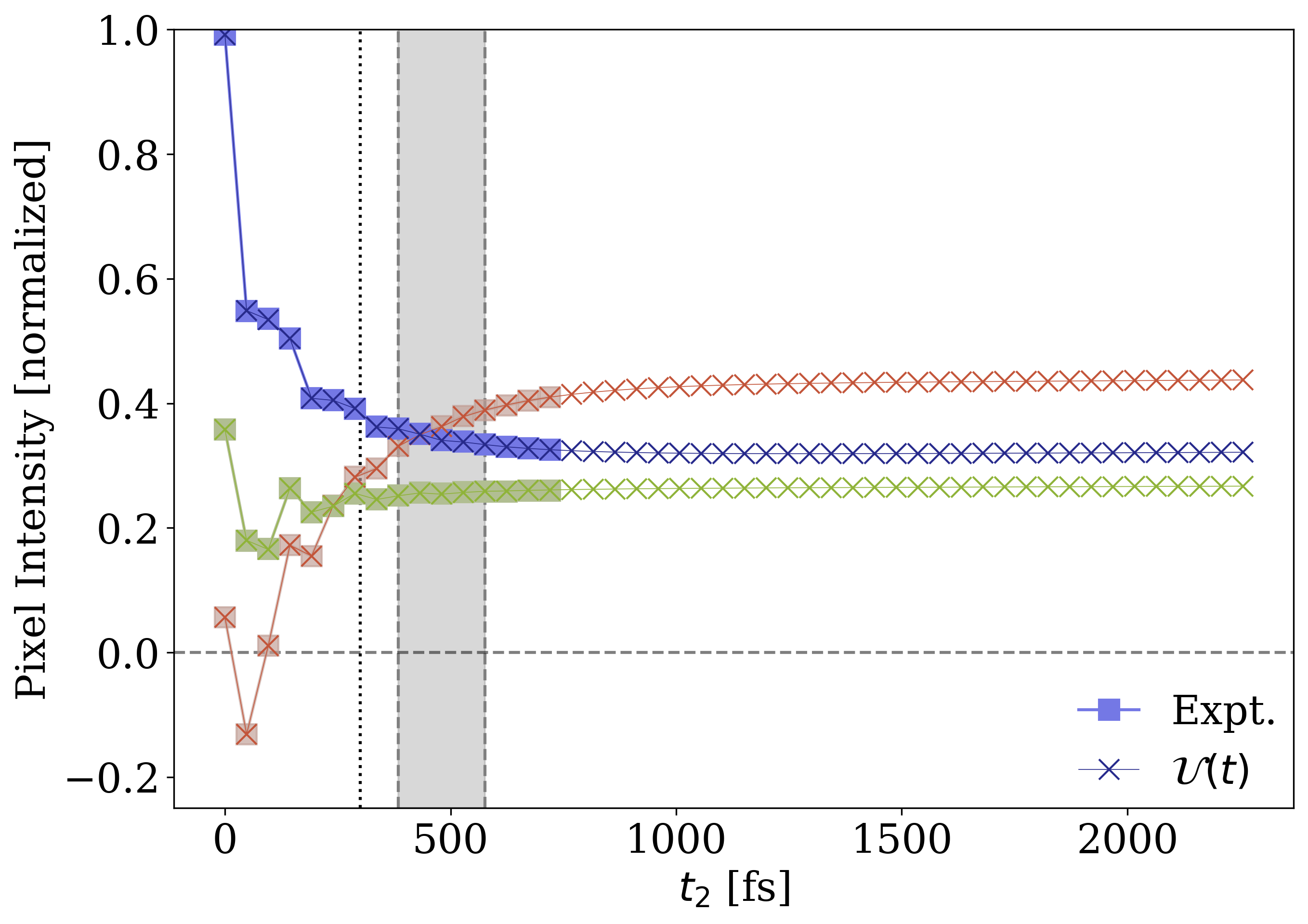}
        \caption{}
    \end{subfigure}
    \caption{Dynamics for two parameterizations of the spectral GME. The pixel dynamics shown here represent the same pixels as those chosen for the HEOM data in Fig.~2 of the main text. \textbf{(a)} Dynamics of pixels with parameters $\{t_u=7 \rmm{ frames}, ~\tU=11 \rmm{ frames}, ~\xi=10^{-4}\}$. \textbf{(b)} Dynamics of the same pixels as (a) with parameters $\{t_u=7 \rmm{ frames}, ~\tU=11 \rmm{ frames}, ~\xi=10^{-3}\}$}
    \label{fig:addit-noise}
\end{figure*}

\textbf{Interplay between noise and error:} Following our protocol in SI Sec.~4, we sweep SVD truncation thresholds across several orders of magnitude and use the same range of memory cutoffs and averaging windows as in the clean HEOM analysis (see SI Sec.~5). In Fig.~\ref{fig:addit-noise}(a), we observe that the spectral GME dynamics become unstable at long times when $\xi =10^{-4}$, despite reproducing the spectrum with only $2.6\%$ error. This instability indicates that noise modes contaminate the $\bmc{U}$ matrix, because the pseudoinverse amplifies any untruncated noise and generates the observed oscillations (see SI Sec.~5). Motivated by the discussion of Fig.~\ref{fig:singular-val-spectra}, we increase the SVD threshold ($\xi=10^{-3}$) to remove these noise modes and suppress the oscillations, producing a result that is still short-time accurate ($3.0\%$ error) and now long-time stable which we show in Fig~\ref{fig:addit-noise}(b). 

\textbf{Takeaway:} The parameters $\{ t_u, \tU, \xi \}$ that minimize the error in Eq.~12 do not necessarily yield the most stable dynamics when working with noisy reference data. In the noisy EET dimer example, slightly larger errors produce more stable predictions because SVD truncation suppresses statistical noise at the level of individual frames. Unlike the experimental noise, this additive noise does not accumulate across $t_2$. Thus, this denoising step complements the suppression of experimental noise that accumulates at longer $t_2$ waiting times discussed in the main text. 

%-------------------------------------------------------------------------
%-------------------------------------------------------------------------
\section{Consistency across experimental replicates}
\label{app:repeats}

\begin{figure*}[!ht]
 \begin{subfigure}[t]{\linewidth}
    \centering
    \resizebox{0.6\textwidth}{!}{\includegraphics{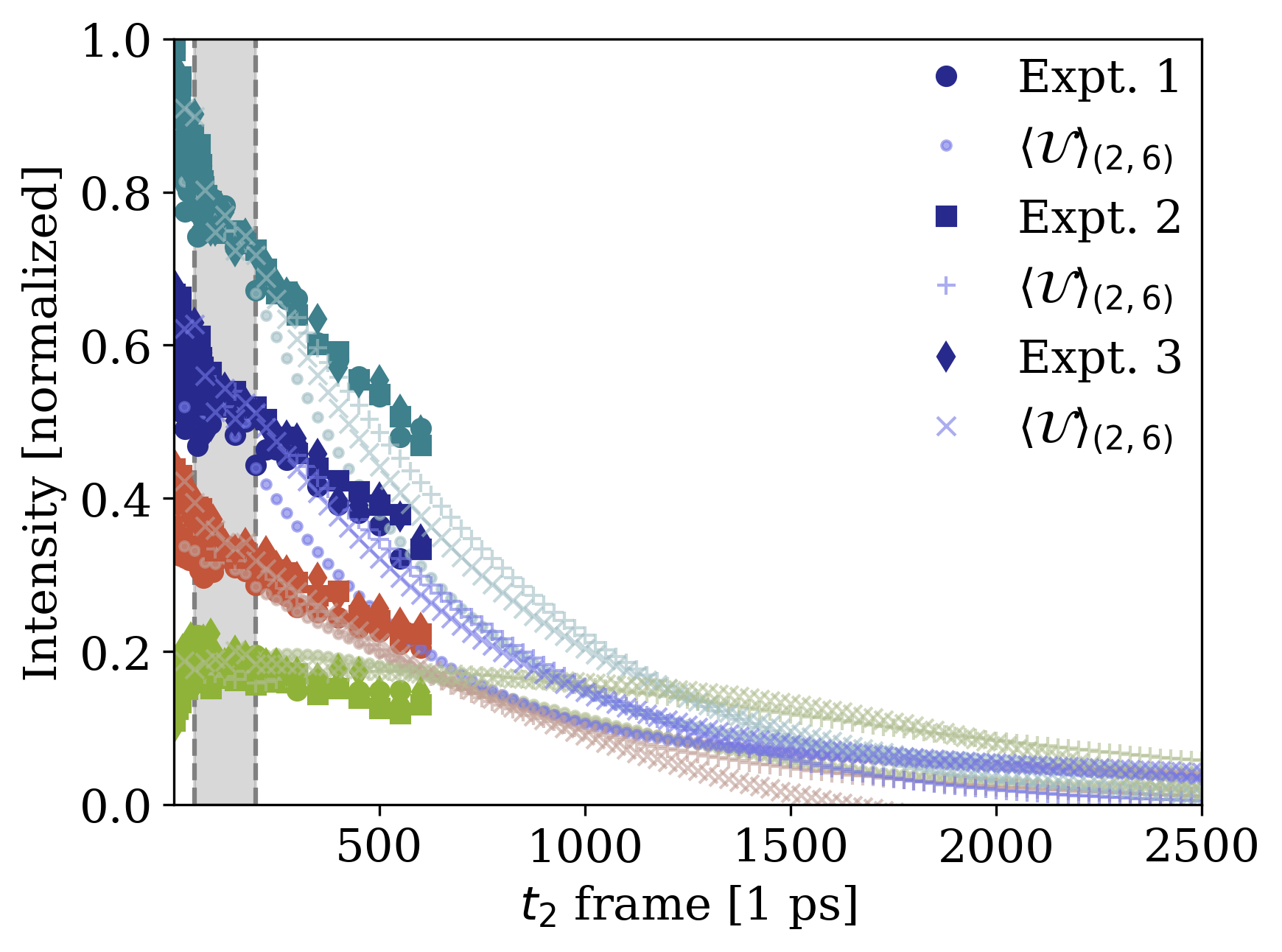}}
    \caption{}
    \label{fig:cy3_repeats_same}
 \end{subfigure}
 
 \begin{subfigure}[t]{\linewidth}
    \centering
    \resizebox{0.6\textwidth}{!}{\includegraphics{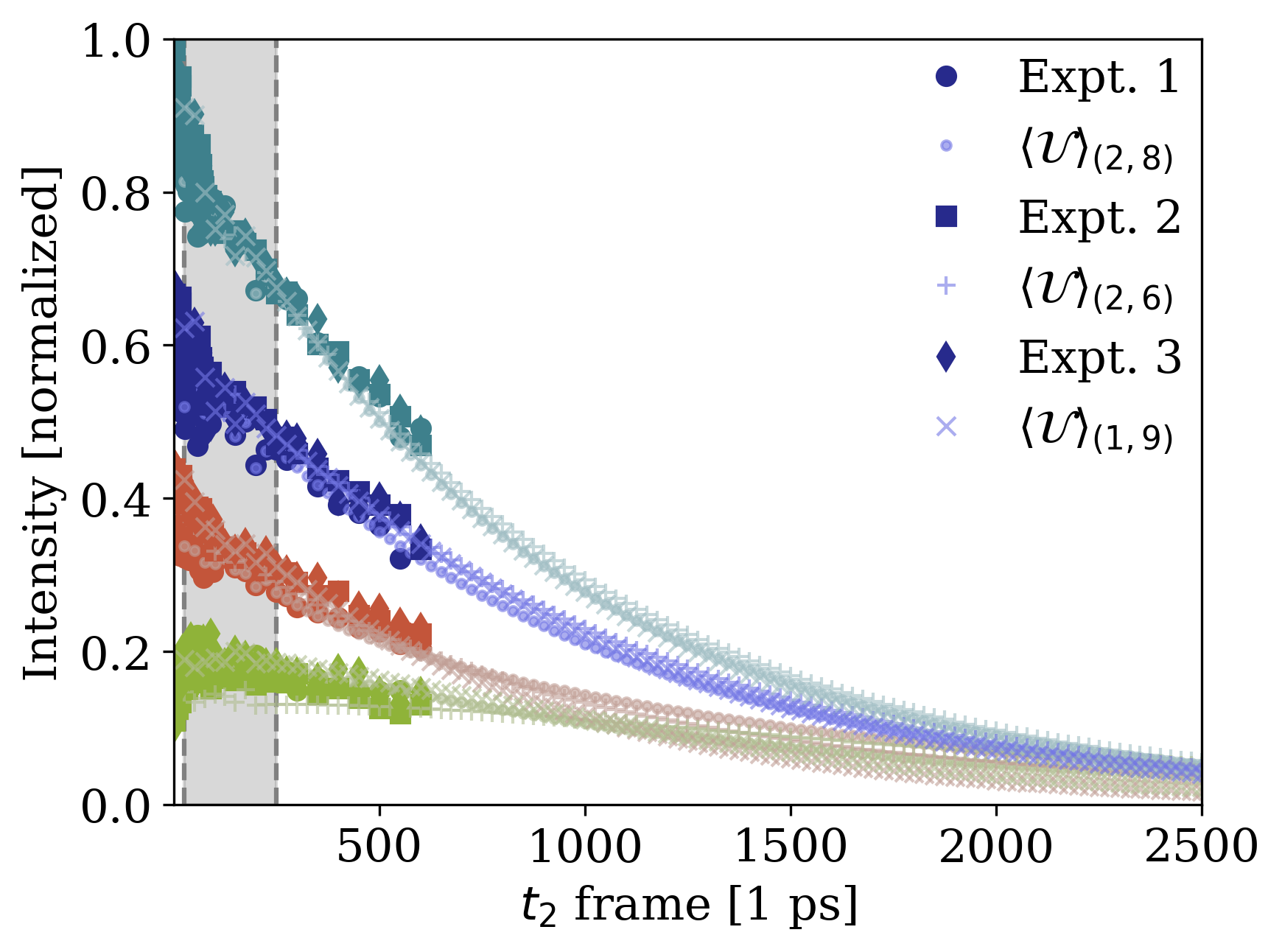}}
    \caption{}
    \label{fig:cy3_repeats_optimized}
 \end{subfigure}
 
 \label{fig:cy3_repeats}
 \caption{Analysis of two repeats of the experiment analyzed in Fig.~2 of the main text. (a) All settings are equal with an SVD threshold of 0.01. The SVD error in Expt.~2 is large, and the Expt.~3 prediction has a qualitatively incorrect (temporarily) increasing intensity for the green pixel. (b) SVD, cutoff, and window are optimized for each experiment. The SVD thresholds are 0.01, 1, and 0.02, respectively.}
\end{figure*}

In this section, we evaluate the consistency of the spectral GME predictions across two additional independent (experimental replicates) 2DES experiments on the Cy3-Cy5 dimers from Fig.3 of the main text. Specifically, we assess whether the output of our spectral GME constructed on one experiment is commensurate with: (1) the outcomes of the other experiments, and (2) with spectral GMEs constructed from the other two experiments on the same system. We show that our spectral GME provides consistent and accurate results across replicates, which reinforces its promise for microscopy and high-throughput experiments.

We begin by building the propagator $\ev{\bmc{U}_\infty (\delta t)}$ with our parameters in Fig.~3 ($\xi=0.01$, $t_u=50$~ps, $\tU=200$~ps) and illustrate the corresponding dynamics in Fig.~\ref{fig:cy3_repeats_same}(a). These pixel dynamics disagree with the experimental measurements for times beyond $\tU$. While these data are experimental replicates, the experiments are quantitatively different and thus require fine-tuning of these parameters. Nevertheless, the identification of the parameters for the first experiment offer a good starting point for fine tuning and expedite the search for optimal parameters in the experimental replicates. 

The three separate experiments display scatter in the plotted pixels; at early times teal, blue, and red pixels are systematically lower for Expt.~1, while the other two are closely clustered. Although not shown in Fig.~\ref{fig:cy3_repeats_same}, the un-normalized raw spectra in the second experiment have relative intensities of the tracked pixels relative to the maximum of only about $60\%$. In Expt.~2, the SVD error is also large. Nevertheless, Fig.~\ref{fig:cy3_repeats_optimized} shows consistent results when we follow the protocol to optimize $\{ t_u, \tU, \xi \}$. By optimizing $\xi$ alone to achieve better agreement with the experimental data before any cutoffs (which is possible in Expt.~3, but only partly successful in Expt.~2), performance improves markedly. However, we obtain the best results by also performing small optimizations to $t_u$ and $\tU$ for each replicate. Through these optimizations, we find proximal solutions that exhibit nearly quantitative agreement across all three replicates (see Fig.~\ref{fig:cy3_repeats_optimized}). The fact that these principled changes result in nearly identical predictions across experiments, even when the differences in the underlying data are so evident, reinforces the robustness of our spectral GME.

%This is because, despite being replicates, the experiments are quantitatively different and require that one perform a shortened version of the optimization workflow on each. 

%-------------------------------------------------------------------------
%-------------------------------------------------------------------------
\section{Pre-interpolation results when the SVD cutoff introduces error}\label{app:interpolation}
\begin{figure}[!h]\label{fig:lukedata_none}
        \resizebox{.6\textwidth}{!}{\includegraphics{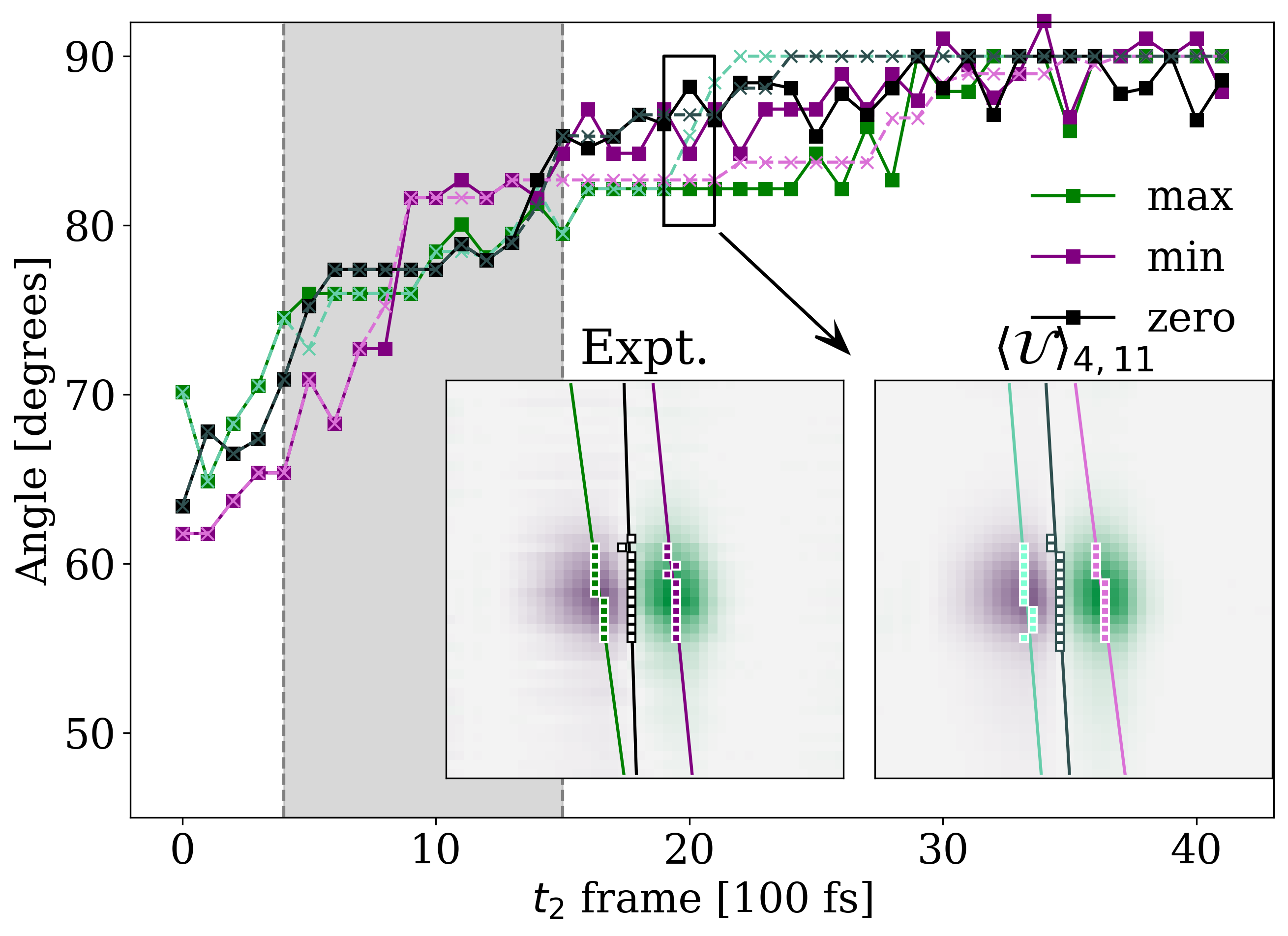}}
    \caption{Same data and analysis as Fig.~5 of the main text, but without any interpolation. The raw data window is a $44 \times 44$ pixel grid.}
\end{figure}

While interpolation aids in the interpretation of CLS data and was also found to improve our ability to predict the dynamics, the spectral GME's performance remains good even without interpolation. That is, we still obtain convincing---although not fully quantitative---CLS predictions for features described at the original resolution of half as many pixels, which we present in Fig.~5 of the main text. This lends further credence to the reliability of the spectral GME, as well as its compatibility with performance augmentations that align with common practices in the analysis of multidimensional spectra.

\FloatBarrier

\bibliography{sgme-refs}